%% file: HC_review4.tex
\documentclass[hyperref]{ws-ijmpa}
\input{HC_proleg}

\begin{document}


%
%

\title{Higgs boson couplings: measurements and theoretical interpretation}

\author{Chiara Mariotti\footnote{chiara.mariotti@cern.ch}}

\address{INFN, Sezione di Torino, Italy}

\author{Giampiero Passarino\footnote{giampiero@to.infn.it}}

\address{Dipartimento di Fisica Teorica, Universit\`a di Torino, Italy\\
         INFN, Sezione di Torino, Italy}

\maketitle

\begin{abstract}
 \noindent
This report will review the Higgs boson properties: the mass, the total width and the couplings 
to fermions and bosons.
The measurements have been performed with the data collected in $2011$ and $2012$
at the LHC accelerator at CERN by the ATLAS and CMS experiments.
Theoretical frameworks to search for new physics are also introduced and discussed.

\keywords{Higgs physics; Standard Model; Effective Field Theory.}

\end{abstract}

\ccode{12.60.-i{}11.10.-z{}14.80.Bn}

 \clearpage
 \small
 \thispagestyle{empty}
 \tableofcontents
 \normalsize
 \clearpage

%

%
\input{NHC_intro}

\section{Higgs boson phenomenology: production and decay \label{Sect1}}
\input{NHCS1}

\section{From discovery to properties \label{Sect2}}
\input{NHCS2} 
\section{Analysis of the measurements \label{Sect3}}
\subsection{The final states \label{Sect31}}
\input{NHCS31}

\section{The original kappa-framework \label{Sect4}}
\input{NHCS4}
\section{Results from Run I \label{Sect5}}
\input{NHCS5}

\subsection{The measurement of the mass \label{Sect51}}
\input{NHCS51}

\subsection{On-shell results \label{Sect52}}
\input{NHCS521}

\input{NHCS522}
\subsection{Off-shell results, experimental constraints on the width \label{Sect53}}
\input{NHCS53}
\section{Theoretical developments \label{Sect6}}
\input{NNHCS6}
\section{Prospects for Run II \label{Sect8}}
%
\input{HC_S8}

\input{HC_run2_exp}

\section{Conclusions \label{Conc}}
\input{HC_conc}


\begin{Acknowledgments}
\noindent
We acknowledge important discussions with Tiziano Camporesi, Andr{\`e} David
and Gino Isidori.
\end{Acknowledgments}

 \clearpage
\bibliographystyle{ws-ijmpa}
\bibliography{HC_review4}

\end{document}

%% file: NHC_intro.tex
\section{Introduction \label{Intro}}
In the Standard Model (SM) when the electroweak symmetry is broken via the so-called 
Brout-Englert-Higgs mechanism (BEH) 
\cite{Englert:1964et,Higgs:1964pj,Guralnik:1964eu,Higgs:1966ev,Kibble:1967sv},  
vector bosons and fermions acquire mass and a new elementary particle with spin zero and 
positive parity appears: the Higgs boson.       
The ATLAS and CMS collaborations (the two general-purpose experiments at LHC) announced in 
July $2012$ the observation of a new resonance in diphoton and $4\,$-leptons final states with 
a mass around $125\UGeV$, whose properties are, to date, compatible within the large 
uncertainties with the Higgs boson predicted by the SM \cite{Aad:2012tfa,Chatrchyan:2012xdj}.

The Higgs boson production and decay rates measured by ATLAS and CMS give a combined signal yield, 
relative to the Standard Model (SM) prediction, of $1.09{\pm}0.11$~\cite{Khachatryan:2016vau}.
The Higgs boson mass is very precisely measured, several decay modes have been
observed with high significance (the $\PAQb\PQb\,$-mode is not far from reaching the sensitivity 
to be observed at 13 $\UTeV$ center of mass energy).
Gluon fusion and vector-boson fusion production modes have been observed, and  $\PV\PH$ and
$\PAQt\PQt\PH$ are not too far to reach the sensitivity to be observed.

The early discovery is certainly based on two pillars: experimental analysis improvements
and theory accuracy improvements. To understand how the last two conspired to allow for the 
Higgs discovery see \Brefs{Dittmaier:2011ti,Dittmaier:2012vm}.

The LHC data are consistent  with the SM predictions for all the parameterisations considered. 
Therefore, after the LHC Run~1, the SM of particle physics has been completed, raising its 
status to that of a full theory. 
However, despite its successes, this standard theory has shortcomings vis-\`a-vis cosmological 
observations. At the same time, there is presently a lack of direct evidence for new physics 
phenomena at the accelerator energy frontier.

No matter what the LHC will uncover in the future, understanding the Higgs boson properties 
is a pillar of the present paradigm. 
Direct searches, thus possibly new physics,  
and precision measurements will have to be consistent with each other. 
The need for a consistent theoretical framework in which deviations from the SM predictions 
can be calculated is necessary.
Such a framework should be applicable to comprehensively describe measurements in all sectors 
of particle physics, not only LHC Higgs measurements but also electroweak precision data, \etc
By simultaneously describing all existing measurements, this framework then becomes an 
intermediate step toward the next SM, hopefully revealing the underlying symmetries.

This report will review the measurements of the Higgs boson properties, mass, width and 
couplings to fermions and bosons, that were performed with the data collected in $2011$ and 
$2012$ (\ie the Run~1) at the LHC accelerator at CERN by the ATLAS and CMS experiment.
It will then introduce a theoretical framework to search for new physics, while measuring 
the Higgs couplings with high precisions.
 

%% file: NHCS1.tex

This Section is not intended to provide a complete phenomenological profile of the Higgs boson
but only to recapitulate few essential informations that are needed to understand the experimental
results and their theoretical interpretations.

At the LHC, the production of the SM Higgs boson occurs via the following processes, 
listed in order of decreasing cross section at $7{-}8\UTeV$ center-of-mass energy in
Tab.~\ref{Pxs}.
\begin{table}[hbt]
\tbl{Higgs boson production processes.}
{\begin{tabular}{@{}lll@{}} \hline\hline
gluon fusion production        & $\Pg\Pg\to \PH$            & Fig.~\ref{fig:feyn_ggFVBF},a \\
vector boson fusion production & $\PQq\PQq \to \PQq\PQq\PH$ & Fig.~\ref{fig:feyn_ggFVBF},b \\
associated production with a $\PW$ boson & $\PQq\PQq \to \PW\PH$ & Fig.~\ref{fig:feyn_prod},a \\ 
or with a $\PZ$ boson                    & $\Pp\Pp \to \PZ\PH$ & Figs.~\ref{fig:feyn_prod},a  \\
including a small~($\sim 8\%$) but less precisely known contribution from & $\Pg\Pg \to \PZ\PH$
& \ref{fig:feyn_prod},b and~\ref{fig:feyn_prod},c\\ 
associated production with a pair of top or bottom quarks & $\PQq\PQq, \Pg\Pg \to \PAQt\PQt\PH$ 
or $\to \PAQb\PQb\PH$ & Fig.~\ref{fig:feyn_ttH} \\
associated production with a single top quark & $\PQq\Pg \to \PQt\PH$ & Fig.~\ref{fig:feyn_tH} \\ 
\hline\hline
\end{tabular} \label{Pxs}}
\end{table}

\begin{figure}[hbt]
\begin{center}
{\begin{tabular}{cc}
\includegraphics[width=0.30\textwidth]{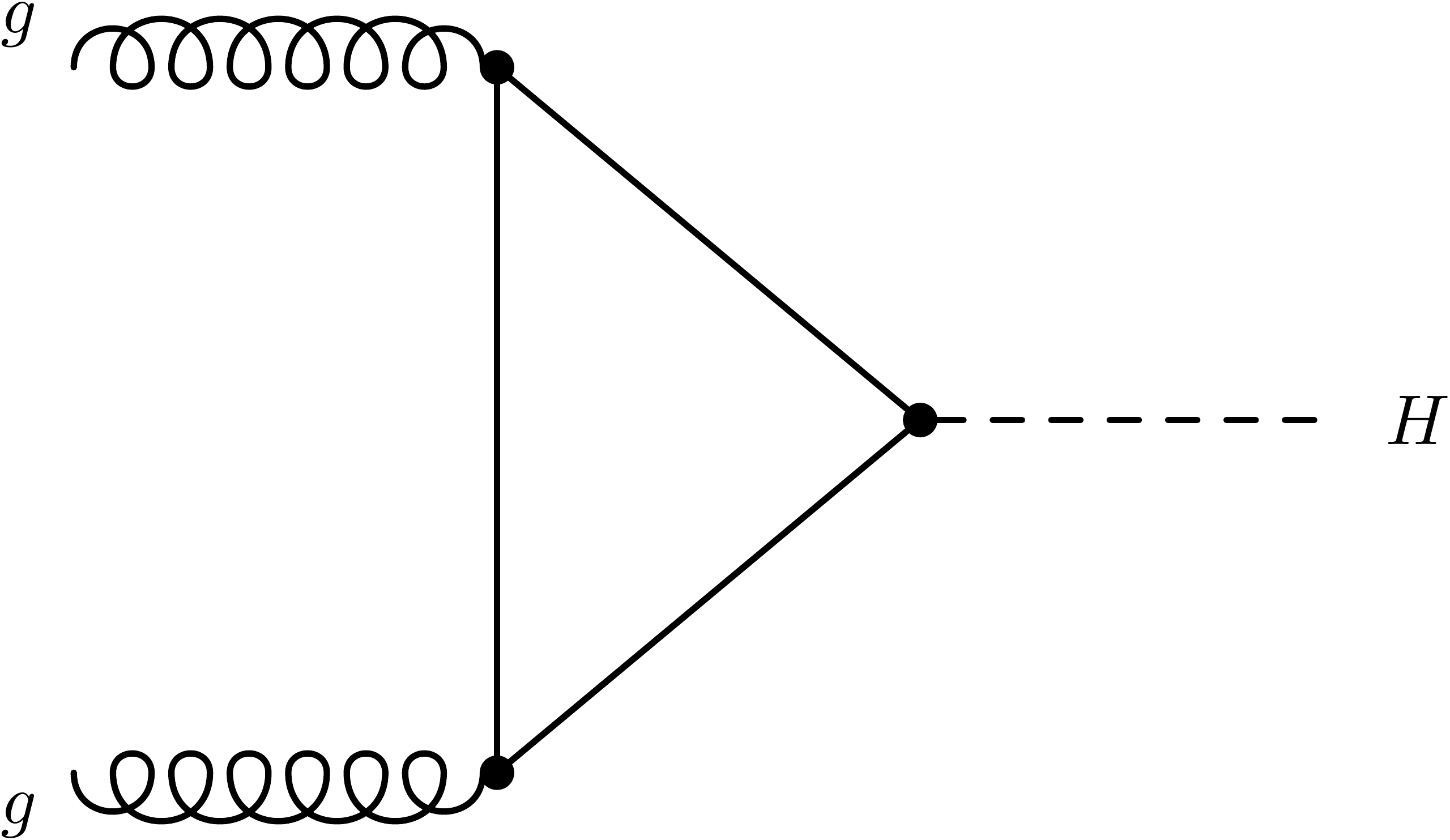} &
\includegraphics[width=0.30\textwidth]{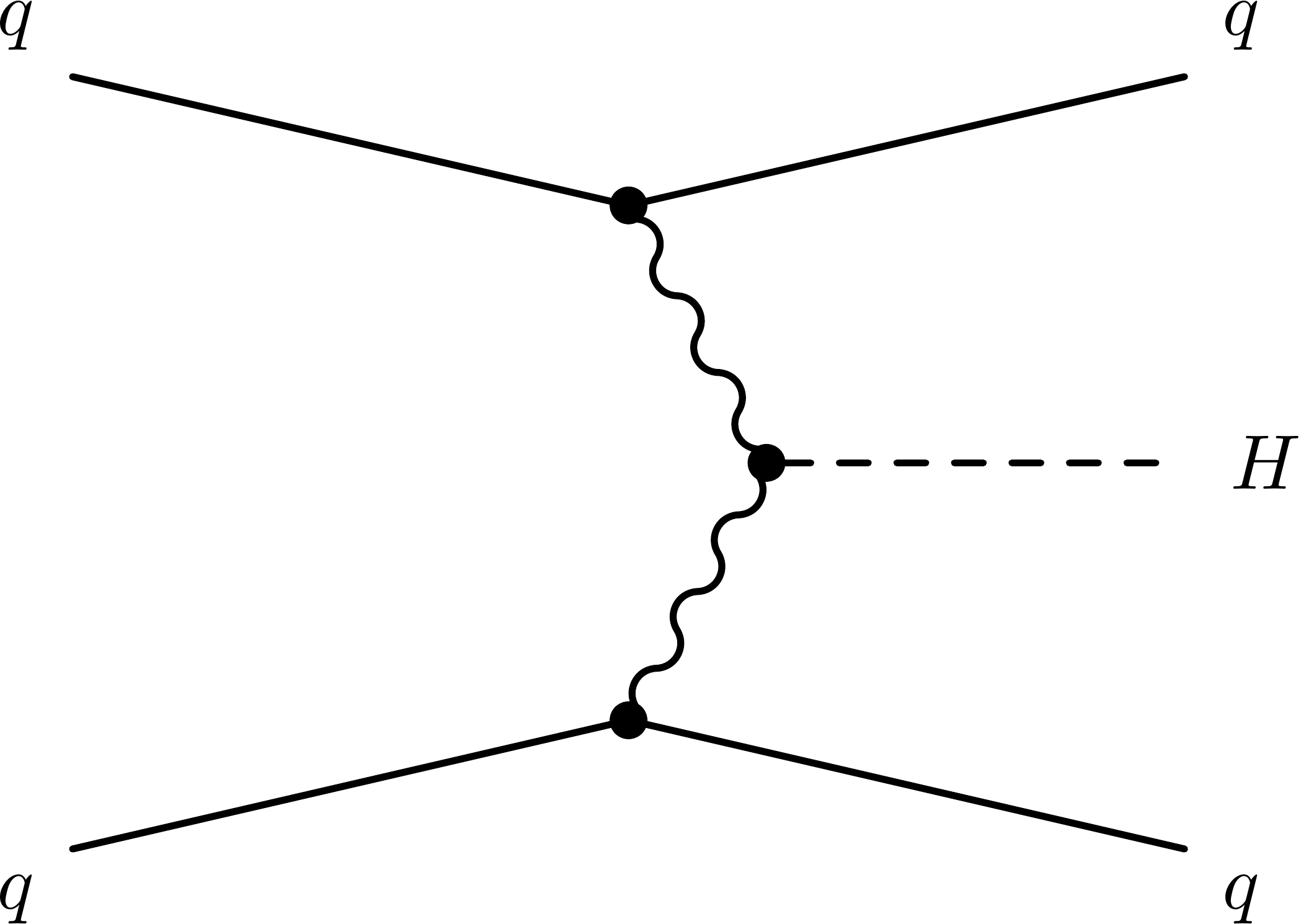}
\\
(a) & (b) 
\end{tabular}}
 \end{center}
\caption[]{Examples of leading-order Feynman diagrams for Higgs boson production via 
the (a) $\Pg\Pg\to \PH$  and (b) $\PQq\PQq \to \PQq\PQq\PH$ production processes.}
\label{fig:feyn_ggFVBF}
\end{figure}
\begin{figure}[hbt]
\begin{center}
{\begin{tabular}{ccc}
\includegraphics[width=0.30\textwidth]{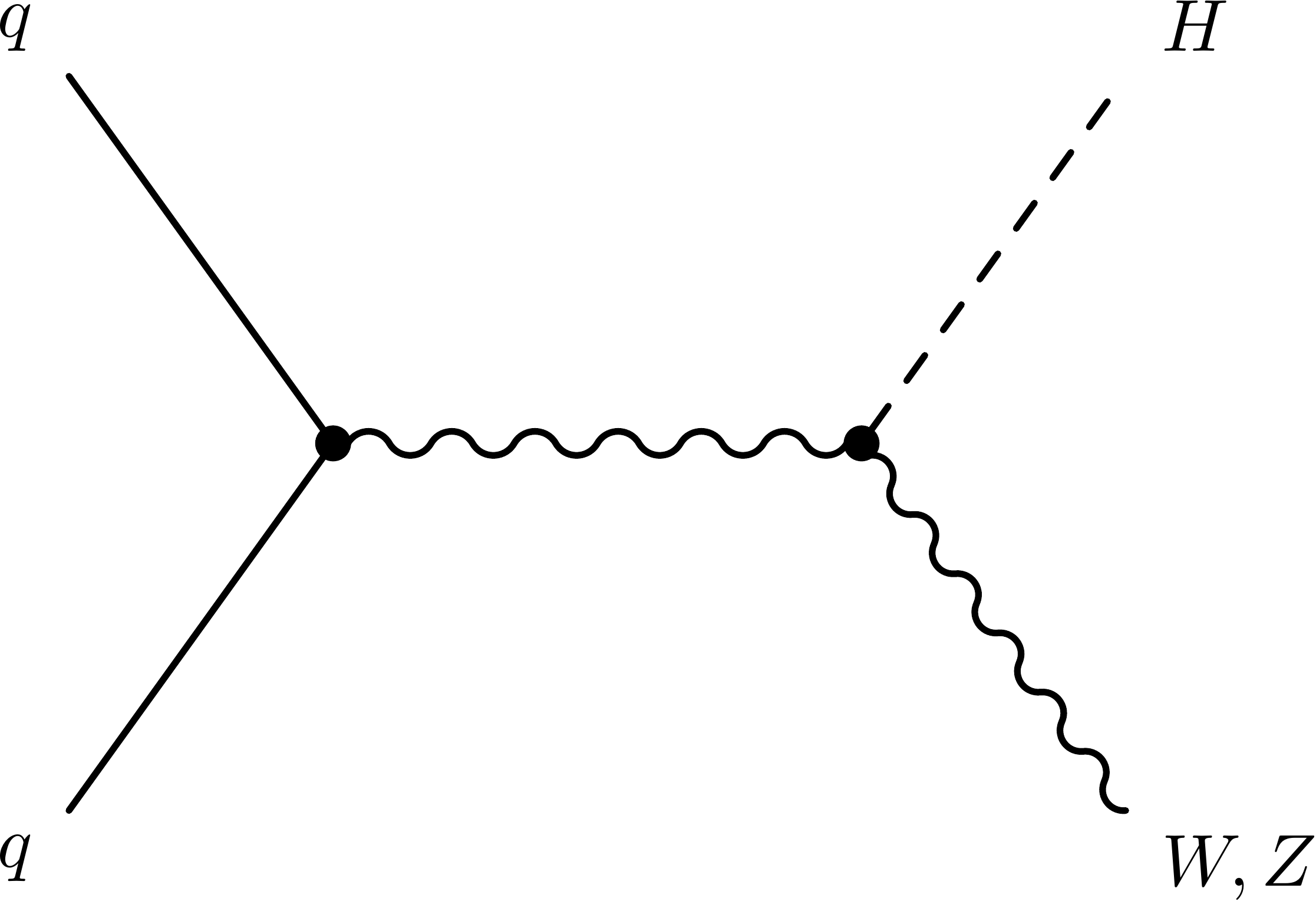} &
\includegraphics[width=0.30\textwidth]{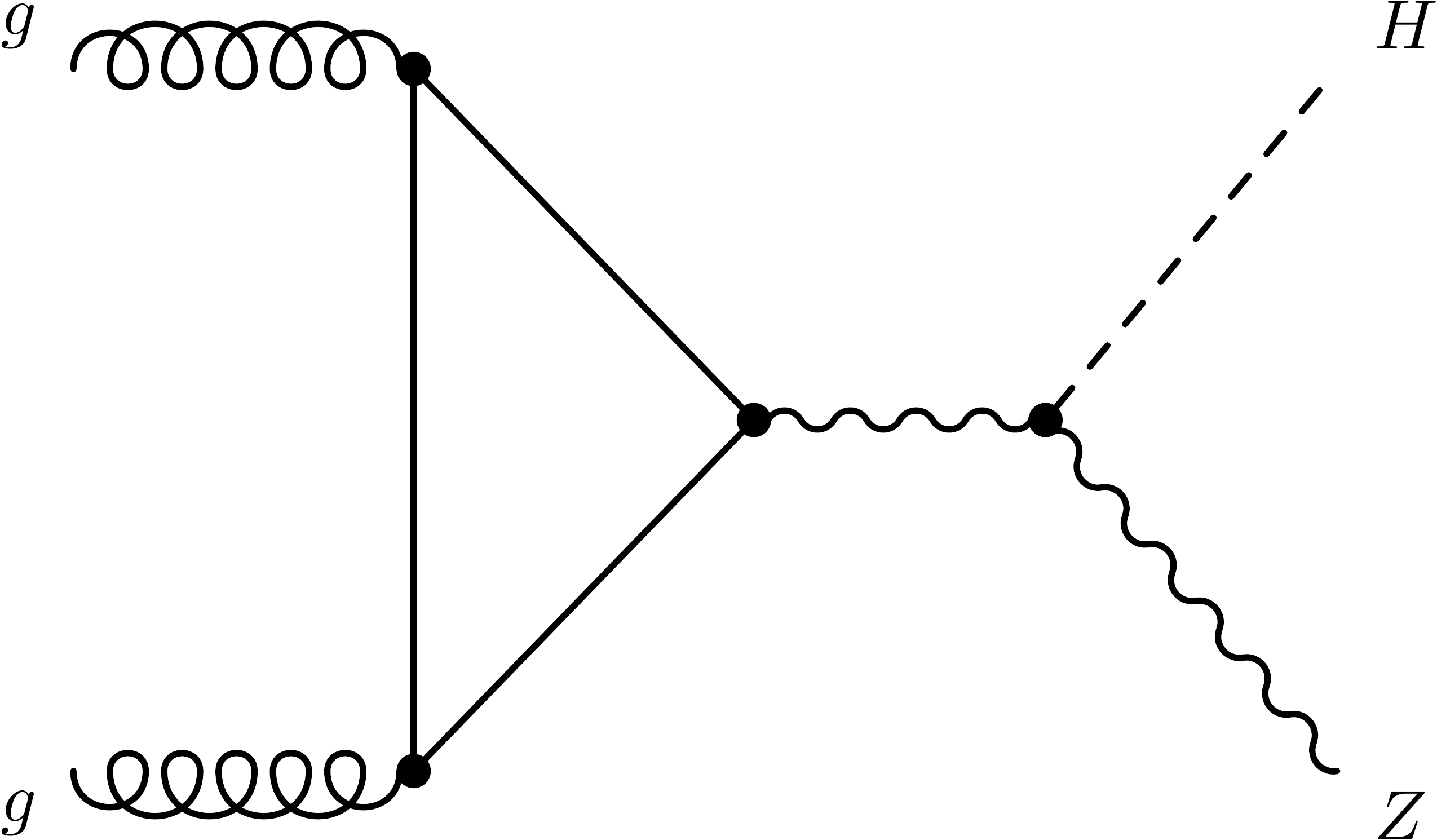} &
\includegraphics[width=0.30\textwidth]{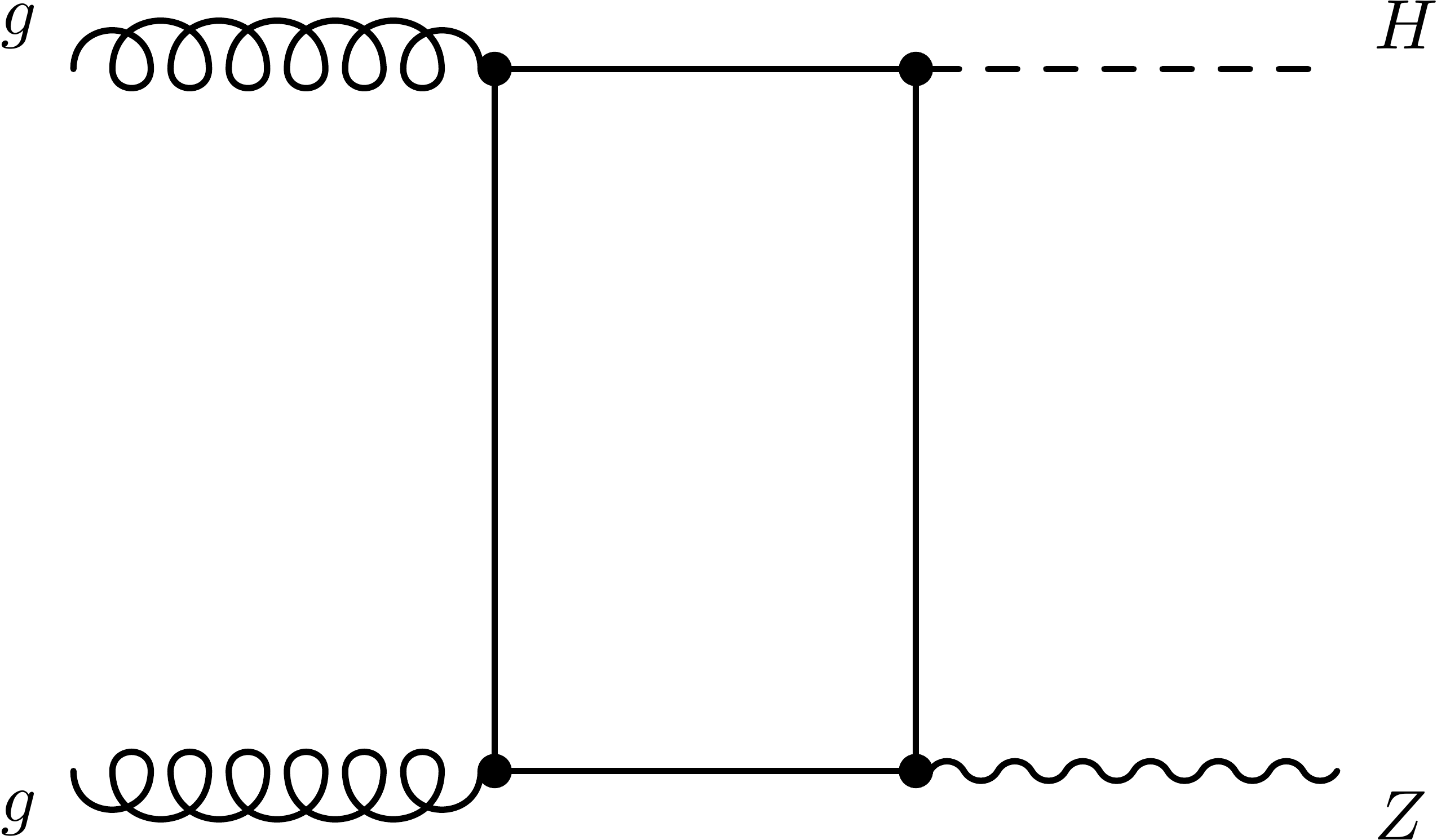}
\\
(a) & (b) & (c) 
\end{tabular}}
\end{center}
  \caption[]{Examples of leading-order Feynman diagrams for Higgs boson production via the (a) 
$\PQq\PQq \to \PV\PH$ and (b,~c)~$\Pg\Pg \to \PZ\PH$ production processes.}
\label{fig:feyn_prod}
\end{figure}
\begin{figure}[hbt]
\begin{center}
{\begin{tabular}{ccc}
\includegraphics[width=0.30\textwidth]{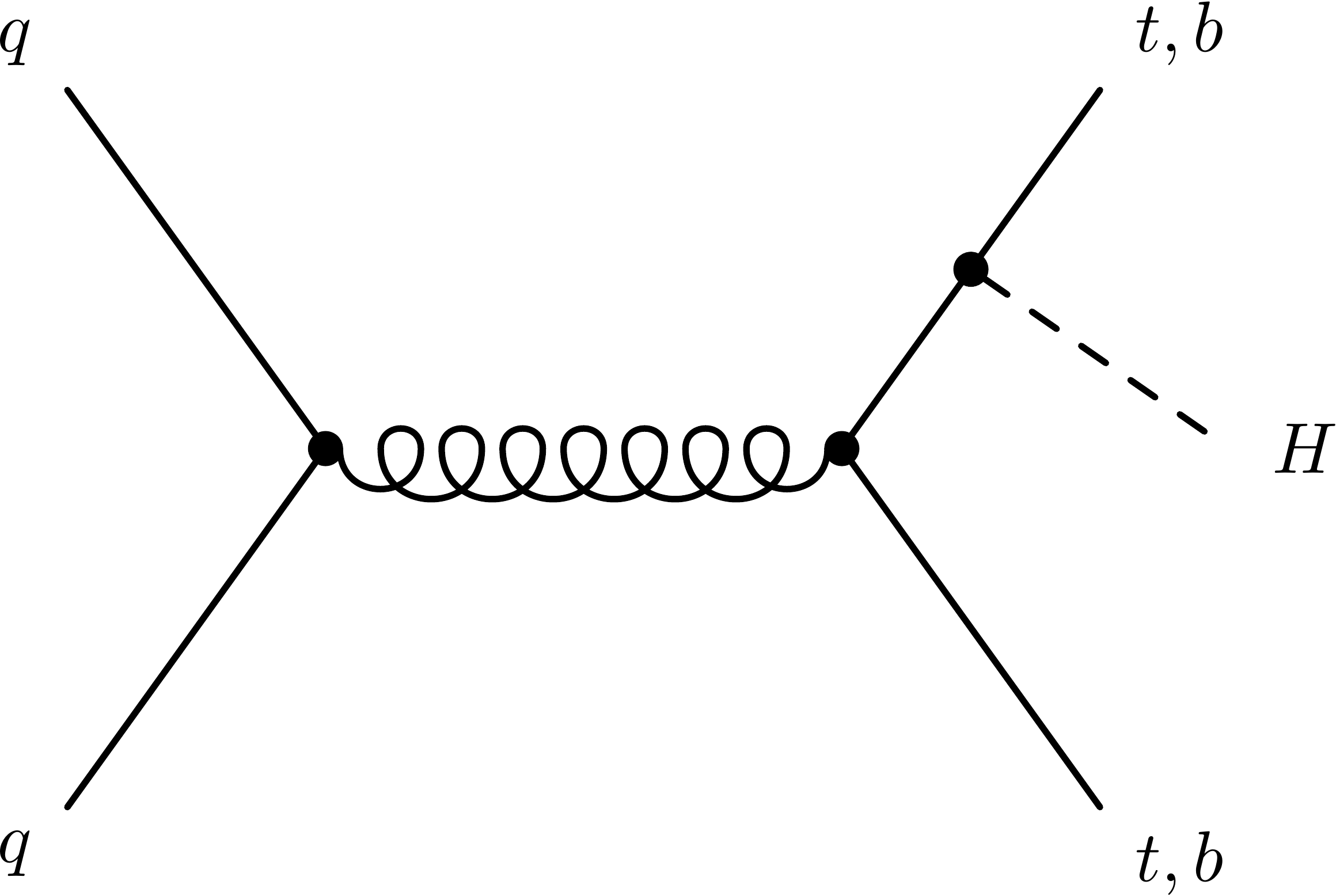} &
\includegraphics[width=0.30\textwidth]{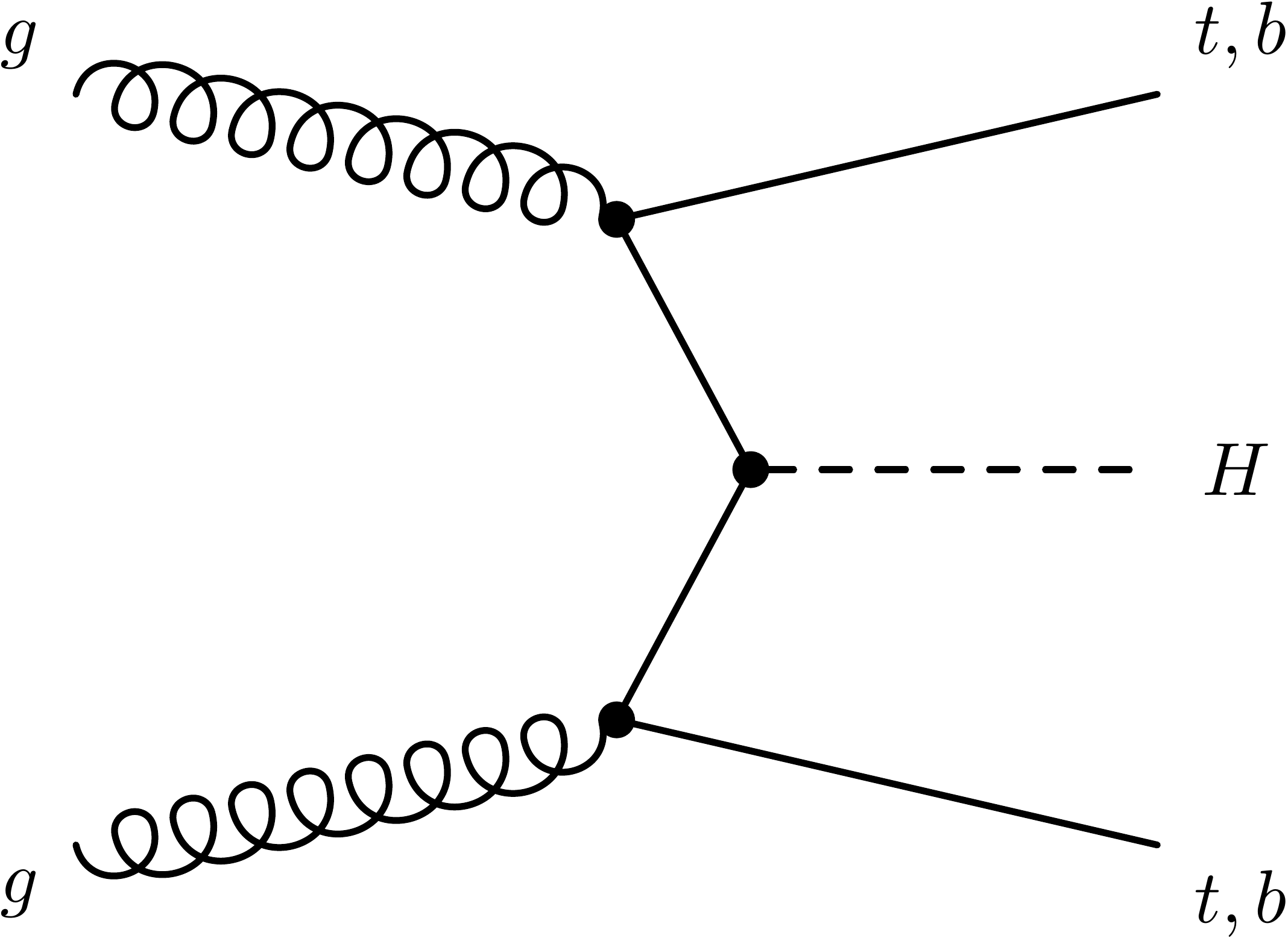} &
\includegraphics[width=0.30\textwidth]{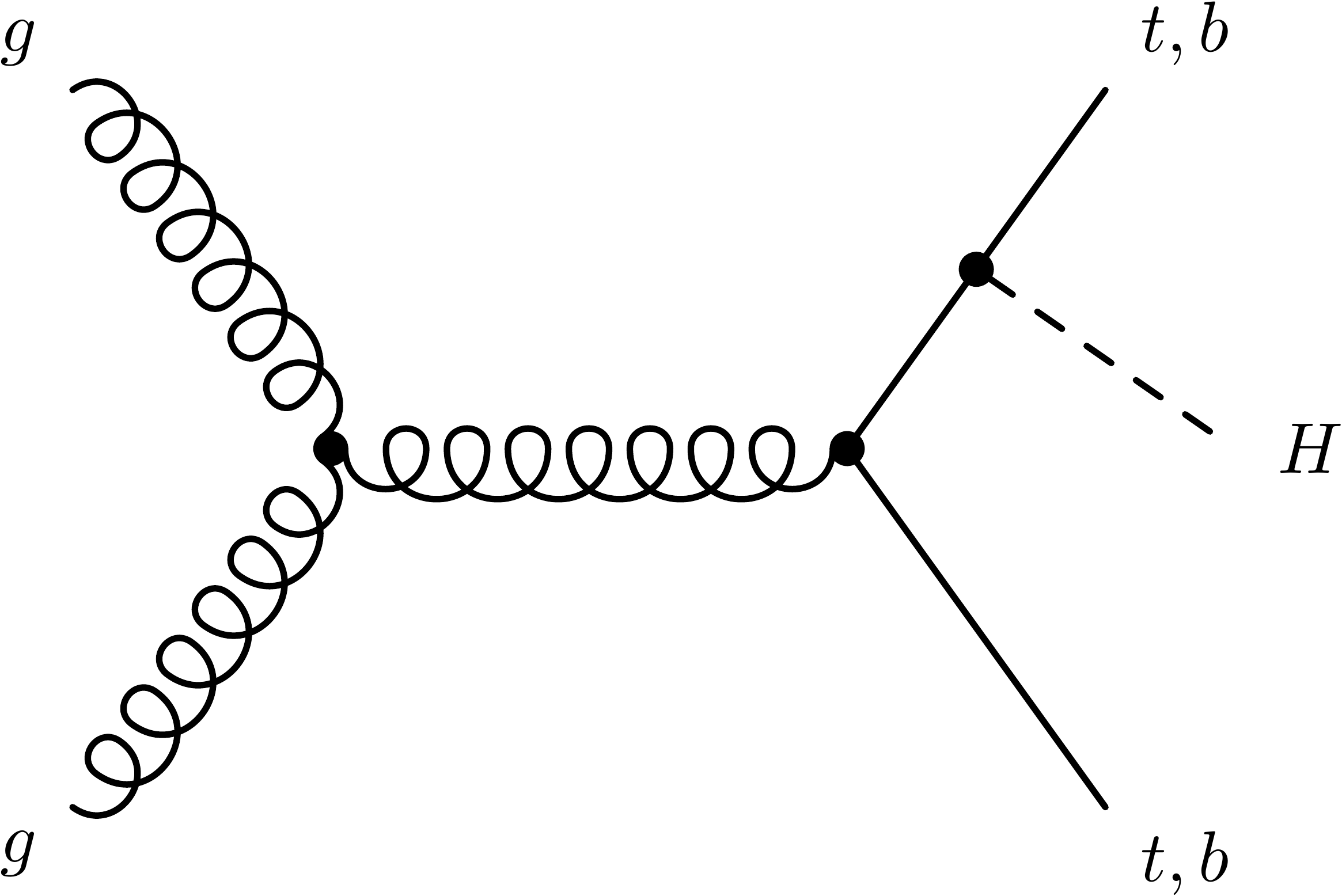} 
\\
(a) & (b) & (c) 
\end{tabular}}
\end{center}
\caption[]{Examples of leading-order Feynman diagrams for Higgs boson production in 
association with $\PAQt\PQt$ or $\PAQb\PQb$}
\label{fig:feyn_ttH}
\end{figure}
\begin{figure}[hbt]
\begin{center}
{\begin{tabular}{cccc}
\includegraphics[width=0.22\textwidth]{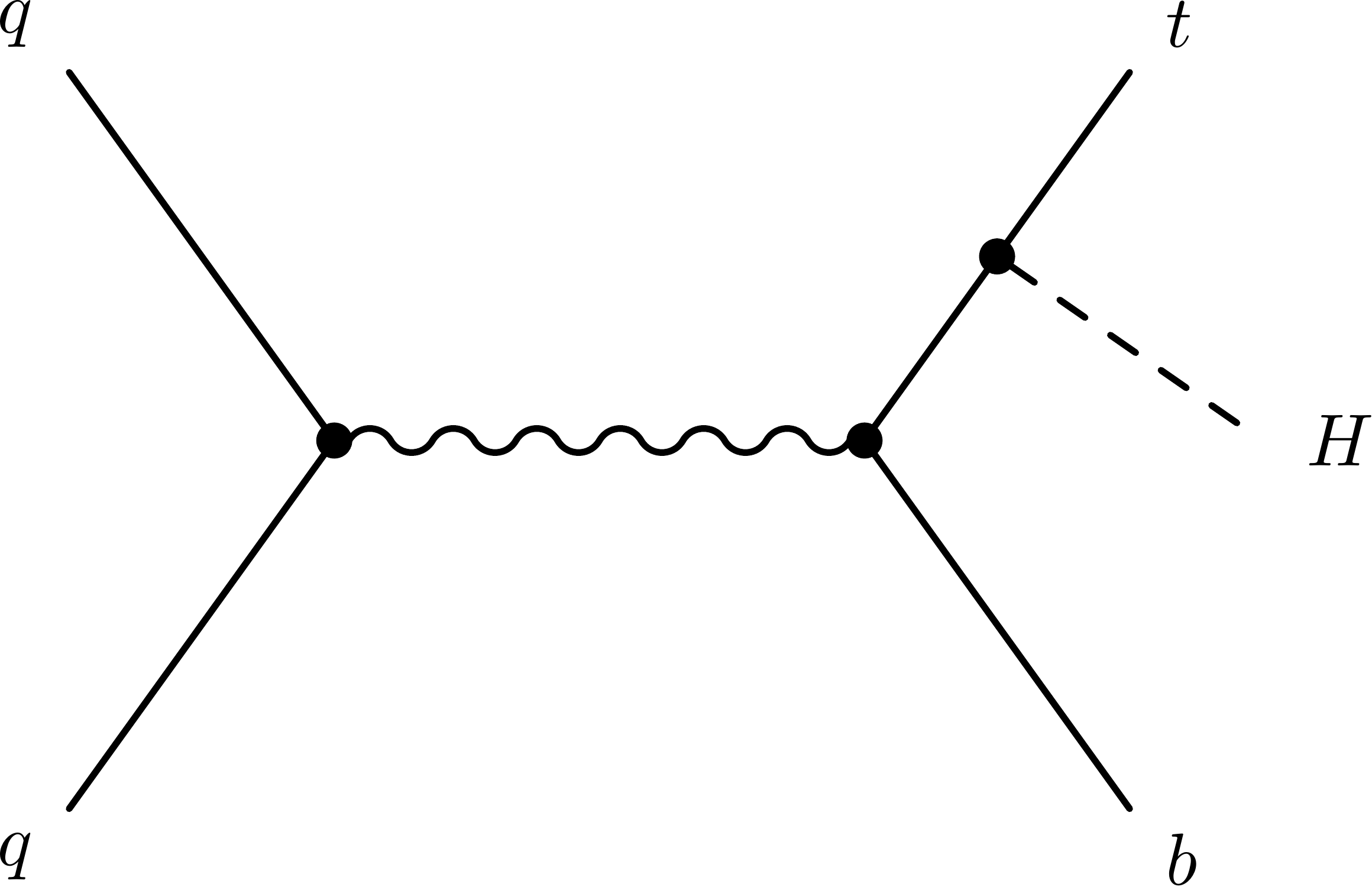} &
\includegraphics[width=0.22\textwidth]{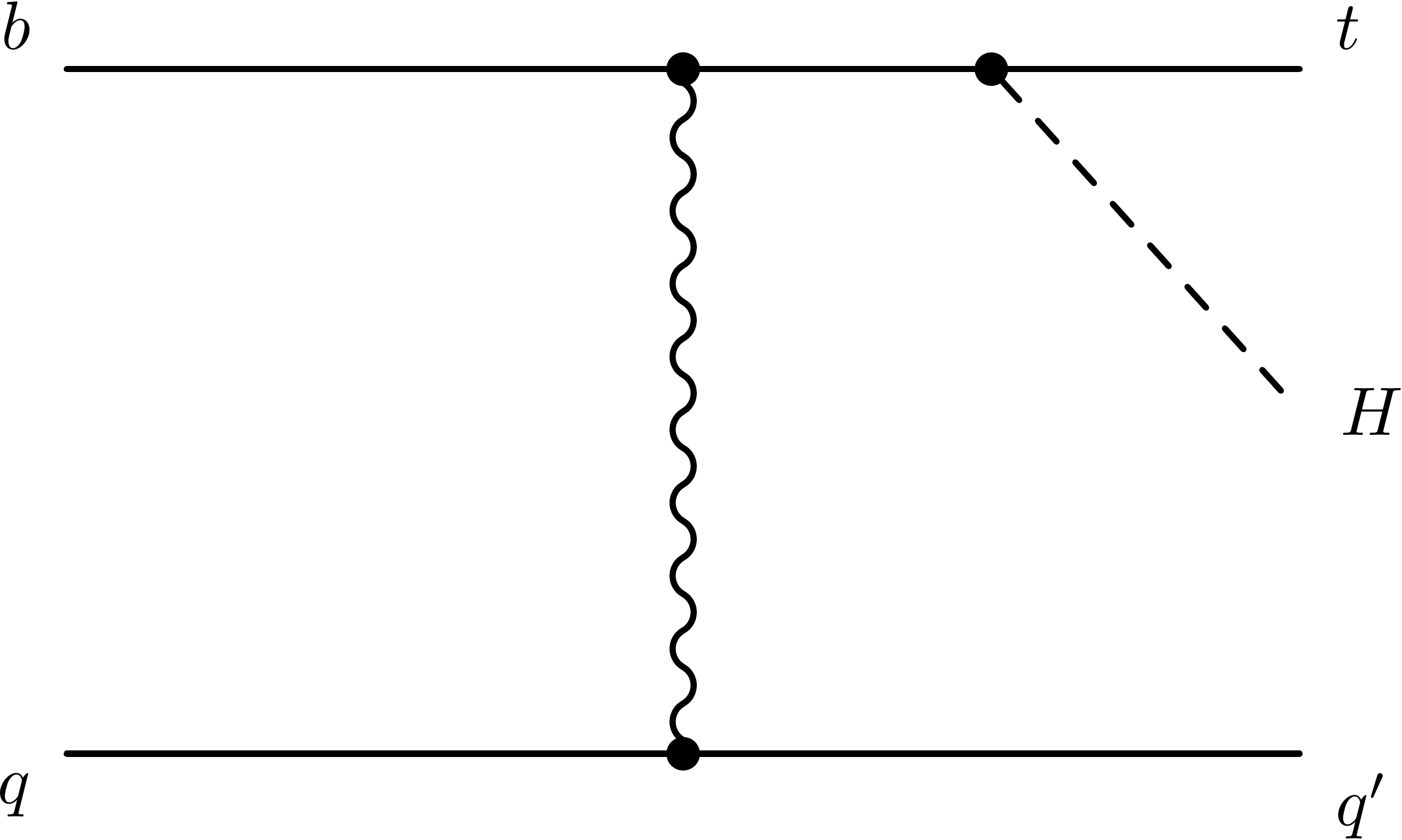} &
\includegraphics[width=0.22\textwidth]{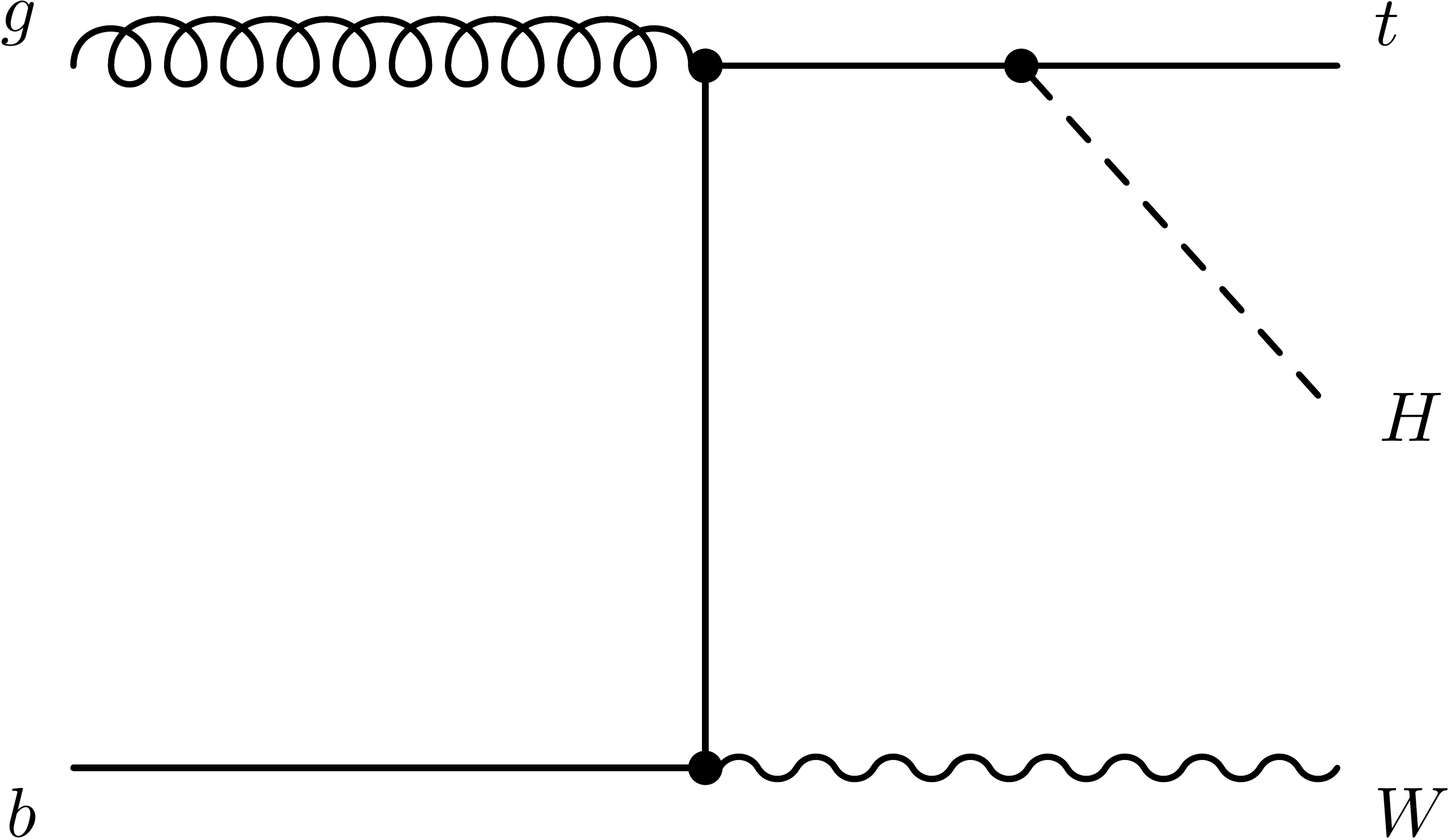} &
\includegraphics[width=0.22\textwidth]{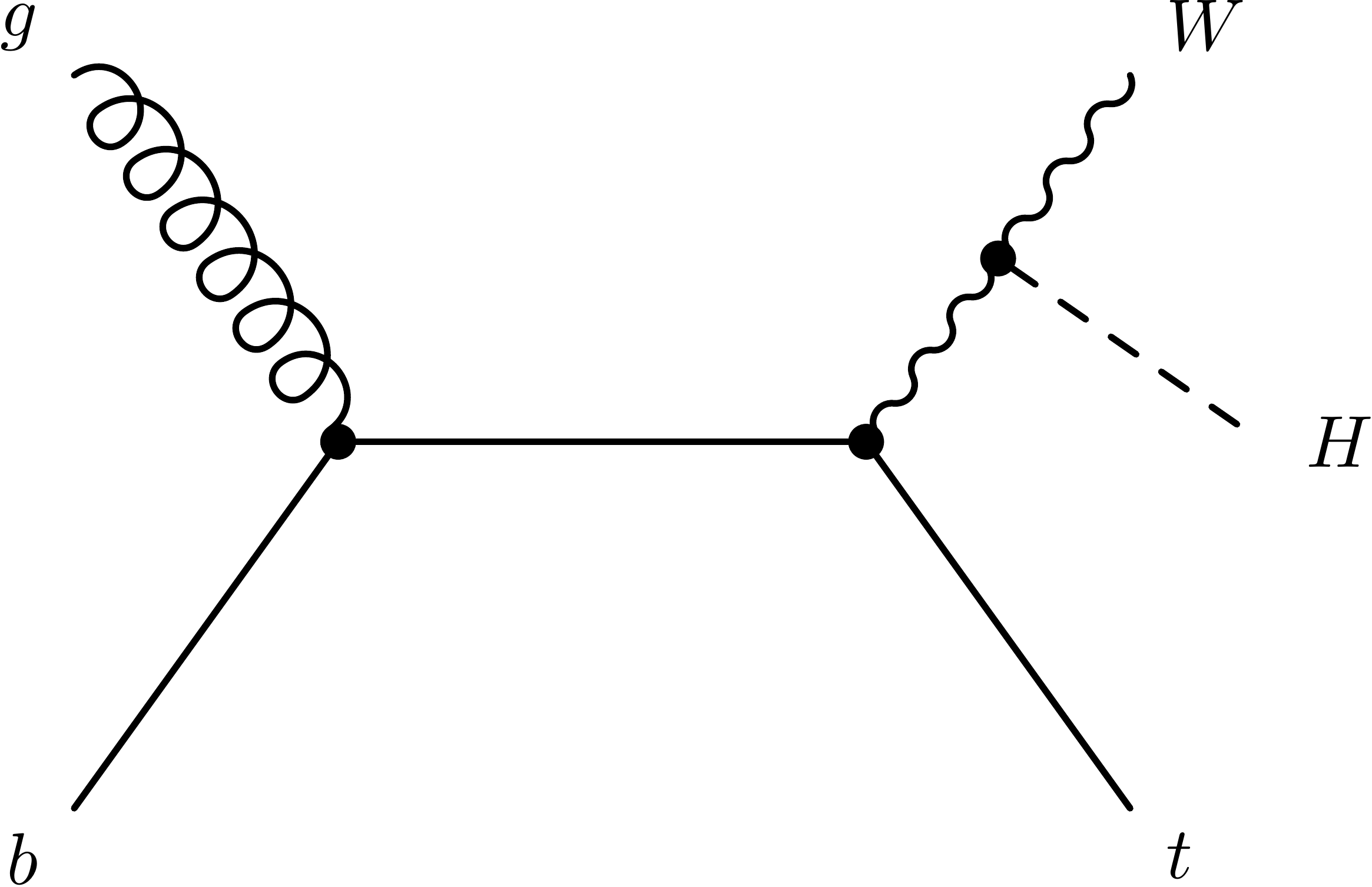} 
\\
(a) & (b) & (c) & (d)
\end{tabular}}
\end{center}
\caption[]{Examples of leading-order Feynman diagrams for Higgs boson production in 
association with a single top quark. }  
\label{fig:feyn_tH}
\end{figure}

The theory has an invariance based on the group 
$\lpar 2_{\ssL}\,\otimes\,2_{\ssR}\rpar = 1\,\oplus\,3$;
the Higgs field, $\PH$, is the custodial singlet contained in a scalar doublet $\upPhi$ 
of hypercharge $1/2$. The field $\Upphi$ develops a vacuum expectation value (VEV) and $\PH$ is 
the quantum fluctuation around the VEV. This has the consequence that the fermion masses and the 
Yukawa couplings are not independent quantities, \ie $\PH$ couples to a fermion-anti-fermion 
pair of the same flavour, with a coupling proportional to the mass of the fermion. Equivalently, 
$\PH$ couples to vector boson ($\PW\PW$ or $\PZ\PZ$) proportionally to their mass. This simple 
fact will have deep consequences when trying to build a model-independent framework for 
SM deviations.

In order to make the results of this Section transparent we stress the conceptual separation
between diagrams, amplitudes, and (pseudo-)observables. 

Diagrams, made of propagators and vertices, describe the couplings and the propagation and, in 
general, are not separately gauge invariant. 

(Sub-)Amplitudes are a set of diagrams, \eg the resonant (often called ``signal'') and the 
non-resonant (often called ``background'') parts of a physical process (which may contain more 
than one resonant part) and, once again, they are not separately gauge invariant. 

Finally, (pseudo-)observables are elements (or are related to elements) of the $\mrS\,$-matrix, 
\eg partial decay widths or production cross sections. Therefore, for a light Higgs boson it 
makes sense to talk about $\PH \to \PV\PV$ or $\PH \to \PAf\Pf$ couplings, but objects like 
partial decay widths, \eg  $\Gamma_{\PH \to \PV\PV}$, (forbidden by kinematics) can only be 
given and interpreted within a certain set of conventions. Actually, it is not only a question of 
kinematics, $\PH$, $\PW$ and $\PZ$ are unstable particles whose theoretical  treatment is far 
from trivial and presents a certain number of subtleties~\cite{Passarino:2010qk,Goria:2011wa}.
Finally, a SM Higgs boson has a very narrow width, more than four orders of magnitude smaller 
than its mass, which means that theoretical ``at the peak'' predictions are provided in the 
so-called ``zero-width-approximation'' (ZWA), equivalent to (on-shell) production cross 
section $\,\times\,$ (on-shell) decay.
In essence, the whole game in determining the Higgs couplings has to do with extracting
vertices from (pseudo-)observables.

Having that in mind, the results presented below have been obtained as 
described below~\cite{Dittmaier:2011ti,Dittmaier:2012vm,Heinemeyer:2013tqa}. The Higgs total width 
resulting from {\tt{HDECAY}}~\cite{Djouadi:1997yw} has been modified according to the prescription:
\bq
\Gamma_{\PH} = \Gamma^{\mathrm{HD}} - \sum_{\PV=\PW, \PZ}\,\Gamma^{\mathrm{HD}}_{\PV} +
\Gamma^{\mathrm{Pr}}_{4\Pf} \spc
\eq
where $\Gamma_{\PH}$ is the total Higgs width, $\Gamma^{\mathrm{HD}}$ the Higgs width obtained 
from {\tt{HDECAY}}, $\Gamma^{\mathrm{HD}}_{\PV}$ stands for the partial widths to $\PZ\PZ$ and 
$\PW\PW$ calculated with {\tt{HDECAY}}, while $\Gamma^{\mathrm{Pr}}_{4\Pf}$ represents the partial
width of $\PH \to 4\,\Pf$ calculated with {\tt{PROPHECY4F}}~\cite{Bredenstein:2007ec}. The latter 
can be split into the decays into $\PZ\PZ$, $\PW\PW$, and the interference,
\bq
\Gamma^{\mathrm{Pr}}_{4\Pf} = \Gamma_{\PH \to \PW^*\PW^* \to 4\,\Pf} +
\Gamma_{\PH \to \PZ^*\PZ^* \to 4\,\Pf} +
\Gamma_{\PW\PW/\PZ\PZ-\mathrm{intf}} \spp
\eq
Whenever $\PV^*$ appears it should be understood as follows: {\tt{PROPHECY4F}} calculations are
consistently performed with off-shell gauge bosons and they are valid above, near and below the 
gauge boson pair thresholds. For instance the definition is such that
\bq
\Gamma_{\PH \to \PW^*\PW^* \to 4\,\Pf} = 
 9\,\Gamma_{\PH \to \PGne\Pep\PGm\PAGnGm} +
12\,\Gamma_{\PH \to \PGne\Pep\PQd\PAQu} +
 4\,\Gamma_{\PH \to \PQu\PAQd\PQs\PAQc} \spp
\eq
These conventions are essential in understanding every statement of the form
``the $\PH$ decays to $\PW$ and $\PZ$ bosons \dots''.

A complete generalisation is represented by the LHC-PO (where PO stands for pseudo-observables)
framework~\cite{Gonzalez-Alonso:2015bha,Greljo:2015sla,Bordone:2015nqa,David:2015waa}; the idea 
of POs has been formalised the first time in the context of electroweak observables around the 
$\PZ$ pole at the LEP time \cite{Z-Pole}.
A list of LHC POs will be introduced and discussed in Sect.~\ref{Sect61}.
Suffice to mention here that the conditions defining POs ensure the generality of the approach
and the possibility to match it to a wide class of new physics (NP) models. 
In brief, POs are experimentally accessible, well-defined from the point of view of QFT and
capture all relevant effects of NP in the absence of new (non-SM) particles close to the Higgs
mass.

Another useful definition concerns the ``leading order'': technically speaking leading-order (LO)
defines the order in perturbation theory where the process starts. Notice that sometimes ``LO''
is used to denote tree level (as opposite to loops).

\begin{figure}[hbt]
\begin{center}
{\begin{tabular}{cc}
\includegraphics[width=0.30\textwidth]{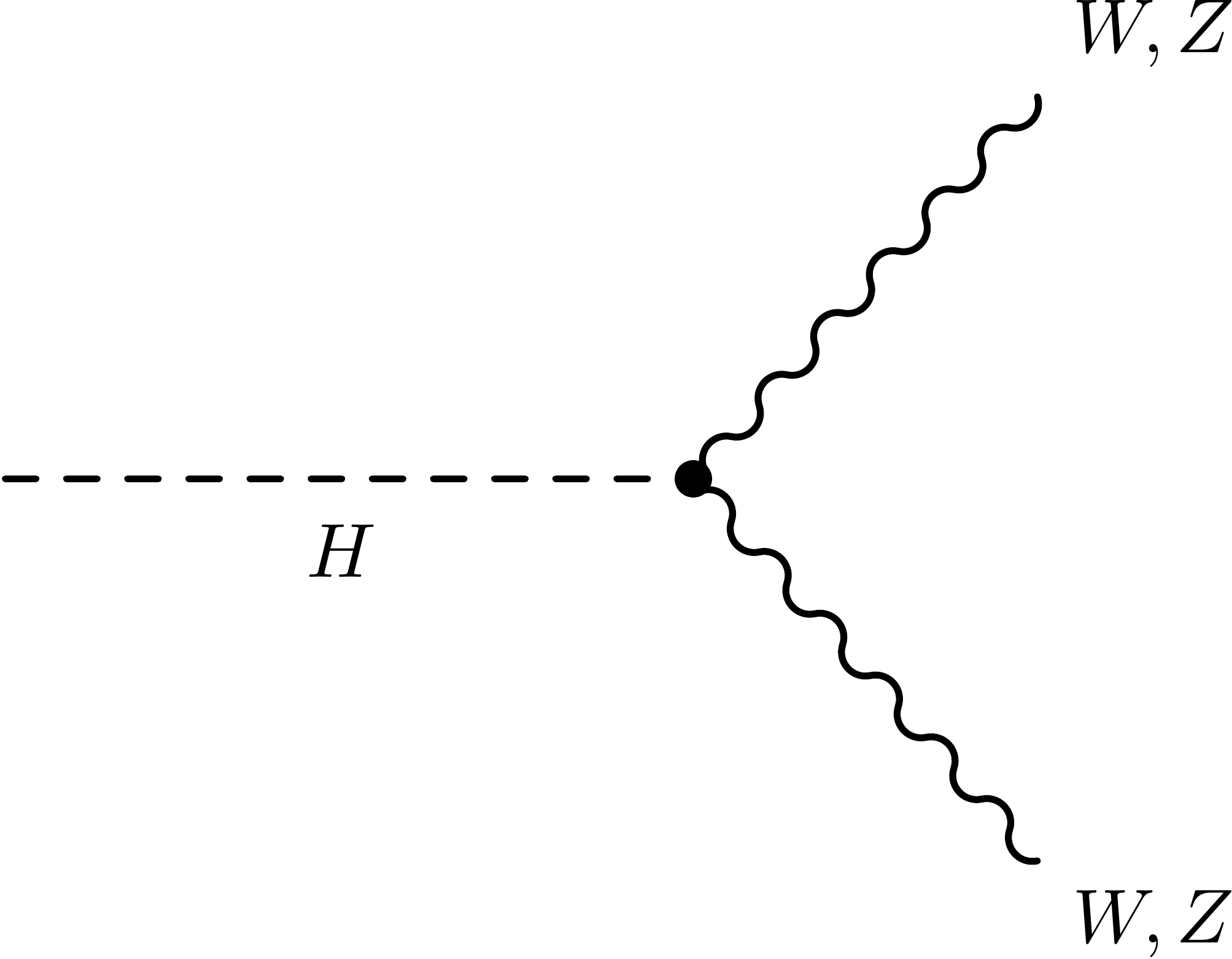} &
\includegraphics[width=0.30\textwidth]{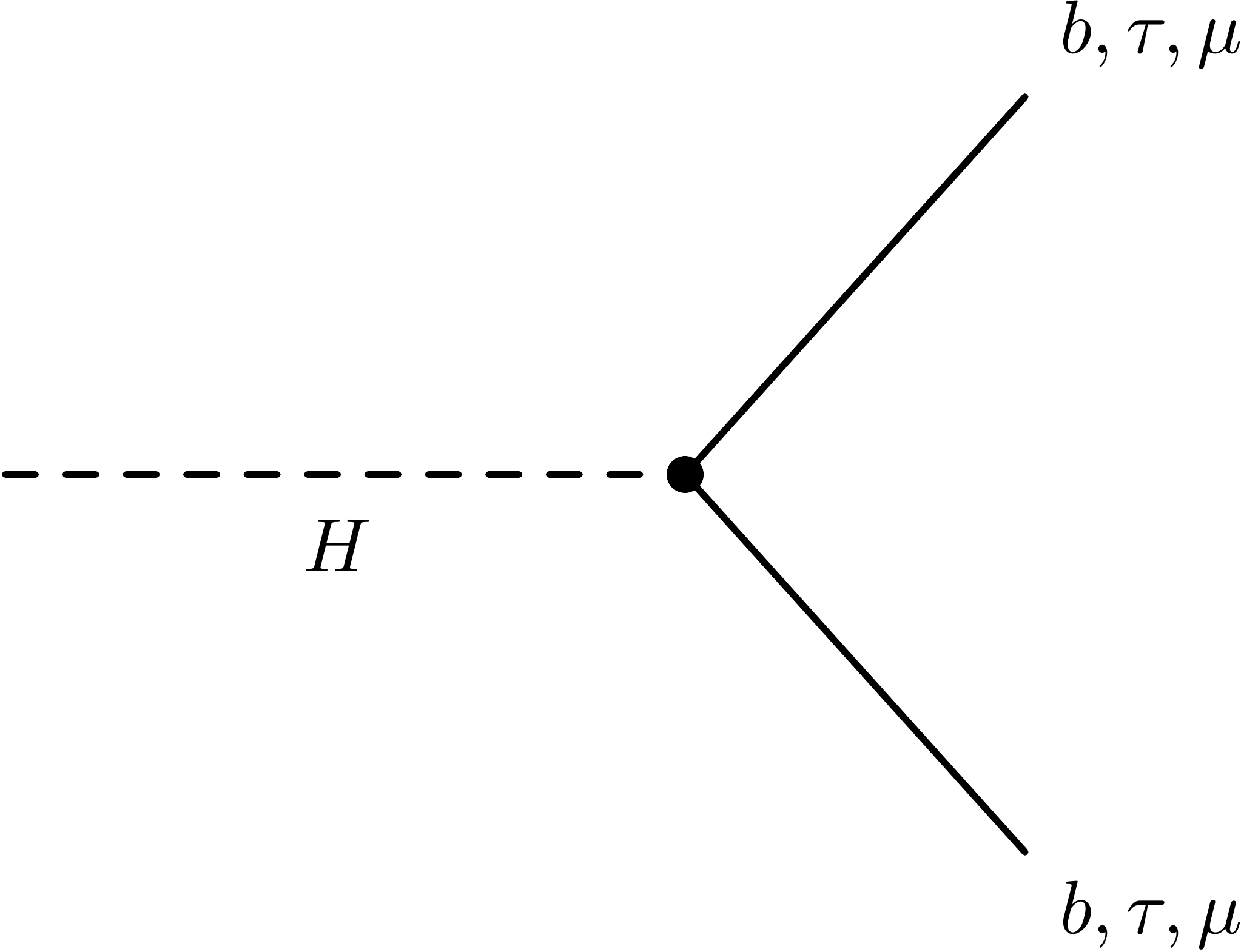}
\\
(a) & (b)
\end{tabular}}
\end{center}
\caption[]{Examples of leading-order Feynman diagrams for Higgs boson decays (a) 
to $\PW$ and $\PZ$ bosons and (b) to fermions ($\Pf=\PQb, \PGt, \PGm$).}
\label{fig:feyn_hVVff}
\end{figure}

\begin{figure}[hbt!]
\begin{center}
{\begin{tabular}{ccc}
\includegraphics[width=0.30\textwidth]{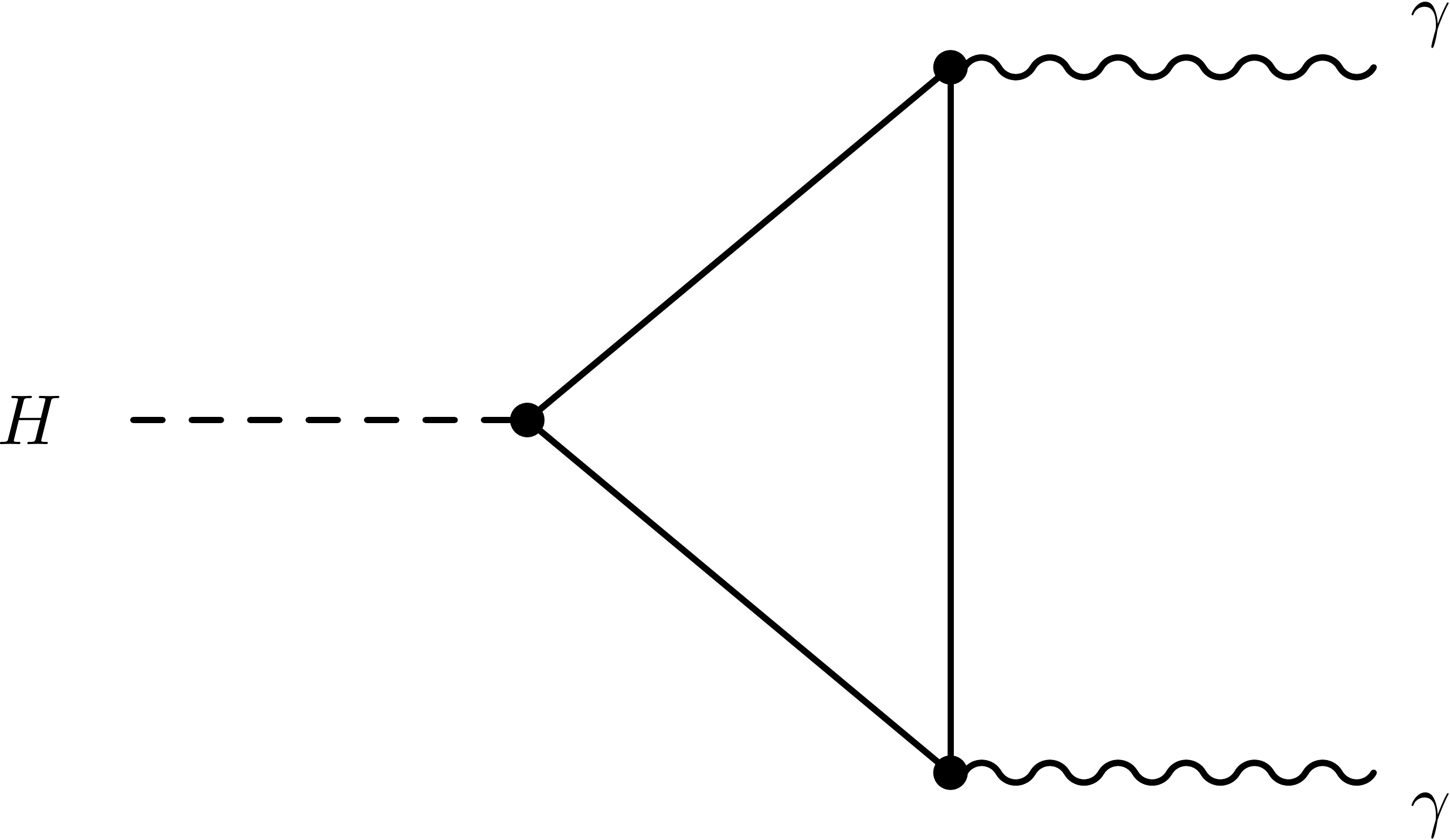} &
\includegraphics[width=0.30\textwidth]{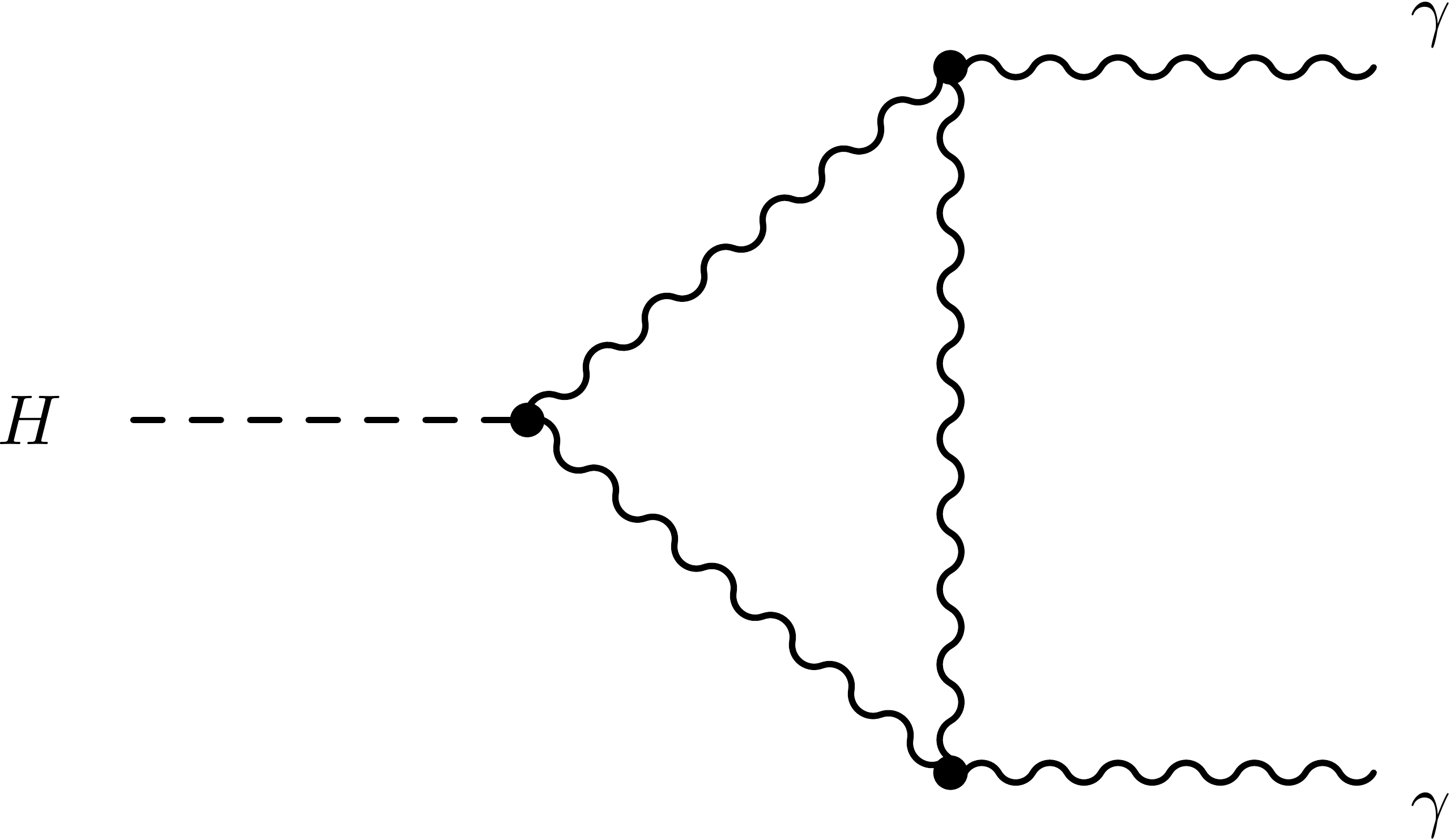} &
\includegraphics[width=0.30\textwidth]{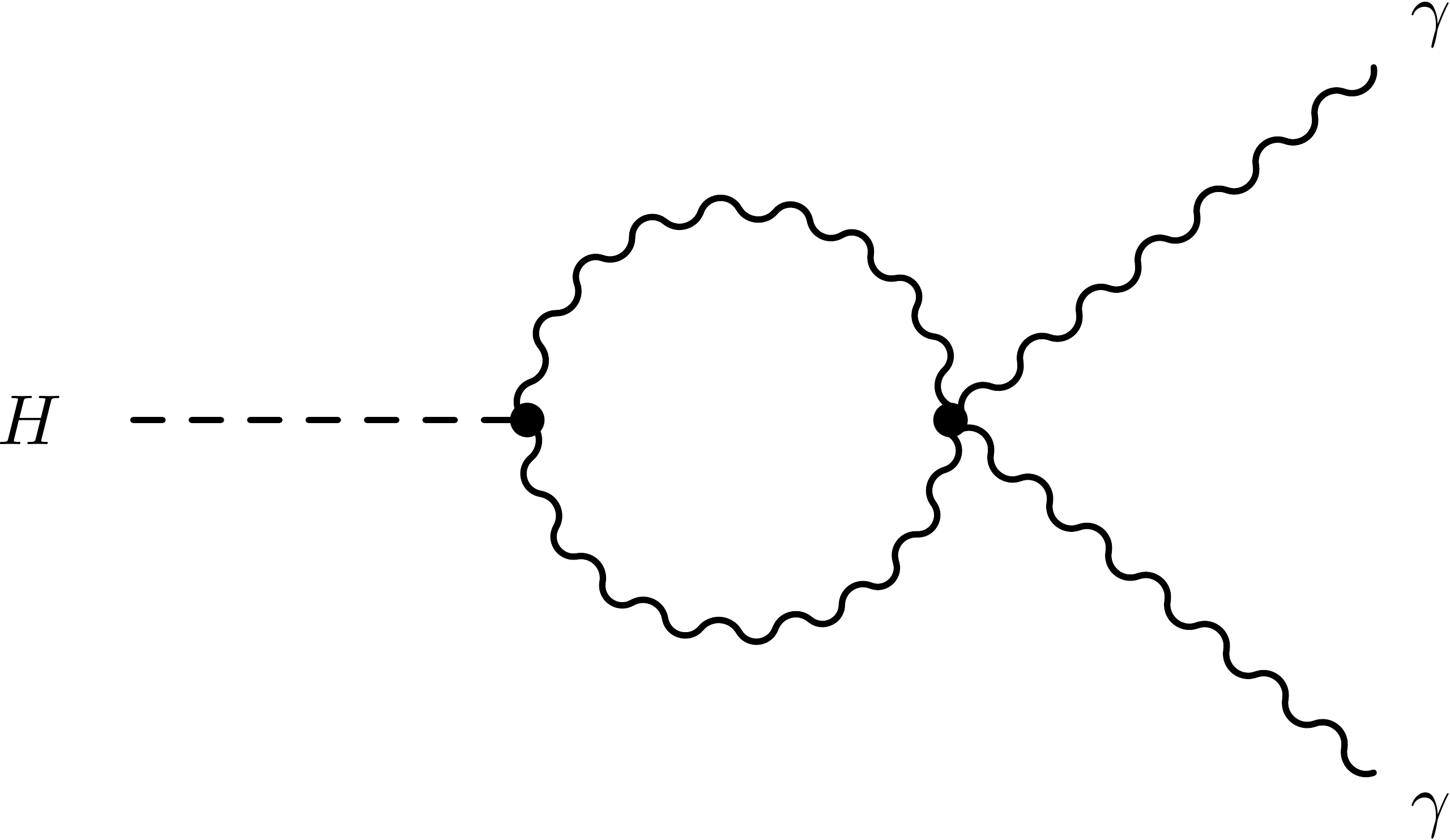}
\\
(a) & (b) & (c) 
\end{tabular}}
\end{center}
\caption[]{Examples of leading-order Feynman diagrams for Higgs boson decays to a pair of photons.}
  \label{fig:feyn_hgg}
\end{figure}

Examples of LO Feynman diagrams for the Higgs boson decays are shown in 
Figs.~\ref{fig:feyn_hVVff} and~\ref{fig:feyn_hgg}. The decays to $\PW$ and $\PZ$ bosons 
(see Fig.~\ref{fig:feyn_hVVff},a) and to fermions (see Fig.~\ref{fig:feyn_hVVff},b) start at tree level 
whereas the $\PH \to \PGg\PGg$ decay starts at one loop, being generated by loops containing 
heavy quark or bosons, see Fig.~\ref{fig:feyn_hgg}.

\begin{figure}[hbt]
\centering
\includegraphics[width=0.48\textwidth]{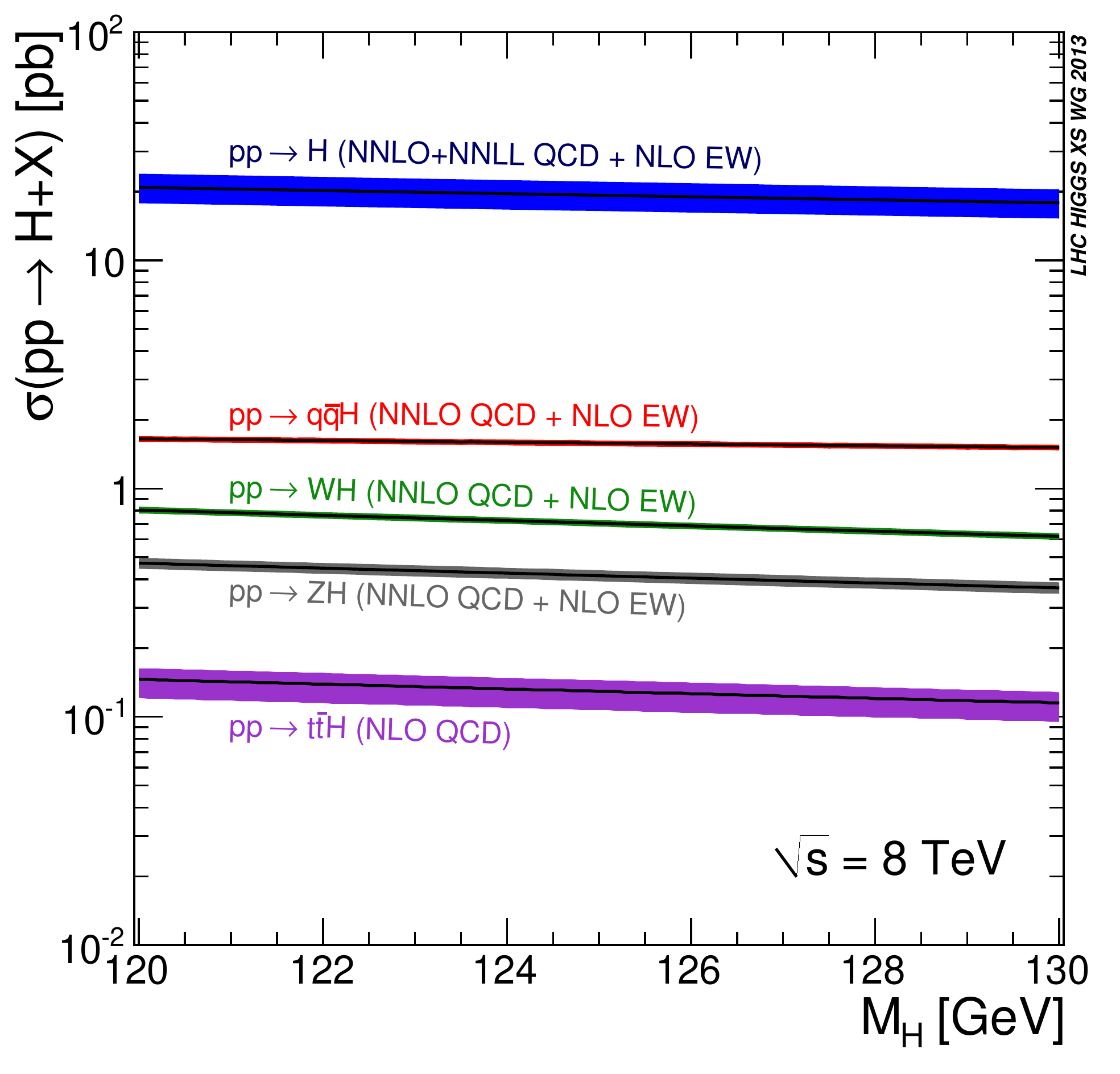}
\includegraphics[width=0.48\textwidth]{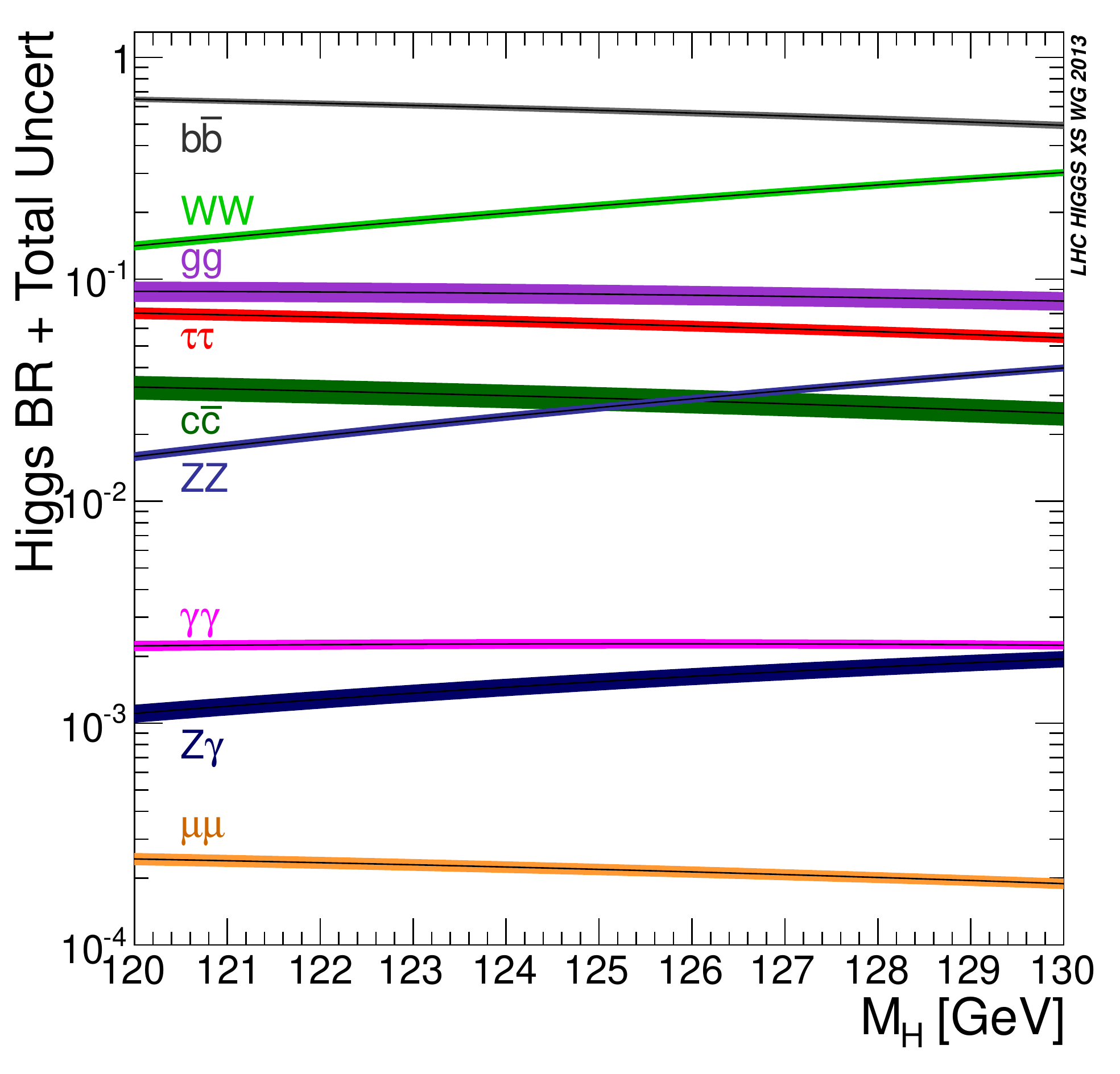}
\caption[]{Standard Model Higgs production cross sections and Branching Ratios at $8\UTeV$ 
center of mass energy }
\label{fig:figures_yr}
\end{figure}

The SM Higgs boson production cross sections and decay branching fractions have been calculated 
in the recent years at high order in perturbation theory.
The many calculations have been compared and then eventually combined and summarised in 
\Brefs{Dittmaier:2011ti,Dittmaier:2012vm,Heinemeyer:2013tqa}, and they are shown in 
Figs.~\ref{fig:figures_yr}. 
Following these calculations, additional and important progress has been made, and many more 
calculations have been performed at higher order, but they will not be reported here, since 
they have not been used in the analysis of the Run~1 data.
Once the mass of the Higgs boson has been measured, all its 
properties are fixed. 
The inclusive cross sections and branching fractions for the most important 
production and decay modes are summarised with their overall uncertainties in 
Tabs.~\ref{tab:SMCrossSections} and~\ref{tab:SMBranchingFractions} for a Higgs boson 
mass of $125.09\UGeV$, the value of the mass measured by the ATLAS and CMS 
collaboration with the Run~1 statistics \cite{Aad:2015zhl}.
\begin{table}[htbp]
\caption[]{Standard Model predictions for the Higgs boson production cross sections together 
with their theoretical uncertainties as of the year 2013. 
The value of the Higgs boson mass is  $m_{\PH}=125.09\UGeV$ as measured by the ATLAS and CMS 
experiments. The predictions  are from \Bref{Heinemeyer:2013tqa}.
The $\Pp\Pp \to \PZ\PH$~cross section, calculated at NNLO in QCD, includes  both the 
quark-initiated, \ie $\PQq\PQq \to \PZ\PH$ or~$\PQq\Pg \to \PZ\PH$, and the 
$\Pg\Pg\to \PZ\PH$ contributions.  
The uncertainties in the cross sections are evaluated as the sum in quadrature of the 
uncertainties resulting from variations of the QCD scales, parton distribution functions, and 
$\alpha_{\text s}$. 
The uncertainty in the $\PQt\PH$ cross section is calculated following the procedure of 
\Bref{Alwall:2014hca}. 
The order of the theoretical calculations for the different production processes is also 
indicated in the table. In the case of $\PAQb\PQb\PH$~production, the values are given for 
the mixture of five-flavour~($5$FS) and four-flavour~($4$FS) schemes as recommended 
in~\Bref{Harlander:2011aa}.
}
\vspace*{-0.2cm}
\begin{center}
{\small
\begin{tabular}{clr@{\hskip 0.4ex}lr@{\hskip 0.4ex}lc} \hline\hline
 &  \multicolumn{1}{c}{Production }  & \multicolumn{4}{c}{Cross section [pb]} & Order of  \\ \cline{3-6}
 &  \multicolumn{1}{c}{process}     &   \multicolumn{2}{c}{$\sqrt{s}=7$ TeV}   &  \multicolumn{2}{c}{$\sqrt{s}=8$ TeV} & calculation  \\ \hline
 &  ggF       &   $15.0$&${} \pm 1.6$      &  $19.2$&${}\pm 2.0$    & {\small\sc NNLO(QCD) + NLO(EW) } \\
 &  VBF       &   $1.22$&${}\pm 0.03$     &  $1.58$&${}\pm 0.04$   & {\small\sc NLO(QCD+EW) + {\it approx}~NNLO(QCD) } \\
 &  WH        &   $0.577$&${}\pm 0.016$   &  $0.703$&${}\pm 0.018$ & {\small\sc NNLO(QCD) + NLO(EW) }  \\
 &  ZH        &   $0.334$&${}\pm 0.013$   &  $0.414$&${}\pm 0.016$ & {\small\sc NNLO(QCD) + NLO(EW) }  \\ 
 & ggZH       &   $0.023$&${}\pm 0.007$    &   $0.032$&${}\pm 0.010$      & {\small\sc NLO(QCD) } \\
 &  ttH       &   $0.086$&${}\pm 0.009$   &  $0.129$&${}\pm 0.014$ & {\small\sc NLO(QCD) } \\ 
 &  tH        &   $0.012$&${}\pm 0.001$   &  $0.018$&${}\pm 0.001$ & {\small\sc NLO(QCD) }   \\ 
 &  bbH       &   $0.156$&${}\pm 0.021$   &  $0.203$&${}\pm 0.028$ & {\small\sc 5FS NNLO(QCD) + 4FS NLO(QCD) }  \\ \hline
 &  Total       &   $17.4 $&${}\pm 1.6$     &  $22.3$&${}\pm 2.0$    &  \\
\hline\hline 
\end{tabular} }
\end{center}
\label{tab:SMCrossSections}
\end{table}
\begin{table}[htbp]
\caption[]{Standard Model predictions for the decay branching fractions of a Higgs boson with 
a mass of $125.09\UGeV$ \cite{Aad:2015zhl}, together with their 
uncertainties \cite{Heinemeyer:2013tqa}. }
\begin{center}
\begin{tabular}{clcr@{\hskip 0.4ex}l}\hline\hline
 &    Decay mode &\multicolumn{3}{c}{Branching fraction [\%]}  \\  \hline
 &     $Hbb$       & &   $57.5$&${}\pm 1.9$    \\
 &     $HWW$       & &   $21.6$&${}\pm 0.9$   \\
 &     $Hgg$       & &   $8.56$&${}\pm 0.86$  \\
 &     $Htt$       & &   $6.30$&${}\pm 0.36$  \\
 &     $Hcc$       & &   $2.90$&${}\pm 0.35$  \\
 &     $HZZ$       & &   $2.67$&${}\pm 0.11$  \\
 &     $H\gamma\gamma$       & &   $0.228$&${}\pm 0.011$  \\
 &     $HZ\gamma$       & &   $0.155$&${}\pm 0.014$  \\
 &     $H\mu\mu$       & &   ~~$0.022$&${}\pm 0.001$  \\ 
\hline\hline
\end{tabular} 
\end{center}
\label{tab:SMBranchingFractions}
\end{table} 

\clearpage


%% file: NHCS2.tex
Before the $2012$ discovery, the hypothesis was that the SM was the correct theory with 
$m_{\PH}$ as the unknown parameter. Therefore, bounds on $m_{\PH}$ were derived through a 
comparison of the SM theory with high-precision data~\cite{Z-Pole}. 
At the LHC, after the discovery, the unknown parameters are the deviations of the measurements 
with respect to the SM predictions, given that the SM is fully specified and experimentally 
constrainable. Of course, the definition of SM deviations requires a characterisation of the 
underlying dynamics.
Notice that, so far, all the available studies on the couplings of the new resonance conclude 
it to be compatible with the Higgs boson of the SM within present 
precision~\cite{Aad:2015gba,Khachatryan:2014jba,Khachatryan:2016vau}, and, as of yet, there is no direct 
evidence for new physics phenomena beyond the SM.

Best-fit results for the production signal strengths for the combination of ATLAS and CMS data
are performed in the so-called kappa-framework. 
The next step will be to identify the optimal framework for SM deviations, \ie a theory  that will replace 
the kappa-framework after the experimental results will confirm a deviation.

We will introduce a few general definitions that will be relevant for understanding 
the experimental results and their interpretation. Technical details will be 
presented in the corresponding sections.
\begin{definition}[The kappa-framework]
The kappa-framework \cite{LHCHiggsCrossSectionWorkingGroup:2012nn} introduced to parametrise 
SM deviations, is a procedure used at LO, partially accommodating factorisable QCD corrections 
(proportional to the SM LO) but not (the non-factorizable) electroweak (EW) corrections. 
It amounts to replace
\bq
\Lag_{\mySM}\lpar \{m\}\,,\,\{g\}\rpar 
\quad \mbox{with} \quad
\Lag\lpar \{m\}\,,\,\{\upkappa_g\,g\}\rpar \spc 
\eq
where $\{m\}$ denotes the SM masses, $\{g\}$ the SM couplings and $\upkappa_g$ the scaling
parameters. This is the framework used during Run~1.
\end{definition}
\begin{definition}[The EFT/SMEFT]
Exact non-perturbative solutions to quantum field theories are rarely known and approximate 
solutions that expand observables perturbatively in a small coupling constant and in a ratio 
of scales are generally developed. Such quantum field theories can be regarded as examples 
of an effective field theory (EFT); the SM effective field theory (SMEFT) is an example. 
The theory is defined by
\bq
\Lag= \Lag_{\mySM} + \sum_{n > 4}\,\sum_{i=1}^{N_n}\,\frac{a^n_i}{\Lambda^{n-4}}\,
      \Ope^{(d=n)}_i \spc
\eq
with arbitrary Wilson coefficients $a^n_i$ which, however, give the leading amplitudes in an 
exactly unitary $\mrS\,$-matrix at energies far below the scale of new physics, $\Lambda$. The 
theory is (strictly) non-renormalisable, which means that an infinite number of higher dimensional operators 
must be included.
Nevertheless, the amplitudes can be expanded in powers of  
$\mrv/\Lambda, E/\Lambda$, where $\mrv$ is the Higgs VEV and $E$ is the typical scale at which
we measure the process.
The expansion is computable to all orders and ultraviolet divergencies can be cured by introducing, order-by-order, 
an increasing number of counter-terms.
A question that is often raised concerns the ``optimal'' parameterisation of the $\mrdim = 6$
basis for the $\Ope^{(d=n)}_i$ operators; all sets of gauge invariant, dimension $d$ operators, 
none of which is redundant, form a basis and all bases are equivalent. For a formal definition 
of redundancy see Sect.~3 of \Bref{Einhorn:2013kja}.

The rationale for constructing the SMEFT, \ie an effective $\mrS\,$-matrix
$\mrS^{\mathrm{eff}}(\Lambda)\,,\;\forall\,\Lambda < \infty$, has been described in 
\Bref{Passarino:2016owu}; the main assumption is that there is no fundamental scale above which 
$\mrS^{\mathrm{eff}(\Lambda)}$ is not defined~\cite{Costello2011}. 
Of course, $\mrS^{\mathrm{eff}}(\Lambda)$ loses its predictive power if a process at $E = \Lambda$ 
requires an infinite number of renormalized parameters~\cite{Preskill:1990fr}.
The basis for NLO calculus of the SMEFT theory has been developed in \Bref{Ghezzi:2015vva} and 
in \Brefs{Hartmann:2015oia,Hartmann:2015aia}, see also \Bref{NLOnote}.

When we compare SMEFT with kappa framework, it is worth noting that, even for QCD, there are 
contributions which induce sizeable corrections unrelated to the SM ones~\cite{Gauld:2016kuu}. 
When considering the $\PH \to \PAQb\PQb$ decay, there are QCD corrections multiplying the SMEFT-modified 
amplitude which alter the vertex at LO but remain proportional to the SM ones, so NLO results can 
be obtained through a simple rescaling of the LO decay rate (i.e. equivalent to kappa framework). On 
the other hand, there are contributions which alter the LO vertex and induce sizeable corrections 
which are unrelated to the SM ones and cannot be easily anticipated. 

We can conclude that, due to the absence of tantalising hints for new physics during Run~1 at LHC, 
the extension of the Higgs sector by dimension-six operators will provide a new standard for 
searches of non-resonant manifestations of the SM. 
\end{definition}
\begin{definition}[Phenomenological Lagrangians]
Any phenomenological approach, \eg an extension of the SM Lagrangian with a limited number 
of interactions (like $\PH\PV\PV$ and $\PH\PAf\Pf$), is a reasonable starting point to describe 
limits on SM deviations. While this outcome is much less desirable than dealing with a consistent 
SMEFT, it is important to recognise that the difference relates to the possibility of 
including an estimate of the uncertainties induced by the truncation of the expansion.
However, one has to mention that Monte Carlo tools are already available for 
phenomenological studies~\cite{Artoisenet:2013puc,Demartin:2014fia,Falkowski:2015wza}.  
\end{definition}
\begin{definition}[Pseudo Observables]
(POs) are a platform between realistic observables and theory parameters, 
allowing experimentalists and theorists to meet half way, without theorists having to run 
full simulation and reconstruction and experimentalists fully unfolding to model-dependent 
parameter spaces. 
Experimenters collapse some ``primordial quantities'' (say number 
of observed events in some pre-defined set-up) into some ``secondary quantities'' which we feel 
closer to the theoretical description of the phenomena.
In other words, POs answer the question ``how to store measurements in order to preserve them 
for a long time?''
Indeed, the original kappa framework cannot explain many SM deviations, and Wilson coefficients in 
any EFT description are deeply rooted in a theoretical framework.
Fiducial cross sections can be performed for a lot of bins. There is the need to extend the kappa parameters to
something in between: theorists may refine their calculations and interpret them against POs;
if SM predictions improve experimentalist may redo POs~\footnote{
A first public tool for POs is available at www.physik.uzh.ch/data/HiggsPO.}.
\end{definition}

To summarize, we can say that
\bei

\item[-] in the kappa framework, SM deviations are a simple rescaling of couplings and there are no
new Lorentz structures; Higgs coupling fits are based on total rates.

\item[-] In the SMEFT there are new kinematical/Lorentz structures;

\item[-] phenomenological Lagrangians contain a subset of the interactions present in the SMEFT.

\eei 


%% file: NHCS31.tex
The following analyses have been performed in the ATLAS and CMS experiments
to first discover and then to measure the Higgs boson properties:
\[
\begin{array}{ll}
\PH \to \PZ\PZ \to 4\Pl & \PH \to \PGg\PGg \\
\PH \to \PW\PW \to \Pl \PGn \Pl \PGn & \PH \to \PGt \PGtp \\
\PH \to \PQb \PAQb & \PH \,\, \mbox{in association with} \,\, \PAQt\PQt \\
\end{array}
\]
The $\PH \to \PGmm\PGmp$ search has been also pursued in the experiments.

\begin{example}[$\PH \to \PZ\PZ \to 4\Pl$] \cite{Aad:2014eva,CMS:H4l}
The signature for this final state is made of $4$ electrons, or $4$ muons, or $2$ electrons and 
$2$ muons of high $\pT$, isolated and coming from the primary vertex. 
The signal is reconstructed as a very narrow peak on top of a smooth background, composed 
by an irreducible part coming from the production of two non resonant $\PZ$ bosons, and a reducible 
part from $\PZ +\,$jets and $\PAQt\PQt$ events, where jets
are originating from heavy quarks, and thus could contain leptons, or are mis-identified 
as leptons.
\end{example}
The cross section of this process is tiny due to the small branching ratio
of $\PH \to \PZ\PZ$ and even smaller branching ratios of $\PZ \to \Plp\Plm$, thus the analysis 
has to conserve the highest efficiency. 
The lepton identification and the lepton reconstruction are extremely pure and of high 
resolution, allowing to reach a $m_{\PH}$ resolution of $1{-}2\%$.

Events within this channel are categorised as VBF-produced if there are $2$ hight $\pT$ jets, 
or as $\PV\PH$ produced if there are additional leptons or $\PV$ bosons, otherwise ggF produced.
\begin{example}[$\PH \to \PGg\PGg$] \cite{Aad:2014eha,CMS:Hgamgam}
The signature for this final state is made by $2$ energetic and isolated photons, that cluster 
in a narrow mass peak on top of a steeply falling spectrum. The background is composed mainly 
by an irreducible component from QCD production of $2$ photons events and by a smaller 
contribution coming from $\PGg +\,$jets events, where a jet has been mis-identified as a photon.
A good fit and a good understanding of the shape of the background allow the analysis to be 
data-driven and not to rely on a perfect Monte Carlo simulation. 

The analysis is very similar in the two experiments: the events are divided into categories of 
different expected signal-to-background ratio and on the presence or not of $2$ jets of high 
invariant mass and high rapidity to select events produced through the Vector Boson Fusion (VBF)
process. The candidate invariant mass is reconstructed with very good resolution of $1{-}2\%$.
\end{example}
\begin{example}[$\PH \to \PW\PW \to \Pl \PGn \Pl \PGn$] \cite{ATLAS:2014aga,ATLAS:VHWW,CMS:HWW}
This is the channel with the highest cross-section. The mass reconstruction is not possible 
due to the presence of $2$ neutrinos, and thus the background is hard to suppress and the 
signal has to be extracted from differences in shapes and from event counting.
The $2$ leptons are expected to have high $\pT$ and small opening angle to conserve the 
V{-}A structure of the theory. The events have large missing transverse energy due to the 
presence of two high $\pT$ neutrinos. 
The analysis is performed on exclusive jet multiplicity ($0$, $1$, $2$-jet) and 
considering also the presence of additional leptons, thus event categories can separate  
the different production modes.

The large background is studied in detail and all the components are estimated with data, 
after having obtained an enriched sample of that specific component. 
The Drell-Yan background is suppressed by  cuts on the invariant mass of the 
2 leptons, $M_{\Pl\Pl}$, and on the transverse missing energy, MET. Particular care has to 
be given to the MET since it is affected by the pile-up. 
The $\PW +\,$jets background (with one jet faking a lepton) is mitigated by a very pure and 
efficient ID. Background coming from top events is suppressed by the $\PQb$-tag veto, or 
rejecting events with additional soft leptons. 
To partially subtract the irreducible $\PW\PW$ background the  $M_{\Pl\Pl}$, $M_{\mrT}$
 (the lepton-neutrino transverse mass) and 
$\Delta(\phi)$ (the angle between the 2 charged leptons) distributions are used.
\end{example}
\begin{example}[$\PH \to \PGtm\PGtp$] \cite{Aad:2015vsa,CMS:Htautau}
The final state with two taus suffers from a low efficiency due to the relatively low $\pT$ of 
the leptons coming from the $\PGt$ decay, from the presence of at least $2$ neutrinos 
from the decay and from the low $\PGtm\PGtp$ invariant mass resolution. 
The events are subdivided into jets categories: the  $0$-jet  (that in CMS is used only 
to constrain the background), the $1$-jet category, subsequently divided into a low $\pT$  
and a high $\pT$ category, and the $2$-jet categories targeting the VBF production mode. Thus the 
VBF, the VH and the ggF production modes can be separated. All the final states for the tau 
decays are considered in ATLAS, and a large fraction is analysed in CMS.
The mass resolution is around $10{-}20\%$.
\end{example}
\begin{example}[$\PH \to \PAQb \PQb$] \cite{Aad:2014xzb,CMS:VHbb}
Due to huge background from $\PAQb \PQb$ QCD events, and due to the poor resolution of the 
$\PAQb{-}\PQb$ invariant mass (of the order of $10\%$) the ggF production mode in this final 
state cannot be considered inclusively. Thus the production mode with the highest signal-over-background is 
the VH one.
In CMS a search of the Higgs boson decaying to $\PAQb \PQb$ produced in VBF has been carried 
out in \Bref{Khachatryan:2015bnx}, but the results have not been considered for the coupling 
analysis that will be presented in this review. 
Final states with zero additional leptons target the $\PZ\PH \to \PGn \PAGn \PQb \PAQb$ 
decay, with one high $\pT$ lepton the $\PW\PH \to \Pl\PGn\PQb\PAQb$ decay and with two high $\pT$ 
leptons the $\PZ\PH \to  \Pl\Pl\PQb\PAQb$ decay.
\end{example}
\begin{example}[$\PH$ produced in association with $\PAQt\PQt$]
\cite{ATLAS:ttHhad,ATLAS:ttHlep,Aad:2014lma,Chatrchyan:2013yea,CMS:ttH}
The final state with two top quark and a Higgs has a very small cross section, a broad 
spectrum of final states and as well as for the previous channel, a very large background.  
The final states with some clean signatures that help in separating the signal from 
the background are the following:
the Higgs decaying into photons accompanied by $2$ top quark signatures, and the Higgs 
decaying into $\PW\PW$, $\PZ\PZ$ and $\PGt\PGtp$, where final states with many leptons 
are selected. 
Although the cross section is not large enough to expect a significant observation, the 
measurements even with very large uncertainties play an important role in the Higgs boson 
coupling extraction.
\end{example}
\begin{example}[$\PH \to \PGmm\PGmp$] \cite{Aad:2014xva,CMS:Hmumu}
The search of the Higgs to $\PGm\PGm$ has been carried out in both experiments. Since the 
branching fraction in the Standard Model is very small, observing a signal would have been a hint for 
new physics. Upper limits have been set by both experiments.
\end{example}

%% file: NHCS4.tex

The original kappa-framework (OKF) has been introduced in 
\Bref{LHCHiggsCrossSectionWorkingGroup:2012nn} as a way to study deviations from the SM. 
To discuss the idea in general terms we consider a process involving the Higgs boson, 
\eg $\PH \to \PGg\PGg$; the SM amplitude starts at one loop comprising fermion and bosonic 
contributions, \ie
\bq
\mrA^{\mySM}_{\PH \to \PGg\PGg} = 
\mrA^{\PQt}_{\PH \to \PGg\PGg} + \mrA^{\PQb}_{\PH \to \PGg\PGg} +  
\mrA^{\bos}_{\PH \to \PGg\PGg}  \spc
\label{SMHAA}
\eq
where light fermions have been discarded. Each contribution is gauge parameter independent
and proportional to the corresponding Higgs coupling. The idea is to modify \eqn{SMHAA}
with ad hoc kappa parameters,
\bq
\mrA^{\upkappa}_{\PH \to \PGg\PGg} = 
\upkappa_{\PQt}\,\mrA^{\PQt}_{\PH \to \PGg\PGg} + 
\upkappa_{\PQb}\,\mrA^{\PQb}_{\PH \to \PGg\PGg} +  
\upkappa_{\PV}\,\mrA^{\bos}_{\PH \to \PGg\PGg}  \spc
\label{KHAA}
\eq
and study their deviation from one. The formalism is simple but suffers of problems of
consistency. Indeed, in a spontaneously broken gauge theory, masses and Yukawa couplings
are not independent quantities. This fact is usually forgotten when dealing with massless
fermions, typically in computing QCD corrections, but is the source of serious inconsistencies
when higher-order EW corrections are included; altering the relation between masses and Yukawa
couplings spoils the gauge invariance of the theory. 

Furthermore, kinematics is not affected by the kappa parameters. Therefore the scheme works at 
the level of total cross-sections, not for differential distributions.
In conclusion, the OKF  is a LO construct, partially accommodating 
factorisable QCD corrections but not EW ones.
\paragraph{Implementation of the kappa framework}
Having in mind \eqn{KHAA} as an example of the ``theoretical'' implementation of the
OKF, we can briefly discuss how the framework is implemented in practice in the experimental 
analysis. For a given production process or decay mode labelled by $\mcO^i$, a kappa parameter 
is defined such that
\bq
\upkappa^2_i = \frac{\mcO^i}{\mcO^i_{\mySM}} \spc
\eq
giving $\upkappa_i = 1$ in the SM. This defines  $\mcO^i_{\mySM}$ as SM equipped with the best available higher-order 
QCD and EW corrections, under the assumption that the dominant higher-order QCD corrections 
factorise.

Contributions from interference effects between the different (gauge invariant) sub-amplitudes
provide some sensitivity to the relative signs of the Higgs boson couplings to different particles.
Therefore, additional coupling modifiers are introduced, \eg $\upkappa_{\PQt}, \upkappa_{\PQb}$,
$\upkappa_{\PV}$ \etc
In this way one has effective and resolved scaling factors, \eg in gluon fusion we have
\bq
\mbox{effective} = \upkappa^2_{\Pg} \spc
\qquad
\mbox{resolved} = \upkappa^2_{\Pg}(\upkappa_{\PQt}, \upkappa_{\PQb}) = 1.06\,\upkappa^2_{\PQt} + 0.01\,\upkappa^2_{\PQb} - 
                  0.007\,\upkappa_{\PQt}\,\upkappa_{\PQb} \spc
\label{asdone}
\eq
where one should observe that, once again, the assumption of higher orders factorisation has 
been made. The generalisation for the resolved modifier case is shown in \eqn{SMEFTc}.
\paragraph{Underlying assumptions}
The OKF as any other framework that aims to study SM deviations is based
on a certain number of additional assumptions.
We consider only one Higgs doublet in the linear representation (a flexible choice)
and the scalar doublet $\Upphi$ (with hypercharge $1/2$) contains $\PH$, the custodial singlet in 
$\lpar 2_{\ssL}\,\otimes\,2_{\ssR}\rpar = 1\,\oplus\,3$. Extensions are possible but
``difficult'', \eg the two-Higgs doublet model (THDM) where
\bq
\Upphi \to \Upphi_i \qquad \Upphi_i = \mathrm{R}_{ij}\lpar \beta \rpar\Uppsi^j \spc
\eq
with the technical hurdle of a diagonalisation of the mass matrix for the CP-even scalars.

New ``light'' degrees of freedom are not included and decoupling from heavy ones is a rigid 
assumption. To examine the consequences, consider the effect of heavy degrees of freedom in 
$\PGg \PGg \to \PH$: to be fully general one has to consider effects due to heavy fermions 
and heavy scalars in arbitrary representations, $R_f$ and $R_s$, of $SU(3)$~\cite{Bonciani:2007ex}. 
Colored scalars decouple from the low energy physics as their mass increases; however, the same 
is not true for fermions. To be more precise, besides decoupling we have other regimes: for a 
given amplitude containing a particle of mass $m$, in the limit $m \to \infty$, we have to 
distinguish three possible cases:
\bei

\item[] Decoupling: $\mrA \sim 1/m^2$ (or more). The corresponding higher-order
operators are called ``irrelevant'' \spc

\item[] Screening: $\mrA \to\,$ const (or $\ln m^2$). The operators are called ``marginal'' \spc

\item[] Enhancement: $\mrA \sim m^2$ (or more). The operators are called ``relevant'' \spp

\eei
It is worth noting that whenever the LO $\uprho\,$-parameter~\cite{Ross:1975fq} is different 
from one, quadratic power-like contributions to its radiative corrections, $\Delta\uprho$, are 
absorbed by the renormalization of the new parameters of the model: in this case $\uprho$ is not 
a measure of the custodial symmetry breaking.
\paragraph{Mixing}
Mixing among scalars is another potential problem: absence of mass mixing of the new heavy 
scalars with the SM Higgs doublet is therefore required since mixings change the scenario.

Consider a model with two doublets and hypercharge $Y = 1/2$ (THDM). These doublets are first rotated 
(with an angle $\beta$) to the Georgi-Higgs basis and successively a mixing-angle $\alpha$ 
diagonalizes the mass matrix for the CP-even states, $\Ph$ and $\PH$. 
The couplings of $\Ph$ to SM particles are almost the same of a SM Higgs boson with the same 
mass (at LO) only if we assume $\sin\lpar \beta - \alpha\rpar = 1$. 
Therefore, interpreting large deviations in the couplings within a THDM should be done only 
after relaxing this assumption.

The interplay between integrating out heavy scalars and the SM decoupling limit 
has been discussed in \Bref{Boggia:2016asg}. In general, decoupling cannot be obtained in terms 
of only one large scale and can only be achieved by imposing further assumptions on the couplings. 
Indeed, there are two sources of deviations with respect to the SM, new couplings and modified
couplings due to VEV mixings, heavy fields. In general, it is not simple to identify only one 
scale for new physics (NP); it is relatively simple in the unbroken phase using weak eigenstates 
but it becomes more complicated when EW symmetry breaking (EWSB) is taken into account and one 
works with the mass eigenstates. In the second case, one should also take into account that 
there are relations among the parameters of the beyond-SM (BSM) model, typically coupling 
constants can be expressed in terms of VEVs and masses; once the heavy scale has been introduced 
also these relations should be consistently expanded. Once again, the SM decoupling limit 
cannot be obtained by making only assumptions about one parameter.

In the top-down approach there is some theory, assumed to be UV-complete or valid on a given high 
energy scale (\eg some BSM model), and the aim is to implement a systematic procedure for getting
the low-energy theory. A typical example would be the Euler-Heisenberg Lagrangian.
Systematic low-energy expansions are able to obtain low-energy footprints of the high energy
regime of the theory.
In the top-down approach the heavy fields are integrated out of the underlying high-energy theory
and the resulting effective action is then expanded in a series of local operator terms.
Even in the top-down EFT approach one has to be careful: for both tree-level and one-loop 
processes, the agreement between the effective Lagrangian and a range of UV-complete models 
depends critically on the appropriate definition of the matching, see \Bref{Freitas:2016iwx}. 
\paragraph{Custodial symmetry}
Finally, let us consider custodial symmetry: it is the set of scalar fields that break EW 
symmetry by developing a VEV. The problem with more VEVs, or one VEV different from 
$(T\,,\,Y)= (\frac{1}{2}\,,\,1)$ ($T$ is isospin and $Y$ is hypercharge), is partially related 
to the rho-parameter which at tree-level is given by
\bq
\uprho_{\myLO} = \frac{1}{2}\,\frac{\sum_i\,\Bigl[ c_i\,\mid \mrv_i\mid^2 + r_i\,\mru^2_i\Bigr]}
{\sum_i\,Y^2_i\,\mid \mrv_i\mid^2}
\quad 
c_i = T_i\,\lpar T_i + 1\rpar - Y^2_i
\quad
\mrr_i = T_i\,\lpar T_i + 1\rpar
\eq 
where the sum is over all Higgs fields and $\mrv_i(\mru_i)$ gives the VEV of a complex (real) 
Higgs field with hypercharge $Y_i$ and weak-isospin $T_i$. The experimental limits on $\uprho - 1$ 
are rather stringent.
The SM Higgs potential is invariant under $SO(4)$; furthermore, 
$SO(4) \sim SU(2)_{\ssL}\,\otimes\,SU(2)_{\ssR}$ and the Higgs VEV breaks it down to the diagonal 
subgroup $SU(2)_{\ssV}$. It is an approximate symmetry since the $U(1)_{\mathrm{Y}}$ is a 
subgroup of $SU(2)_{\ssR}$ and only that subgroup is gauged. 
Furthermore, the Yukawa interactions are only invariant under 
$SU(2)_{\ssL}\,\otimes\,U(1)_{\mathrm{Y}}$ and not under $SU(2)_{\ssL}\,\otimes\,SU(2)_{\ssR}$ and 
therefore not under the custodial subgroup.
Therefore, if we require a new CP-even scalar, which is also in a custodial 
representation of the group, the $\PW/\PZ\,$-bosons can only couple to a singlet or a 
$5\,$-plet~\cite{Low:2012rj}. 

If $(N_{\ssL}\,,\,N_{\ssR})$ denotes a representation of $SU(2)_{\ssL}\,\otimes\,SU(2)_{\ssR}$
we see that the usual Higgs doublet scalar is a $(2\,,\,\mybar{2})$, while 
the $(3\,,\,\mybar{3}) = 1\,\oplus\,3\,\oplus\,5$ contains the Higgs-Kibble 
ghosts (the $3$), a real triplet (with $Y = 2$) and a complex triplet (with $Y = 0$).
The Georgi-Machacek model, \Bref{Georgi:1985nv} has EWSB from both a $(2\,,\,\mybar{2})$ and 
a $(3\,,\,\mybar{3})$.
Custodial symmetry is a statement on the $\uprho$ parameter but translation to
$\PS\PV\PV$ couplings requires care:
when a single source of EWSB is present, custodial symmetry implies
$\frac{g_{\PS^0\PW\PW}}{g_{\PS^0\PZ\PZ}} = \frac{\mws}{\mzs}$.
In general $\frac{g_{\PS\PW\PW}}{g_{\PS\PZ\PZ}} = \lambda\,\frac{\mws}{\mzs}$,
\eg $\lambda = - 1/2$ for a $5\,$-plet (already excluded).

%% file: NHCS5.tex
In this Section we discuss results from Run~1, including the measurement of the Higgs mass, 
constraints on the Higgs width, pole observables and off-shell (tail) observables. 
The results presented here have been published by ATLAS and CMS, and then combined 
together in \Brefs{Khachatryan:2014jba,Aad:2014aba,Aad:2015gba,Aad:2015zhl,Khachatryan:2016vau}.

%% file: NHCS51.tex

The most important parameter of a particle is its mass, although the definition of mass and
width for an unstable particle requires particular care and is not unique. Here we discuss
on-shell quantities.

In any theory the parameters of the Lagrangian cannot be predicted but have to be related to 
quantities measured experimentally, the so-called input parameter set (IPS).
In the SM, once the Higgs boson mass (the ``on-shell'' mass) is known, the IPS is complete and 
all properties of the particle can be computed with high precision.

The mass can be measured with very high precision from the $\PH \to \PZ \PZ$~\footnote{With
the usual caveats in the interpretation of this decay mode.} and 
$\PH \to \PGg \PGg$ decays, since muons, electrons and photons are reconstructed  
with high precision, see Refs. \cite{Aad:2014aba,Khachatryan:2014jba}.

The energy scale, the momentum scale and resolution of muons, electrons and photons are excellent 
in both the experiments. Well-known particles like the $\PZ, \PGU$, $\PJGy \to 2\Pl$ are  
used to calibrate the detectors.
The decay $\PZ \to 4\Pl$ is used to validate the procedure.
The systematic uncertainty is of $\pm 0.1\%/ \pm 0.3\%$ for the  muon/electron  
momentum scale in CMS, and $\pm 0.3\%/ \pm 0.1\%$ for for the muon/electron momentum scale in 
ATLAS. In CMS the mass measurement is performed with a $3$D fit using four-lepton invariant 
mass $m_{4\Pl}$, associated per-event mass uncertainty $\delta m_{4\Pl}$, kinematic discriminant 
$KD$, see \Bref{CMS:H4l}.
         
The invariant mass from the di-photon system is given by:
\bq
m_{\PGg\PGg} = 2 \times \mrE_{\PGg_1}\,\mrE_{\PGg_2}\,(1 - \cos(\theta_{12})) \spc
\eq
thus not only the energy of the photons has to be measured with high precision, but also their
directions. The determination of the primary vertex is thus affecting the precision of 
$\theta_{12}$. 
In ATLAS, a likelihood discriminant has been developed combining the information on the axis 
of the shower from the calorimeter, photon conversion, and track recoil.
In CMS two Boosted Decision Tree (BDT) have been developed using all the information of the event
and then carefully calibrated and cross-checked with $\PZ \to \Pem\Pep$ events. 

The photon energy calibration is the dominant systematic in the mass reconstruction.
The energy scale is determined using $\PZ \to \Pem\Pep$ events, then a correction is applied to
 account  for the $\Pe \to \PGg$ difference and subsequently an extrapolation is performed
 in order to move from  the energy scale of the $\PZ$ to the energy scale of  the $\PH$.
In summary, the systematic uncertainties in the mass measurement from the $\PGg\PGg$ channel 
are due to the knowledge of the material in front of the electromagnetic calorimeter, the non-linearity 
of the calorimeter, the calibration of the detector, and the differences between electron and photons.

\begin{figure}[hbt]
\centering
\includegraphics[width=1.\textwidth]{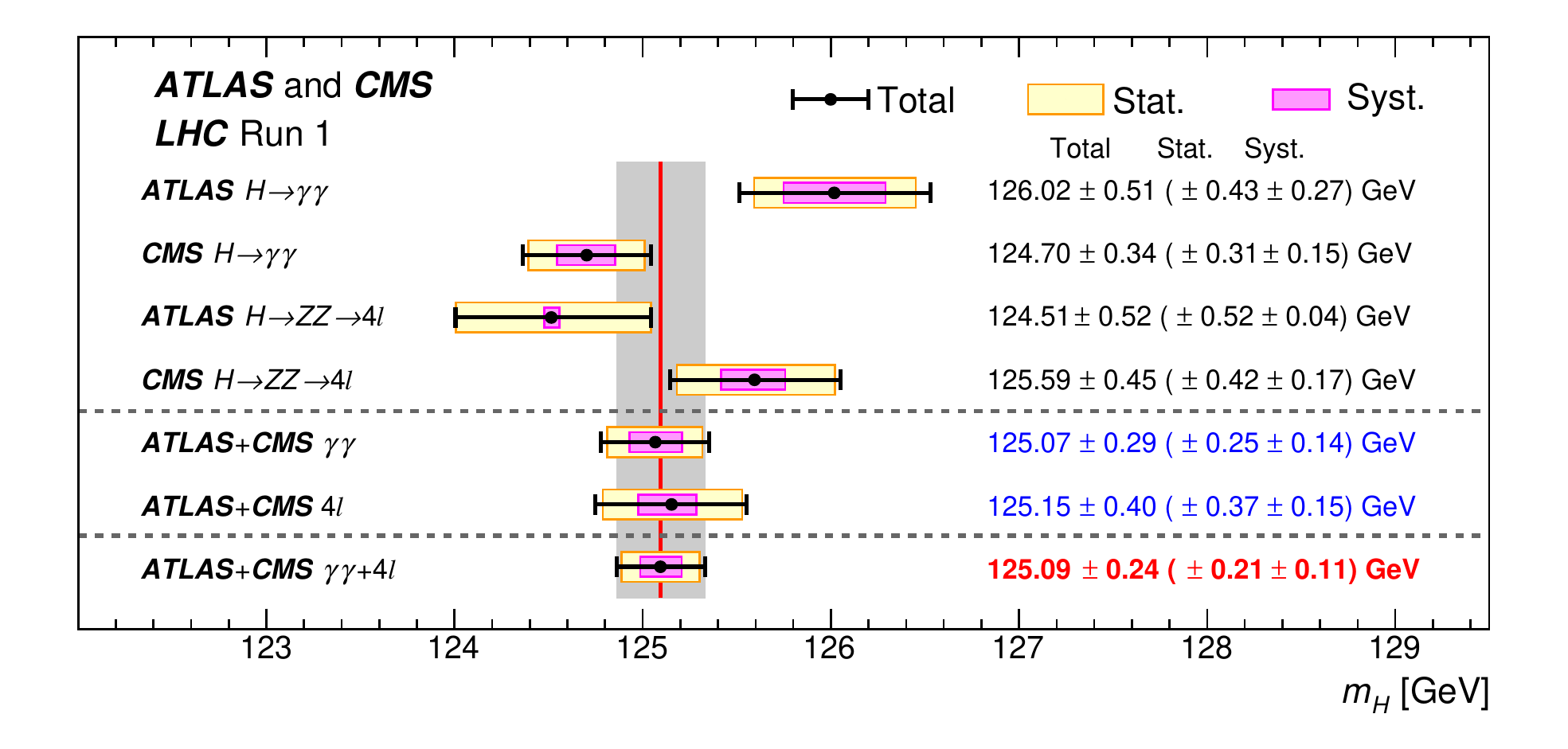}
\caption[]{Summary of Higgs boson mass measurements from the individual analyses of ATLAS and 
CMS and from the combination. The figure is from \Bref{Khachatryan:2016vau}.}
   \label{fig:mass_table} 
\end{figure}

\begin{figure}[hbt]
\centering
\includegraphics[width=1.\textwidth]{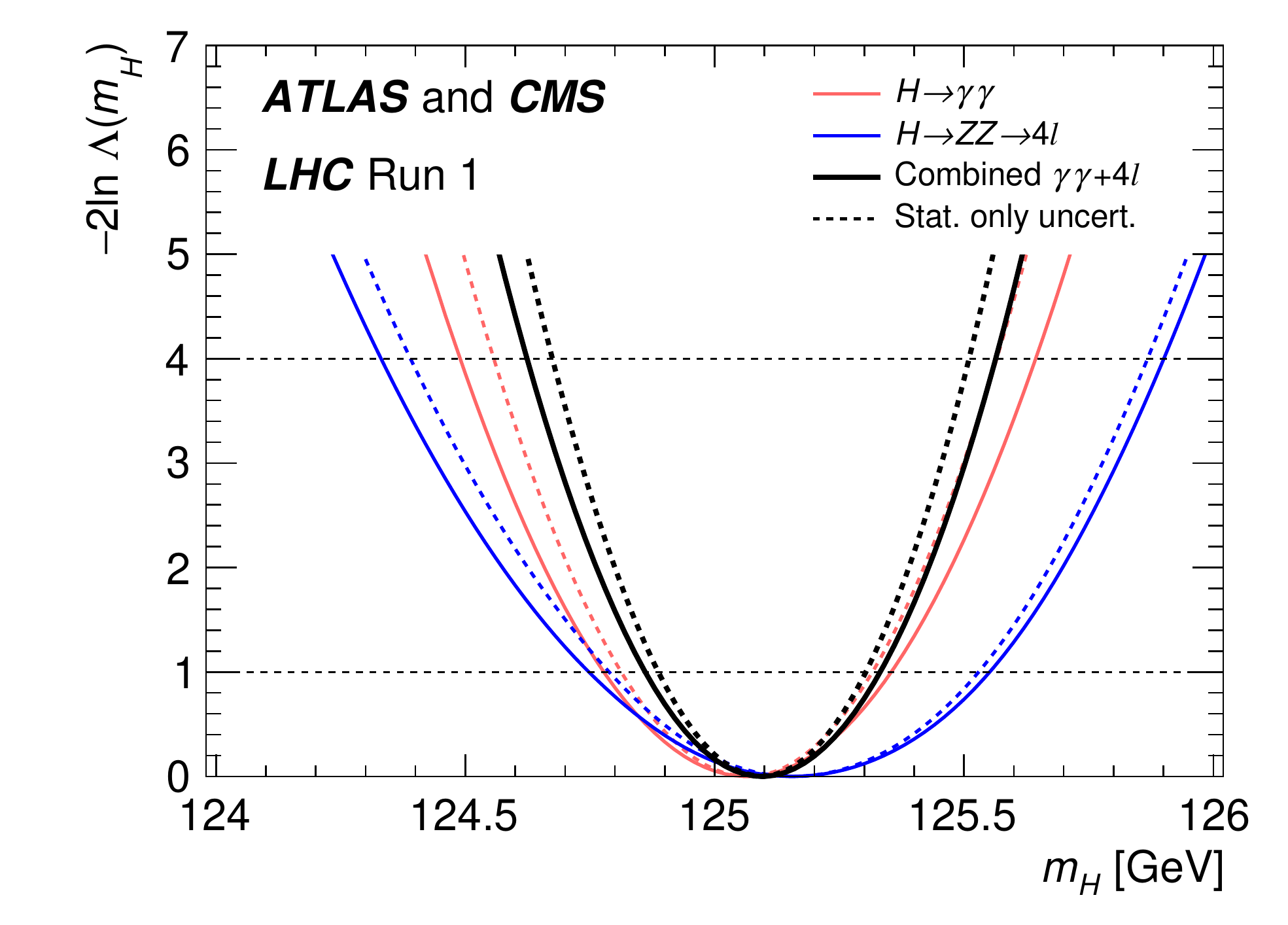}
\caption[]{The scans of twice the negative log-likelihood ratio $2 q {\it ln} \Lambda(m_\PH)$ 
as a function of the Higgs boson mass $m_{\PH}$ for the ATLAS and CMS combination of the 
$\PH \to \PGg\PGg$ (red), $\PH \to \PZ\PZ \to 4\Pl $ (blue),  and combined (black) channels. 
The figure is from \Bref{Khachatryan:2016vau}}
   \label{fig:mass} 
\end{figure}

\begin{figure}[hbt]
\centering
\includegraphics[width=1.\textwidth]{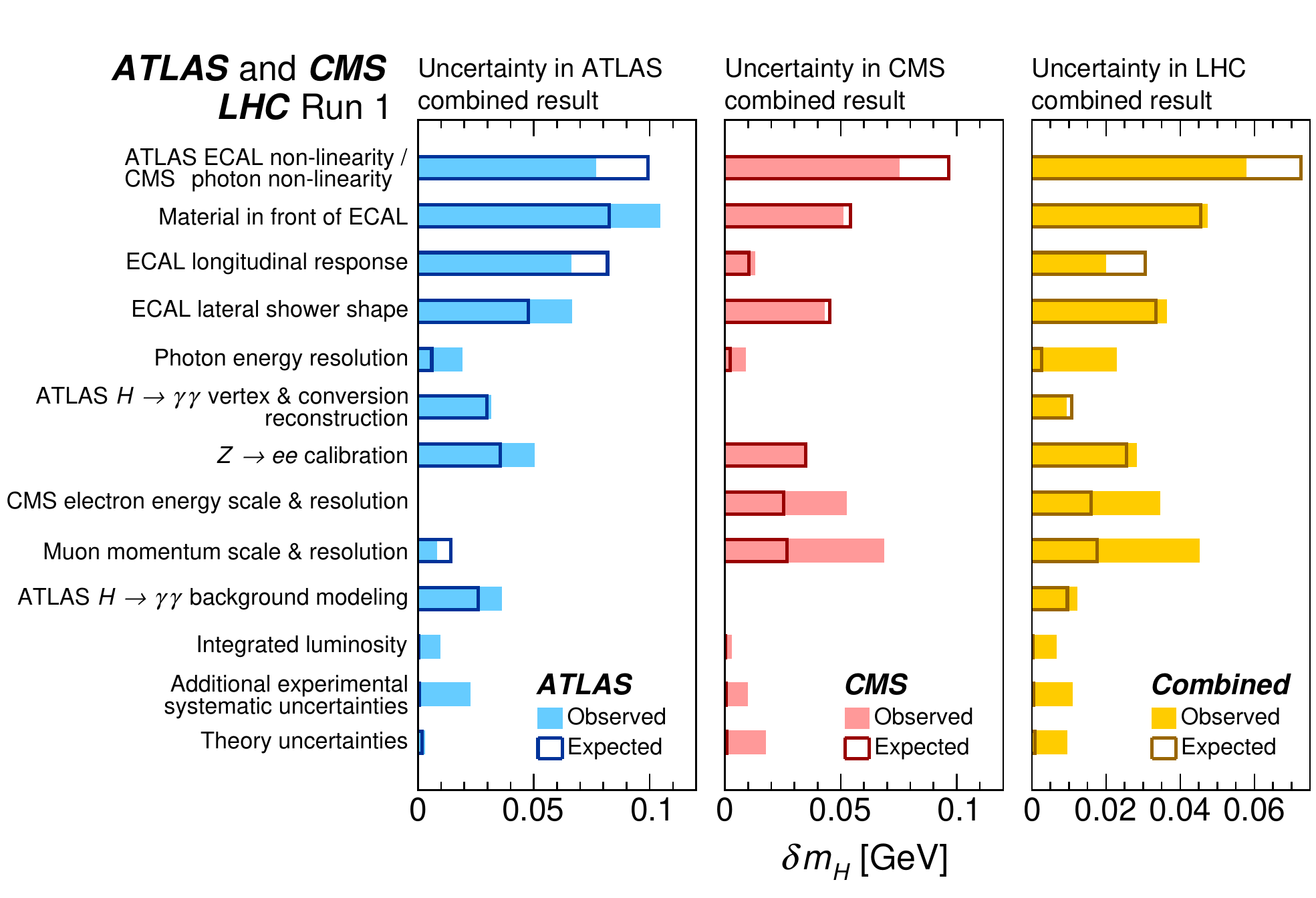}
\caption[]{Systematic uncertainties on the mass for the ATLAS (left), CMS (center), and 
combined (right). The observed (expected) results are shown by the solid (empty) bars.
The figure is from \Bref{Khachatryan:2016vau}}
   \label{fig:mass_syst} 
\end{figure}

In Fig.~\ref{fig:mass_table} the summary of Higgs boson mass measurements from the individual 
analyses of ATLAS and CMS~\cite{Aad:2014aba,Khachatryan:2014jba} and from the combined analysis 
are presented~\cite{Aad:2015zhl}. The systematic 
uncertainties (narrower, magenta-shaded bands), the statistical uncertainties 
(wider, yellow-shaded bands), and total uncertainties (black error bars) are indicated. 
The (red) vertical line and corresponding (gray) shaded column indicate the central value 
and the total uncertainty of the combined measurement, respectively.
In Fig.~\ref{fig:mass} the scans of twice the negative log-likelihood ratio 
$2 ln \Lambda(m_{\PH})$ as a function of the Higgs boson mass $m_{\PH}$ for the ATLAS and CMS 
combination of the $\PH \to \PGg\PGg$ (red), $\PH \to \PZ\PZ \to 4\Pl$ (blue),  and combined 
(black) channels are shown. 
The dashed curves show the results accounting for statistical uncertainties 
only, with all nuisance parameters associated with systematic uncertainties fixed to their 
best-fit values. The $1$ and $2$ standard deviation intervals are indicated by the intersections 
of the horizontal lines at $1$ and $4$, respectively, with the log-likelihood scan curves.
In Fig.~\ref{fig:mass_syst} the systematic uncertainties are shown for the measurement in 
ATLAS, in CMS, and their combination.
The mass of the Higgs boson as measured from the first run at LHC at $7$ and $8\UTeV$ 
center of mass energy is:
\bq
m_{\PH} = 125.09 \pm 0.24\UGeV = 125.09 \pm 0.21(stat) \pm 0.11(syst)\UGeV \spc
\eq
where the total uncertainty is dominated by the statistical term, with the systematic 
uncertainty dominated by the non linearity of the electromagnetic calorimeter and by the 
knowledge of the material in front of them and by the lepton energy/momentum scale 
uncertainty. Compatibility tests are performed to ascertain whether the measurements are 
consistent with each other, both between the different decay channels and between the two 
experiments. All tests on the combined results indicate consistency of the different 
measurements within $1\,\sigma$, while the four Higgs boson mass measurements in the two 
channels of the two experiments agree within $1.3\,\sigma$.

%% file: NHCS521.tex

%
%

The results presented in this section are a selected summary of the combined analysis of the ATLAS and CMS 
Higgs boson data, as published in Ref.~\cite{Khachatryan:2016vau} 

\subsection{The measurement of $\mu$}

The signal strengths $\mu_i^f$ are defined as the ratios of cross sections and branching fractions to the corresponding SM predictions such that:
\begin{equation}
  \mu_i^f =  \frac{\sigma_i\cdot BR^f}{(\sigma_i)_\SM \cdot (BR^f)_\SM} = \mu_i \times \mu^f,
\label{eq:muif}
\end{equation}
where the subscript $i$ and superscript $f$ indicate the production mode and decay channel, respectively.
By definition all $\mu_i^f$ are equal to 1 for the SM Higgs boson.

The simplest and most restrictive signal strength parameterisation is to assume that the 
$\mu_i$ and the $\mu^f$ values are the same for all production processes and decay channels. 
In this case, the SM predictions of signal yields in all categories are scaled by a global 
signal strength $\mu$. Such a parameterisation, a very special case of the kappa-framework
described in Sect.~\ref{Sect4}, provides the simplest test of the compatibility of the 
experimental data with the SM predictions. 

A fit to the combined ATLAS and CMS data at 
$E_{\mathrm{CM}} = 7,8\UTeV$, with $\mu$ as the parameter of interest, results in the best-fit 
value:
\bq
\mu = 1.09^{+0.11}_{-0.10} = 1.09 \pm 0.07 \mbox{(stat)} \pm 0.04  \mbox{(expt)} 
\pm 0.03 \mbox{(th-bkgd)}^{+0.07}_{-0.06} \mbox{(th-sig)},
\eq
where the breakdown of the uncertainties into their four main components is done as described 
in the following:
\begin{itemize}

\item uncertainties, labelled as ``stat'',  are statistical in nature.
These include in particular the statistical uncertainties on background control regions and 
fit parameters used to parameterise the backgrounds measured from data; 

\item theory uncertainties affecting the Higgs boson signal, labelled as "th-sig";

\item theory uncertainties affecting background processes only, labelled as "th-bkgd";

\item all other experimental uncertainties, labelled as "expt", including those related to 
      the finite size of the MC simulation samples.

\end{itemize}
The overall systematic uncertainty of $+0.09\,,\,-0.08$ is larger than the statistical uncertainty 
and its largest component is the theoretical uncertainty on the ggF cross section. 
This result is consistent with the SM expectation of $\mu=1$ within less than $1\,\sigma$ 
and the p-value of the compatibility between the data and the SM prediction is $34\%$. 

As a further step we can measure signal strengths for the different production modes and 
decays modes.
The production processes can be divided into two subgroups: the production via strong 
interactions as in Fig.~\ref{fig:feyn_ggFVBF} (a) and Fig.~\ref{fig:feyn_ttH} (a) where the 
coupling is a fermion coupling, or the production via EW production where the coupling is a 
vector boson coupling as in Fig.~\ref{fig:feyn_ggFVBF} (b) or Fig.~\ref{fig:feyn_prod} (b).

In the experiments, in order to disentangle production mechanisms, activities in the candidates 
events are analysed:
\begin{itemize} 

\item the associated production with a $\PZ$ or a $\PW$ is identified if high $\pT$ leptons, 
or large missing transverse energy, or low-mass dijets are present and compatible with a 
electroweak boson in association with a candidate $\PH$ boson;

\item the vector boson production (VBF) is identified if two high $\pT$ jets with 
high invariant mass and large pseudo-rapidity separation are present in the event together 
with the Higgs candidate.

\item the production in association with two top quarks is identified if two top quarks are 
reconstructed, thus if leptons, large missing transverse energy, multi-jets or $\PQb\,$-tagged 
jets are present in the event and compatible with a top quark decay.

\item finally, all the other remaining Higgs candidate events are mostly produced by the  
gluon-gluon fusion production process.

\end{itemize}
These differences can be exploited using advanced techniques to enhance the separation between processes, 
like  Boosted Decision Trees.

\begin{figure}[hbt]
\centering
\includegraphics[width=1.\textwidth]{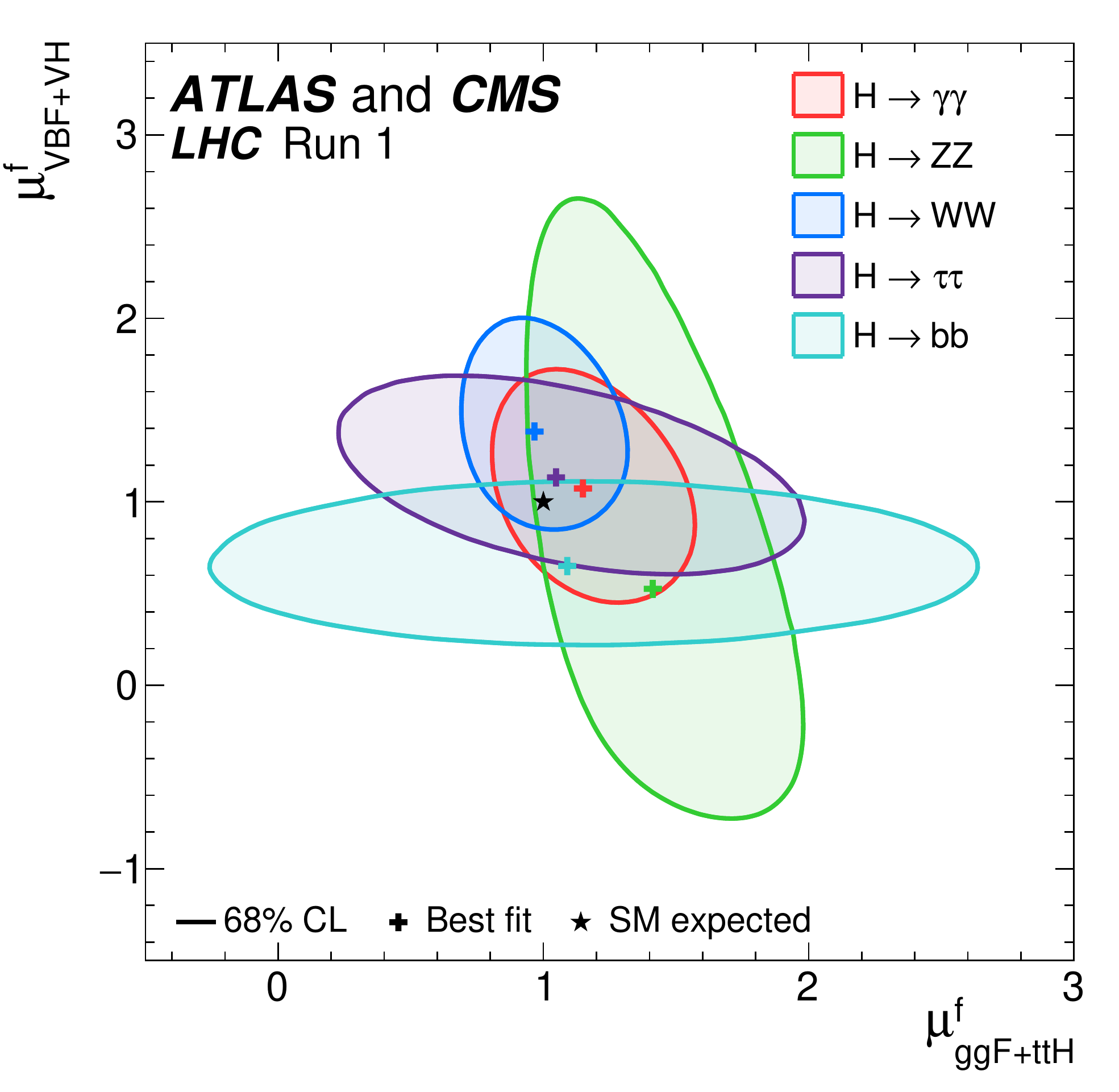}
\caption[]{Negative log-likelihood contours at $68\%$ CL in the 
($\mu_{ggF+tt},\mu_{VBF+VH}$) plane for the combination of ATLAS and CMS, for each of the 
final state analysed $\PH \to \PZ\PZ$, $\PH \to \PW\PW$, $\PH \to \PGg\PGg$, 
$\PH \to \PGt\PGt$,  $\PH \to \PQb\PQb$, and their combination. The SM expectation is also 
shown as a black star. The figure is from \Bref{Khachatryan:2016vau}.}
\label{fig:production_2d} 
\end{figure}

\begin{figure}[hbt]
\centering
\includegraphics[width=1.\textwidth]{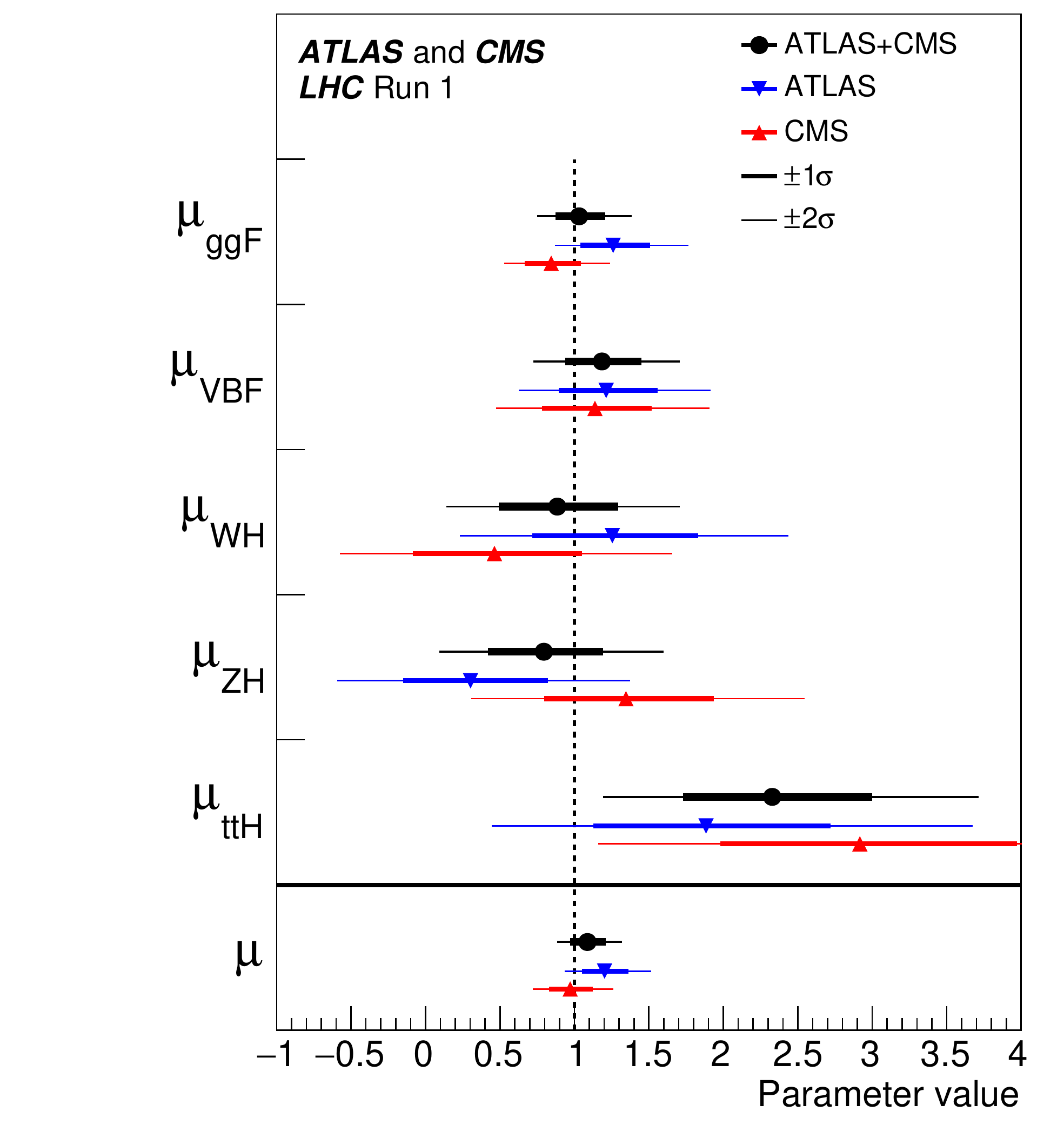}
\caption[]{Best fit results for the production signal strengths for the combination of 
ATLAS and CMS data. Also shown are the results from each experiment. The error bars indicate 
the $1\,\sigma$ (thick lines) and $2\,\sigma$ (thin lines) intervals. The measurements of the 
global signal strength $\mu$ are also shown. The figure is from \Bref{Khachatryan:2016vau}.}
\label{fig:production_mu} 
\end{figure}

\begin{figure}[hbt]
\centering
\includegraphics[width=1.\textwidth]{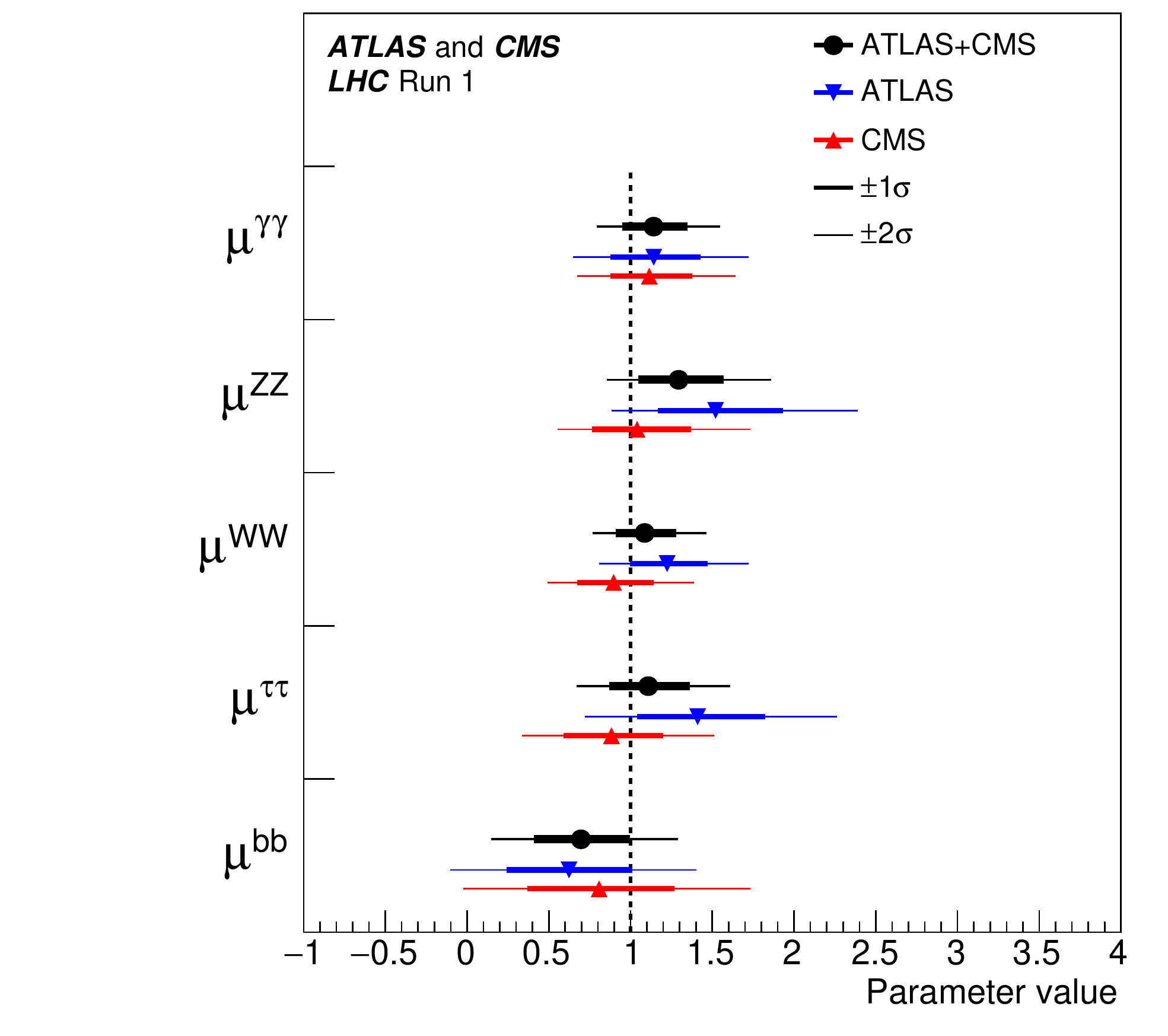}
\caption[]{Best fit results for the decay signal strengths for the combination of ATLAS and 
CMS data. Also shown are the results from each experiment. The error bars indicate the $1\,\sigma$ 
(thick lines) and $2\,\sigma$ (thin lines) intervals. The figure is from \Bref{Khachatryan:2016vau}.}
\label{fig:decay_mu} 
\end{figure}

An interesting result is given by the plot of the signal strength for the ``strong production'' 
modes as a function of the signal strength for the "electroweak production" mode, as shown in 
Fig.~\ref{fig:production_2d}.
Alternatively we can plot the individual signal strengths for each production mode and by decay 
process: Fig.~\ref{fig:production_mu} and Fig.~\ref{fig:decay_mu}.
From these figures we can draw the following conclusions:

\begin{itemize} 
\item the ggF production process is well established, 
\item there is clear observation of the VBF production mode; the $\mu$ value is $5.4\,\sigma$ 
larger than zero.
\item there are indications of the existence of $\PW\PH$ and $\PZ\PH$ production modes; when
combining them together we reach more than $3\,\sigma$ evidence.
\item there is not yet sensitivity to the  $\PQt\PQt\PH$  production mode.
\end{itemize}

Furthermore, we observe that

\begin{itemize}
 \item the $\PH \to \PZ\PZ$, $\PH \to \PW\PW$, $\PH \to \PGg\PGg$
       production modes are well established
 \item the $\PH \to \PGt\PGt$ decay mode is observed with more than $5\,\sigma$ significance when combining the results of the two experiments.
 \item there is not yet evidence for the $\PH \to \PAQb\PQb$ decay mode.
\end{itemize}

\subsection{The couplings}
As a subsequent step we want to disentangle production and decay processes to measure 
the individual vertices to test the SM and search for new physics. We are not measuring 
couplings directly, but their ratio with the SM predictions, \ie the $\upkappa$ 
parameters as explained in Sect.~\ref{Sect4}.
We will consider the vertices of Fig.~\ref{fig:feyn_hVVff} and Fig.~\ref{fig:feyn_hgg}.

\begin{figure}[hbt]
\centering
\includegraphics[width=1.\textwidth]{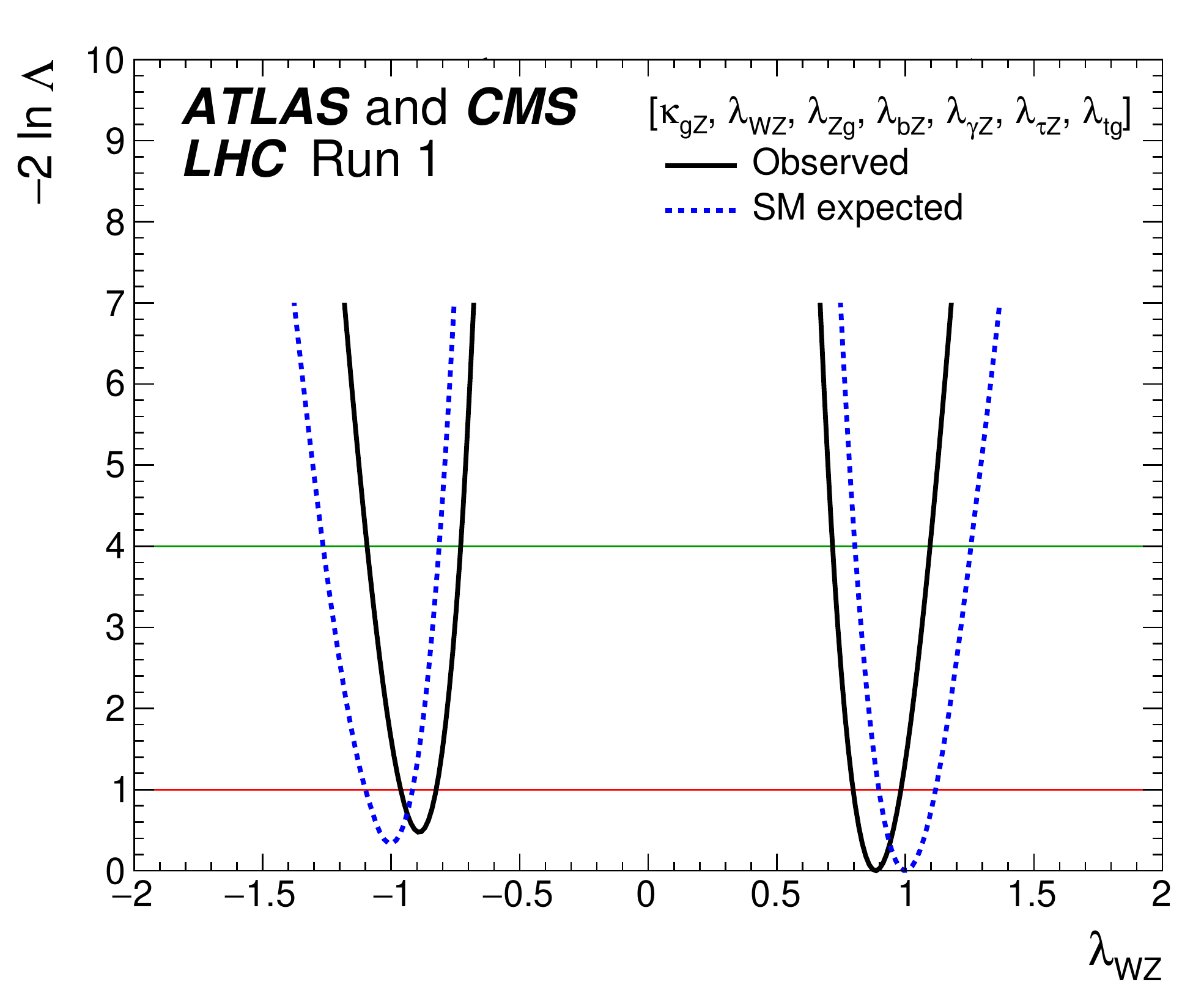}
\caption[]{Observed (solid line) and expected (dashed line) negative log-likelihood scans for
$\lambda_{\PW\PZ} $. All the other parameters of interest from the list in the legend are also 
varied in the minimisation procedure. The red (green) horizontal lines at the $ - 2 ln \Lambda$ 
value of $1 \, (4)$  indicate the value of the profile likelihood ratio corresponding to a 
$1 \,\sigma$  ($2\,\sigma$) CL interval. The figure is from \Bref{Khachatryan:2016vau}.}
\label{fig:custodial} 
\end{figure}

{\bf{Custodial Symmetry in the kappa framework}} As explained in Sect.~\ref{Sect4} one of the 
first important tests of the SM is to validate the Custodial Symmetry. 
The parameter 
\bq
\rho = M^2_{\PW}/M^2_{\PZ} \cdot cos^2 \theta_{\PW} 
\eq
is $1 $ at tree level.
At LEP the experiments have measured: 
$ \rho =1.005 \pm 0.001$,  \ie $5$~sigma away from $1$, but in perfect agreement with the 
theoretical value of $\rho = 1 + \Delta \rho$ when radiative corrections are correctly taken into 
account. Measuring the $\PW$ to $\PZ$ coupling ratio from $\PH$ decays, means deriving
the ratio between boson masses, thus $\rho$; it will tell us if the object produced
is a (minimal) SM-Higgs boson like. 
In the $\upkappa$ framework we measure $\lambda_{\PW\PZ} = \upkappa_{\PW}/\upkappa_{\PZ}$ 
that is expected to be $1$ in the SM. The result is shown in Fig.~\ref{fig:custodial} and it is 
$\lambda_{\PW\PZ}=0.89^{ +0.10}_{-0.09}$, \ie consistent with one within $1\,\sigma$. Thus we 
can, from now on, assume $\upkappa_{\PW}= \upkappa_{\PZ}$.
 
\begin{figure}[hbt]
\centering
\includegraphics[width=1.\textwidth]{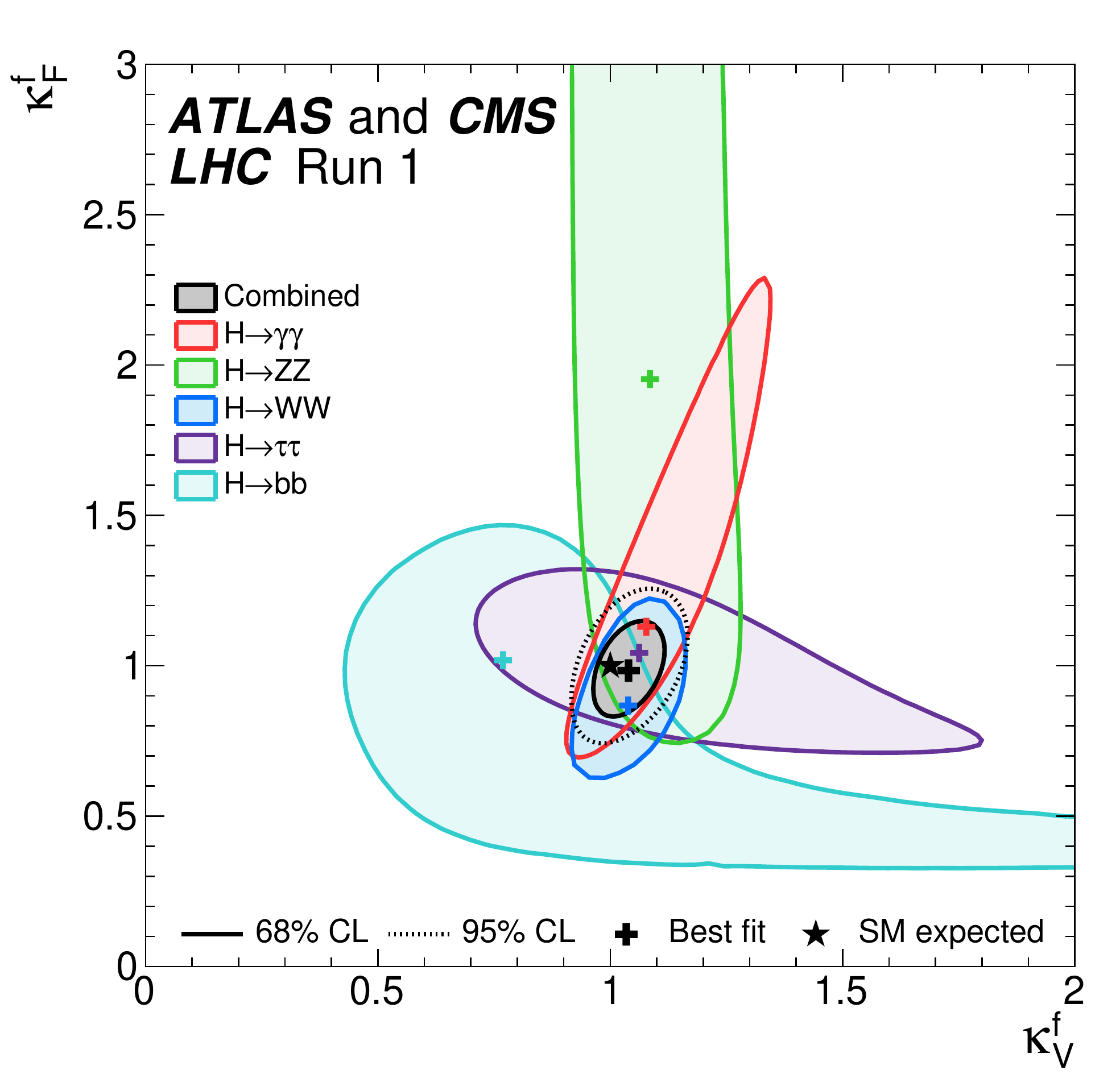}
\caption[]{Negative log-likelihood contours at $68\%$ CL in the 
($\upkappa_{\Pf}, \upkappa_{\PV}$) plane for the combination of ATLAS and CMS and for the individual decay 
channels as well as for their global combination, assuming that all coupling modifiers are 
positive. The figure is from \Bref{Khachatryan:2016vau}.}
\label{fig:kf_vs_kV} 
\end{figure}

{\bf{Vector and fermion Higgs couplings in the kappa framework}}. The further step is to assume 
that all fermion couplings scale as $\upkappa_{\Pf}$ while all vector boson couplings scale as 
$\upkappa_{\PV}$. The result is shown in Fig.~\ref{fig:kf_vs_kV}.
The figures shows the $5$ different final states and their combination. The shapes of the various
contours can be easily understood by writing the cross-section formula as a function of 
$\upkappa_{\Pf}$ and $\upkappa_{\PV}$.
\bq
(\sigma \times  \text{BR})_{( ii \to \PH, \PH \to  jj )} = 
\sigma_{ii} \times \Gamma_{jj}  /  \Gamma_{\PH} = 
\sigma_{SM}(ii \to \PH) \times \text{BR}(\PH \to jj) \times   
\upkappa_{i} \cdot \upkappa_{j} /  \upkappa_{\PH}
\eq
In the denominator we have the width of the Higgs, that for a Higgs of $125\UGeV$ is dominated 
by the $\PAQb\PQb$ decay channel, \ie by $\upkappa_{\Pf}$. If the production mode is ggF then 
the initial state is contributing with $\upkappa_{\Pf}$, while if the production is via VBF or VH, 
in the equation there will be a $\upkappa_{\PV}$ in the numerator.
As an example: the $\PH\PZ\PZ$ channel is dominated by the ggF production mode, thus it will behave
as $\upkappa_{\Pf} \upkappa_{\PV} / \upkappa_{\Pf}$ , thus it will depend on $\upkappa_{\PV}$.
In $\PH \to \PGg\PGg$ the $\PH$ boson does not couple directly to the photon, but via the 
diagrams of Fig.~\ref{fig:feyn_hgg}; it will behave as $\upkappa_i \times 
(8.6 \upkappa_{\PV} -1.8 \upkappa_{\Pf})/\upkappa_{\Pf}$, where $i=\Pf$ for ggF and $i=\PV$ for 
VBF. Thus the behaviour of the red region in the plot.

{\bf{Model with 9 parameters}}. As a subsequent step, the individual couplings, actually 
the strength modifier $\upkappa$ for each of the couplings, can be extracted. 
The rate of the Higgs boson production is inversely proportional to the Higgs boson width, which is
sensitive to invisible or undetected Higgs boson decays predicted by many BSM theories. 
To directly measure the individual coupling modifiers, an assumption on the Higgs boson width is
necessary. 
Two scenarios are considered: the first one assumes that the Higgs boson does not have
any BSM decays, $\text{BR}_{\text{BSM}} = 0$, while the second one leaves 
$\text{BR}_{\text{BSM}}$ free, but assumes $\upkappa_{\PW}  \le 1$, $\upkappa_{\PZ} \le1$ (\ie 
$\upkappa_{\PV} \le 1$) and  $\text{BR}_{\text{BSM}} \ge 0$. 
Notice that these latter constraints are compatible with a wide range of BSM physics models. 
BSM physics can also contribute in the loop-induced processes for the $\Pg\Pg \to \PH$ 
production and and $\PH \to \PGg\PGg$ decay. A dedicated measurement of these two processes 
will also be presented.
BSM physics will also appear as a deviation from $1$ of the individual coupling modifiers 
$\upkappa_i$.
The parameters of interest of the fits to the data are thus the seven independent coupling 
modifiers: $\PGg$, $\Pg$, $\PZ$, $\PW$, $\PQb$, $\PQt$, and $\PGt$, \ie one for each SM particle 
involved in the production processes and decay channels studied, plus $\text{BR}_{\mathrm{BSM}}$ in 
the case of the second fit.
 
\begin{figure}[hbt]
\centering
\includegraphics[width=1.\textwidth]{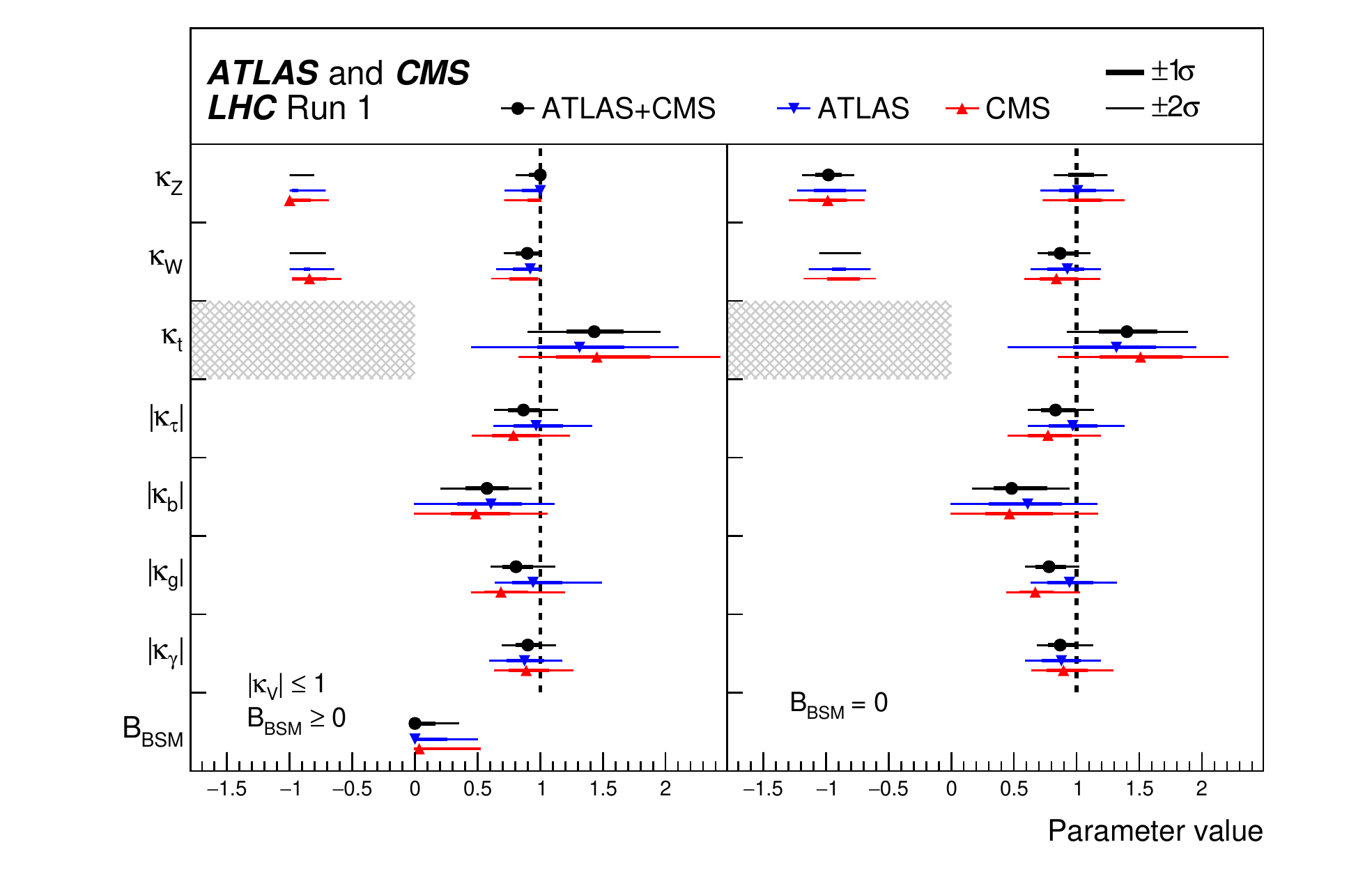}
\caption[]{The fit results for two parameterisations: the first one for
BR$_{\mathrm{BSM}} \ge $0 and $\upkappa_{\PV} \le 1$, and the second one for  
BR$_{\mathrm{BSM}}$= 0. The measured  results for the combination of ATLAS and CMS are reported 
together with their uncertainties, as well as the individual results from each experiment. The 
error bars indicate the $1\,\sigma$  and $2\,\sigma$ (thin lines) intervals. The hatched areas 
show the non-allowed regions for the $\upkappa_{\PQt}$ parameter, which is assumed to be positive 
without loss of generality. The figure is from \Bref{Khachatryan:2016vau}.}
\label{fig:fit_k_8} 
\end{figure}

In Fig.~\ref{fig:fit_k_8}, the fit results for the two parameterisations: the first one for
BR$_{\mathrm{BSM}} \ge 0$ and $\upkappa_{\PV} \le 1$, and the second one for  
BR$_{\mathrm{BSM}} = 0$. 
The measured results for the combination of ATLAS and CMS are reported together with their
uncertainties, as well as the individual results from each experiment. The error bars indicate 
the $1\,\sigma$  and $2\,\sigma$ (thin lines) intervals. 

\begin{figure}[hbt]
\centering
\includegraphics[width=1.\textwidth]{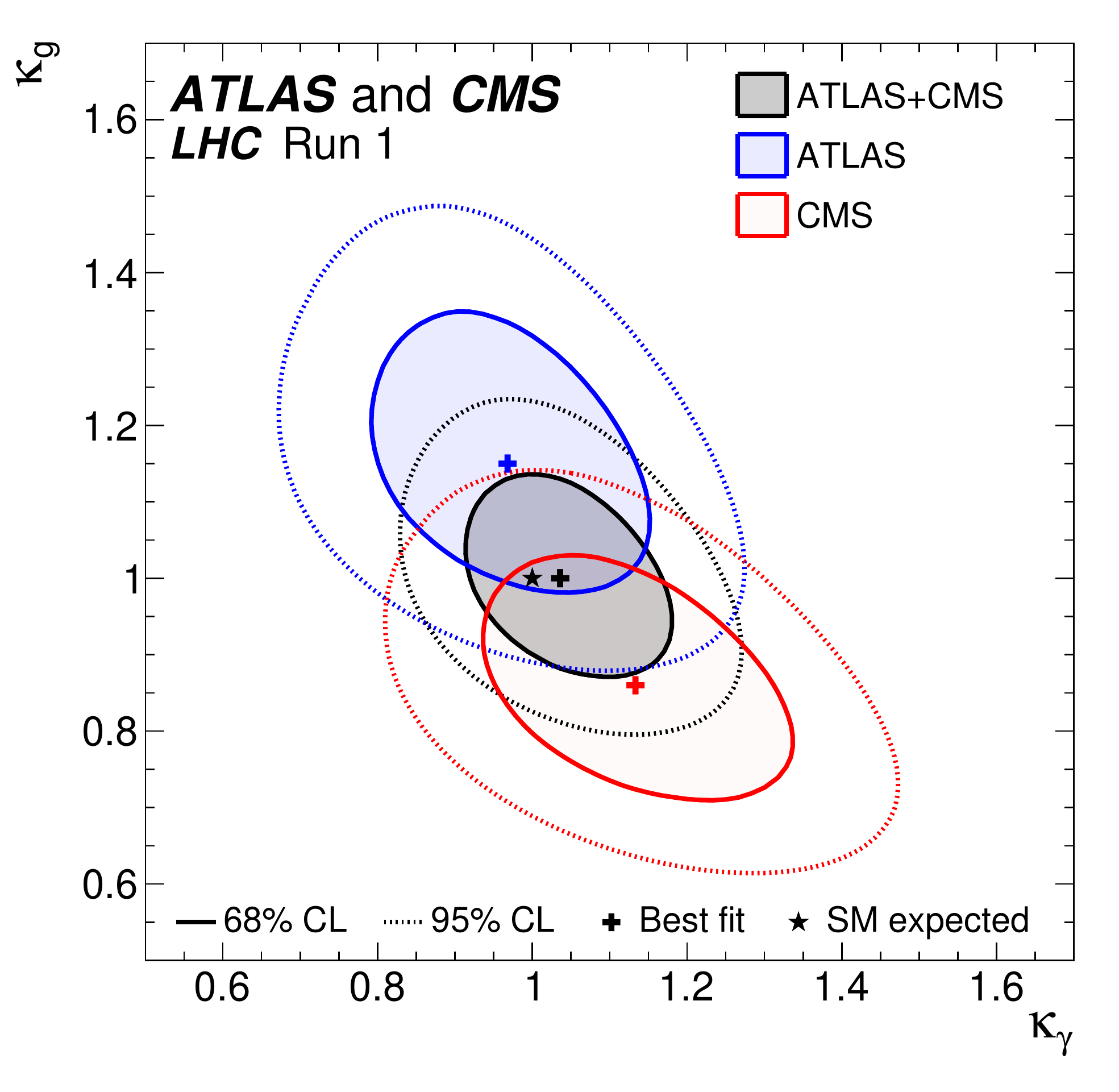} 
\caption[]{Negative log-likelihood contours at $68\%$ and $95\%$ CL in the 
($\upkappa_{\PGg}$,$\upkappa_{\Pg}$) plane for the combination of ATLAS and CMS and for each
experiment separately, as obtained from the fit to the parameterisation constraining all the other
coupling modifiers to their SM values and assuming $\text{BR}_{\text{BSM}} = 0$. The figure is from \Bref{Khachatryan:2016vau}.}
\label{fig:kg_kgamma} 
\end{figure}

{\bf The gluon and photon loops}. A scenario were new heavy particles contribute to loop-induced 
processes in Higgs boson production or decay, and all the couplings to SM particles are the same 
as in the SM,   and thus $\text{BR}_{\mathrm{BSM}} = 0$, could be tested. In this case only the 
gluon-gluon  production and decay loops in the $\PH \to \PGg\PGg$ could be affected by the 
presence of additional particles. The results of this fit, in which only the effective coupling 
modifiers $\upkappa_{\PGg}$ and $\upkappa_{\Pg}$ are the free parameters, and with all the other 
coupling modifiers fixed to their SM value of unity, is shown in Fig.~\ref{fig:kg_kgamma}. 
The point ($\upkappa_{\PGg}$= 1,$\upkappa_{\Pg}$=1) lies within the $68\%$ CL contour and the 
p-value of the compatibility between the data and the SM predictions is $82\%$.

\begin{figure}[hbt]
\centering
\includegraphics[width=1.\textwidth]{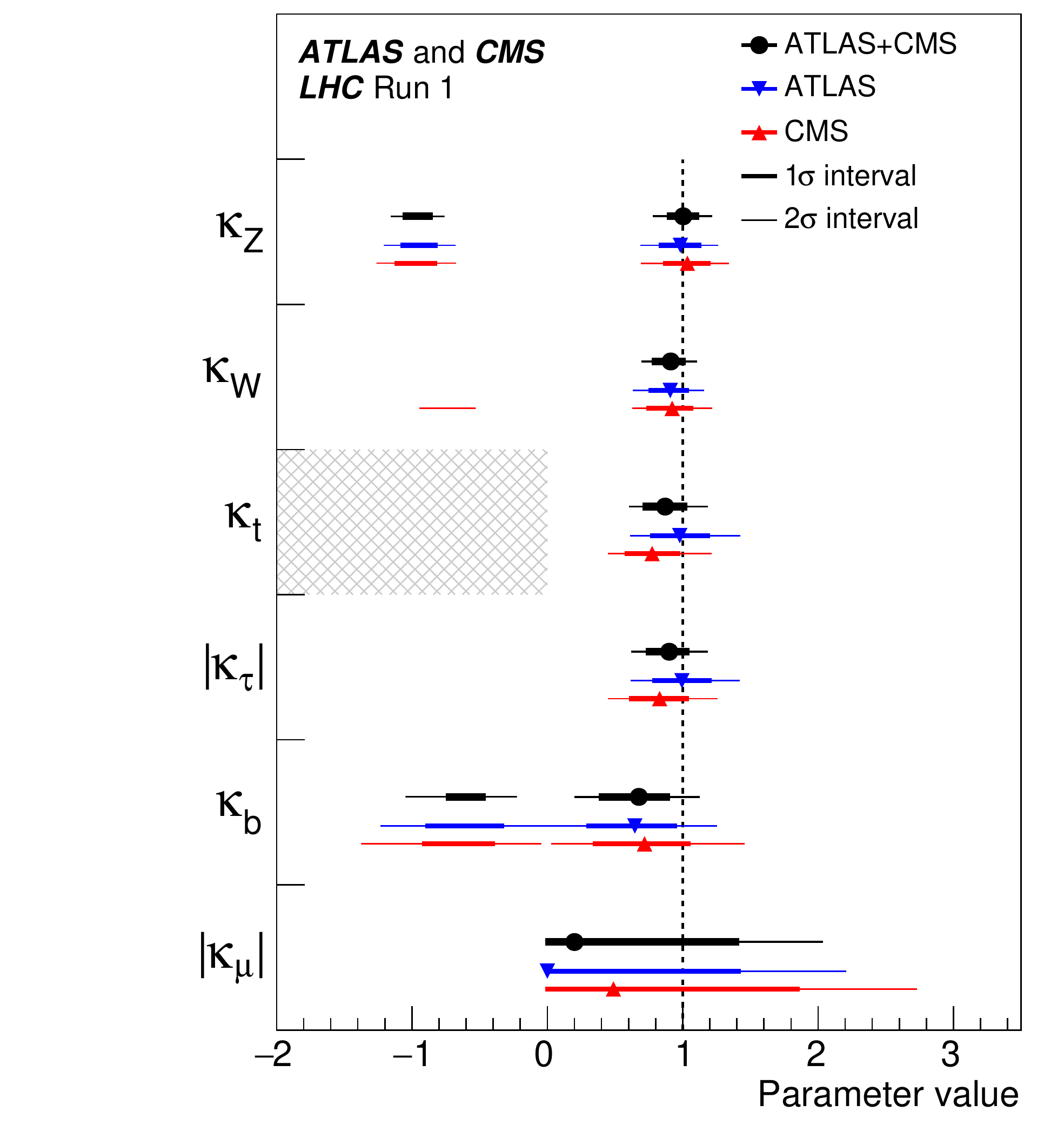} 
\caption[]{Best fit values of parameters for the combination of ATLAS and CMS , and separately for
 each experiment, for the parameterisation assuming the absence of BSM particles in the loops. 
The hatched area indicates the non-allowed region for the parameter that is assumed to be positive
 without loss of generality. The error bars indicate the $1\,\sigma$ (thick lines) and $2\,\sigma$ 
(thin lines) intervals. For the parameters with no sensitivity to the sign, only the absolute 
values are shown. The figure is from \Bref{Khachatryan:2016vau}.}
\label{fig:fit_k_6} 
\end{figure}

{\bf{Model with 6 parameters}}. Given that the effective coupling modifiers
$\upkappa_{\Pg}$  and $\upkappa_{\PGg}$ are measured to be consistent with the SM expectations,
we assume in the following that there are no new particles in these loops. 
The SM relations for the loops are used with their respective coupling modifiers. This
leads to a parameterisation with six free coupling modifiers: $\PW$, $\PZ$,  $\PQt$, $\PQb$, 
$\PGt$ and $\PGm$. 
The results of the $\PH \to \PGm\PGm$ analysis are included for this specific case. 
In this more constrained fit, it is also assumed that $\text{BR}_{\mathrm{BSM}}= 0$.
Fig.~\ref{fig:fit_k_6} shows the results of the fit for the combination of ATLAS and CMS and 
separately for each experiment.  
From the comparison of these results with those of the fitted decay signal strengths of 
Fig.~\ref{fig:decay_mu} it is evident that the 6 parameters fit results in lower values 
of the coupling modifiers than the SM expectation. 
This is a consequence of the low value of $\upkappa_{\PQb}$, as measured by the experiments. 
A low value of $\upkappa_{\PQb}$ reduces the total Higgs boson width through the dominant 
$\PH \to \PQb\PQb$ partial decay width, and, as a consequence, the measured values of all the 
coupling modifiers are reduced. 

{\bf{Lepton vs quark, up-fermion vs down-fermion}}. 
Models of new physics beyond the SM (as the THDM or MSSM) predict differences in the coupling 
modifications for up-type fermions versus down-type fermions or for leptons versus quarks. 
The parameter of interest are  $\lambda_{\PQd \PQu} = \upkappa_{\PQd} / \upkappa_{\PQu}$,   
for the up- and down-type fermion symmetry test, and 
$\lambda_{\Pl \PQq} = \upkappa_{\Pl} / \upkappa_{\PQq}$ for the lepton and quark symmetry test. 

The combined experimental result for the up and down quark symmetry test is: 
$\lambda_{\PQd\PQu} = 0.91^{+0.12}_{-0.11}$,
where the down-type fermion couplings are mainly probed by the $\PH \to \PQb\PQb$ 
and $\PH \to \PGt\PGt$ decays.

The combined experimental result for the lepton and quark symmetry test is: 
$\lambda_{\Pl \PQq} = 1.06^{+0.15}_{-0.14}$, 
where the quark couplings are mainly probed by the ggF process, the $\PH \to \PGg\PGg$ and 
$\PH \to \PQb\PQb$ decays, and to a lesser extent by the $\PQt\PQt\PH$ process; while the lepton 
couplings are probed by  the $\PH \to \PGt\PGt$ decays.  
The results are expected to be insensitive to the relative sign of the couplings because there 
is no sizeable lepton-quark interference in any of the relevant Higgs boson production and decay 
processes.


\subsection{Ratios of cross sections and branching ratios}

The measured Higgs boson rates are sensitive to the product of the cross sections times the 
branching ratios. Thus, from the measurements of the rate of a single process, the cross sections 
and decay branching ratios cannot be separately determined in a model-independent way. 
Using more processes, ratios of cross sections and branching ratios can be extracted, from a 
combined fit to the data. This could be achieved by normalising the yield of any specific channel 
$i \to \PH \to \Pf$ to the reference process $\Pg\Pg \to \PH \to \PZ\PZ$. This channel has been 
chosen by the experiments because the combined value for~$\sigma( \Pg\Pg \to \PH \to \PZ\PZ)$ has 
the smallest systematic and one of the smallest overall uncertainties. 

Expressing the measurements through ratios of cross sections and branching ratios has the advantage
that the ratios are independent of the theoretical predictions on the inclusive production cross 
sections and decay branching ratios of the Higgs boson. In particular, they are not subject to 
the dominant signal theoretical uncertainties on the inclusive cross sections for the various 
production processes.
The remaining theoretical uncertainties are the ones due to the acceptances and selection 
efficiencies in the various categories, for which SM~Higgs boson production and decay kinematics 
are assumed in the simulations.
       
\begin{figure}[hbt]
\centering
\includegraphics[height=.7\textheight,width=.7\textwidth]{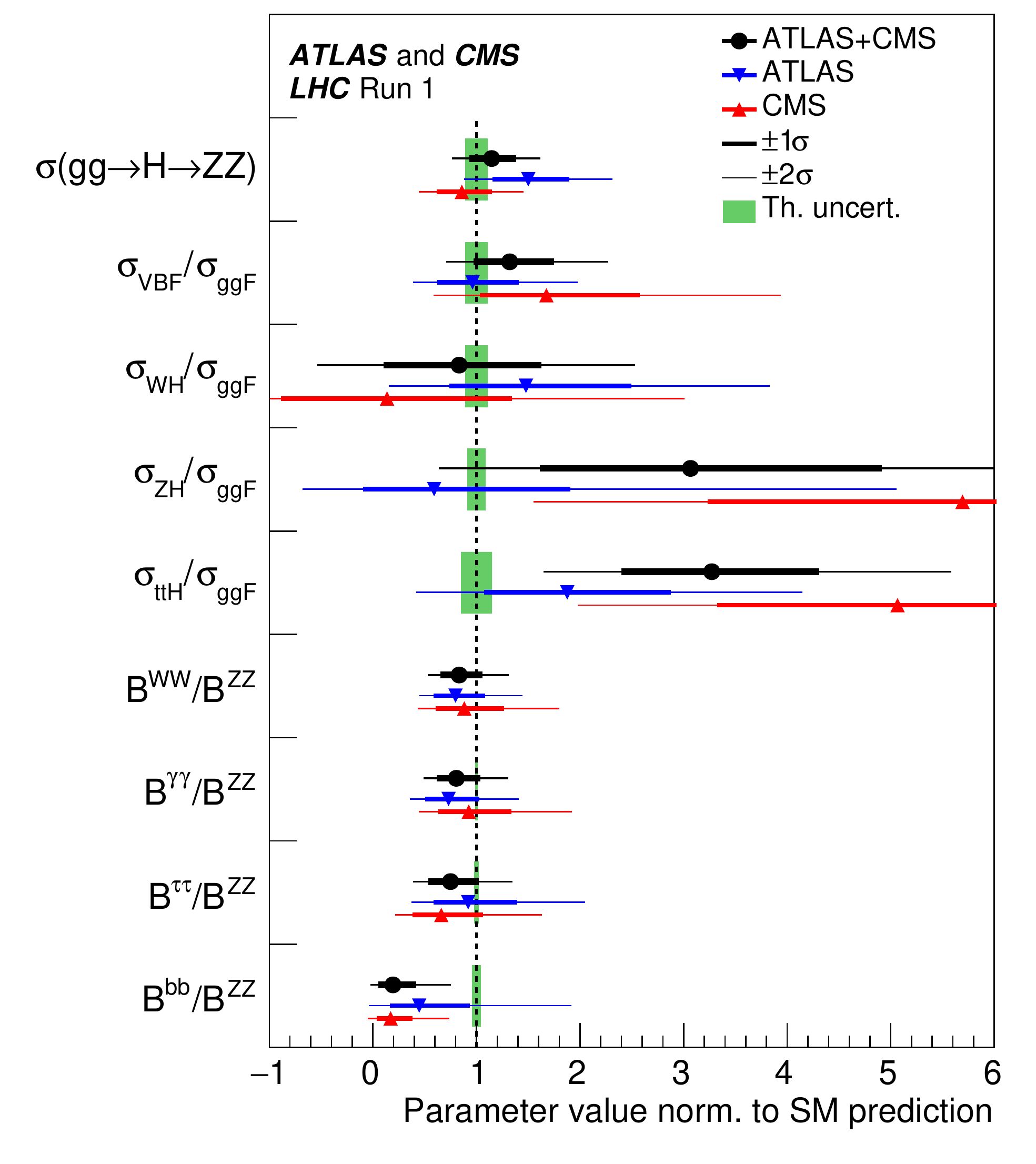}
\caption[]{Best-fit values of the $\sigma(\Pg\Pg \to \PH \to \PZ\PZ)$ cross section 
and of ratios of cross  sections and branching ratios, as obtained from the generic 
parameterisation described in the text for the combination of ATLAS and CMS measurements and 
for each experiment individually. The error  bars indicate the 
$1\sigma$~(thick lines) and $2\sigma$~(thin lines) intervals. In this figure, the fit
   results are normalised to the SM predictions for the various parameters and the shaded bands
    indicate the theory uncertainties on these predictions. The figure is from \Bref{Khachatryan:2016vau}.}
\label{fig:sigma_br_ratios}
\end{figure}

The product of the cross section and the branching ratio of $i\to \PH \to \Pf$ can then be 
expressed using the ratios as:
\bq
\sigma_i\cdot \text{BR}^{\Pf} = 
\sigma( \Pg\Pg \to \PH \to \PZ \PZ) \times \left(\frac{\sigma_i}{\sigma_{\text{ggF}}}\right)
\times \left(\frac{\text{BR}^{\Pf}}{\text{BR}^{\PZ\PZ}}\right),
\label{eq:sigratio}
\eq
where $\sigma(\Pg\Pg \to \PH \to \PZ \PZ) = \sigma_{ggF}\cdot \text{BR}^{\PZ\PZ}$ and the narrow 
width approximation is assumed. Since the cross section $\sigma(\Pg\Pg \to \PH \to \PZ\PZ)$ is 
constraining the normalisation, the ratios in~Eq.~\ref{eq:sigratio} can be determined separately, 
based on the five production processes (ggF, VBF, $\PW\PH$, $\PZ\PH$, and $\PQt\PQt\PH$) 
and five decay modes ($\PH\PZ\PZ$,  $\PH\PW\PW$, $\PH\PGg\PGg$, $\PH\PQt\PQt$, and 
$\PH\PQb\PQb$). 
The combined fit results is presented as a function of nine parameters of interest: the reference
cross section times branching ratio,~$\sigma(gg \to \PH \to \PZ\PZ)$, four ratios of production 
cross sections,~$\sigma_i/\sigma_{ggF}$, and four ratios of branching 
ratios,~$\text{BR}^{\Pf}/\text{BR}^{\PZ\PZ}$, as shown in Fig.~\ref{fig:sigma_br_ratios}. In this figure 
the fit results are normalised to the SM predictions for the various parameters and the shaded 
bands indicate the theory uncertainties on these predictions.
The theory uncertainties on the ratios of branching ratios are very small, and therefore almost not visible. 
The combination of $7$ and $8\,\UTeV$ data is carried out under the assumption that the ratios 
of the production cross sections with respect to the SM predictions are the same at $\sqrt{s}=7$ 
and $8\UTeV$. 

The total relative uncertainty on~$\sigma(\Pg\Pg \to \PH \to \PZ\PZ)$ is approximately~$19\%$, 
where the dominant  contribution is  the statistical one. The total relative systematic 
uncertainty is ~$\sim4\%$. 
The ratio of cross sections $\sigma_{VBF}/\sigma_{ggF}$ and the ratios 
$\text{BR}^{\PW\PW}/\text{BR}^{\PZ\PZ}$ and~$\text{BR}^{\PGg\PGg}/\text{BR}^{\PZ\PZ}$
are measured with a relative uncertainty of approximately~$30\%$, while the 
$\text{BR}^{\PGt\PGt}/\text{BR}^{\PZ\PZ}$ ratio is measured with a relative accuracy 
of approximately~$40\%$.    

The $p$-value of the compatibility between the data and the SM predictions is~16\%. 
The most precise measurements are all consistent with the SM predictions within less than
~$2\,\sigma$. The production cross-section ratio 
$\sigma_{\PQt \PQt \PH} / \sigma_{ggF}$ 
relative to the  SM~ratio,  is measured to be $3.3^{+1.0}_{-0.9}$, 
corresponding to an excess compared to the SM prediction of approximately $2.3\,\sigma$. 
This excess is mainly due to the multi-lepton categories.
The ratio of branching ratios $\text{BR}^{\PQb\PQb}/ \text{BR}^{\PZ\PZ}$ relative to the
SM ratio is measured to be $0.19^{+0.21}_{-0.12}$. 
In this parameterisation, the high values found for the production cross-section ratios for the 
$\PZ\PH$ and $\PQt\PQt\PH$ processes induce a low value for the 
$\PH\PQb\PQb$ decay branching ratio because the $\PH\PQb\PQb$~decay channel 
does not contribute to the observed excesses. The result is  an overall deficit
 compared to the SM~prediction of approximately $2.5\,\sigma$.

\subsection{A summary plot}

 \begin{figure}[hbt]
\centering
\includegraphics[width=1.0\textwidth]{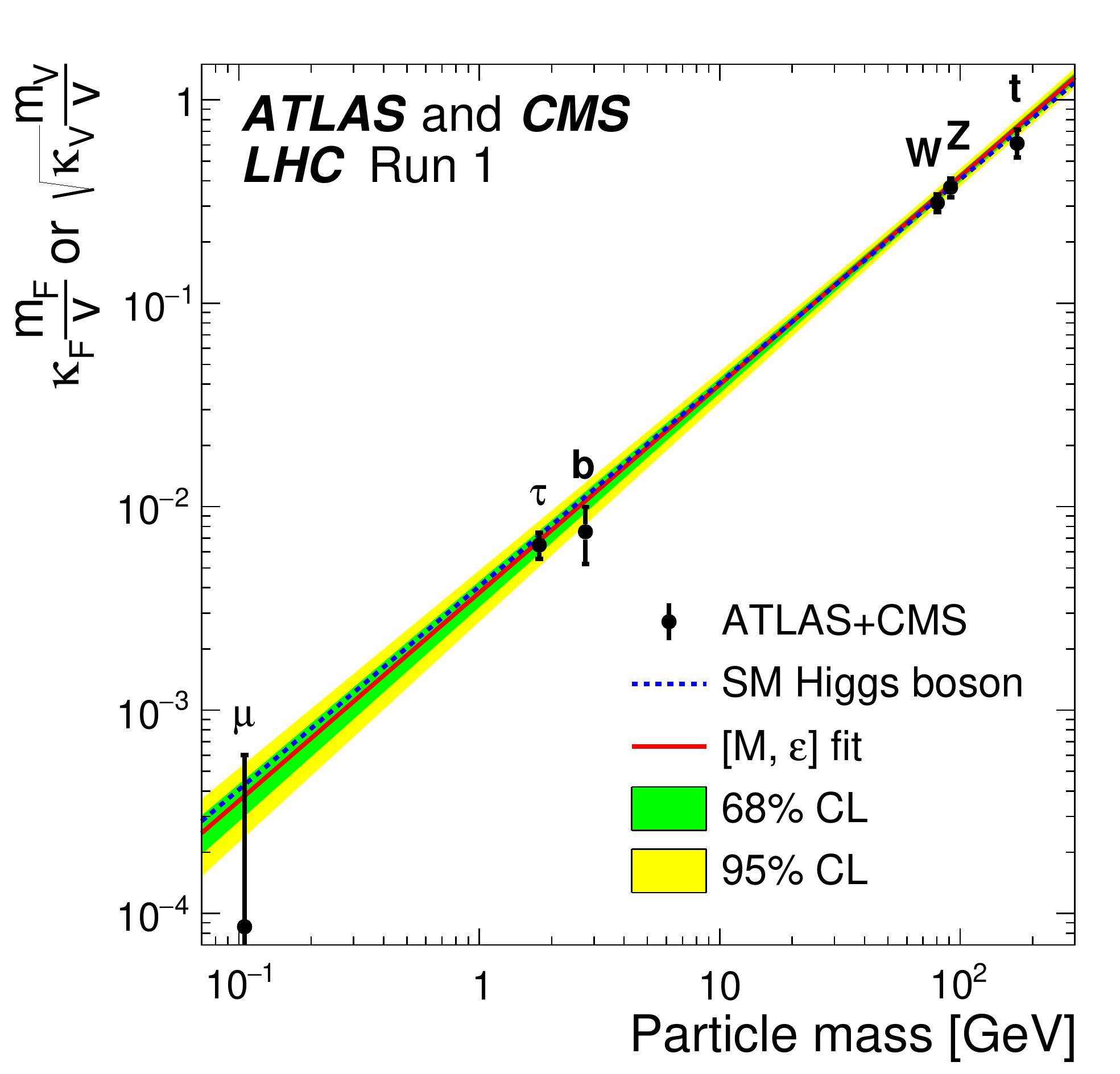}
\caption[]{Fit results as a function of the particle mass  in the case of the parameterisation with
 reduced coupling modifiers $y_{V,i}$ for the weak vector bosons, and  $y_{f,i}$ for
the fermions,  as explained in the text, for the combination of ATLAS and CMS. 
The dashed line indicates the predicted  dependence on the particle mass for the SM Higgs boson.
The solid (red) line indicates the best fit result to the [M,$\epsilon$] phenomenological model of 
\Bref{ref-mass} with the corresponding $68\%$ and $95\%$ CL bands in green and yellow. The figure is from \Bref{Khachatryan:2016vau}.}
\label{fig:couplings_vs_mass}
\end{figure}

The relation between the coupling modifiers and the SM predictions can be presented as a function 
of the mass of the particles to which the $\PH$ boson is coupling. 
The coupling of the Higgs to vector bosons of mass $m_{\PV}$ when expressed as a function on 
$\upkappa_{\PV}$ is: 
\bq 
y_{V, i} = \sqrt {\upkappa_{\PV,i} {\frac{g_{\PV,i}}{2\nu}}} = 
\sqrt {\upkappa_{\PV,i}} \frac{m_{\PV,i}}{\nu}
\eq
where $g_{\PV,i}$ is the absolute Higgs boson coupling strength and $\nu = 246 \UGeV$  
is the vacuum expectation value of the Higgs field.

The coupling of the Higgs to the fermions of mass $m_{\Pf}$ when express as a function on 
$\upkappa_{\Pf}$ is: 

\bq 
y_{f, i} = \upkappa_{\Pf,i} \frac{g_{\Pf,i}}{\sqrt{2}} = \upkappa_{\Pf,i} \frac{m_{\Pf,i}}{\nu}.
\eq

The linear scaling of the reduced coupling modifiers as a function of the particle masses is 
shown in Fig.~\ref{fig:couplings_vs_mass} and indicates the consistency of the 
measurements with the SM.

%% file: NHCS522.tex
\subsection{Theoretical perspectives}
In the previous Section different variants of the original kappa-framework have been presented
and discussed in relation to experimental data. Here we briefly summarise theoretical
perspectives on improving the experimental strategy.

One possibility to go  beyond the kappa framework is represented by the SM effective field
theory: SMEFT will be described in details in Section \ref{Sect6} and it is based on the Lagrangian of
\eqn{SMEFTLag}, which is the SM Lagrangian with the addition of $\mrdim = 6$ operators.

Several (theoretical) analyses have been performed with the available Run~1 data, as summarised in
\Bref{Englert:2015hrx}, see also \Brefs{Englert:2014uua,Ellis:2014jta,Buckley:2015lku}.
These analyses always use a subset of the full set of $\mrdim = 6$, gauge invariant operators
(\eg the so-called ``Warsaw'' basis) and show a good agreement, with differences due to different 
sets of assumptions. The results can be summarised by saying that current measurements show 
good agreement with the SM hypothesis. In practice, the predicted number of events for a given 
final state $\PF$ is obtained as
\bq
\mrN_{\mathrm{th}} = \sigma\lpar \PH + \PX \rpar \times \mathrm{BR}\lpar \PH \to \PY\PY \rpar
 \times \mathrm{BR}\lpar \PX\,,\PY \to \PF \rpar  \times \mathcal{L} \spp
\eq
The ``number of events'' is 
\bq
\mrN_{\mathrm{ev}} = \mrN_{\mathrm{th}}\,\varepsilon_p\,\varepsilon_d \spc
\eq
where $\varepsilon_{p,d}$ are the efficiencies to measure production and decay and
$\mathcal{L}$ is the luminosity.
Theoretical uncertainties for production and decay channels are shown in Tab.~\ref{tubd} while
the order of available calculations is given in Tab.~\ref{tab:SMCrossSections}. In  Tab.~\ref{tubd} 
(differently from Tab.~\ref{tab:SMCrossSections} as used by the experimental analyses) 
the uncertainty is  given by the linear sum of the QCD scale variation  and of the PDF uncertainty as presented 
in \Brefs{Dittmaier:2011ti,Dittmaier:2012vm}. 
\begin{table}[t]
\caption{Theoretical uncertainties for production and decay channels in $\%$}
\hspace{0.3cm}
\begin{center}
\begin{tabular}{@{}llll@{}} \hline\hline
Production & Decay \\ \hline
$\Pp\Pp \to \PH$                 & $14.7$ & $\PH \to \PAQb\PQb$       & $6.1$ \\
$\Pp\Pp \to \PH + \mathrm{j}$    & $15$   & $\PH \to \PGg\PGg$        & $5.4$ \\
$\Pp\Pp \to \PH + 2\,\mathrm{j}$ & $15$   & $\PH \to \PGtp\PGtm$      & $2.8$ \\
$\Pp\Pp \to \PH\,\PZ$            & $5.1$  & $\PH \to 4\,\Pl$          & $4.8$ \\
$\Pp\Pp \to \PH\,\PW$            & $3.7$  & $\PH \to 2\,\Pl\,2\,\PGn$ & $4.8$ \\
$\Pp\Pp \to \PAQt\PQt\PH$        & $12$   & $\PH \to \PZ\PGg$         & $9.4$ \\
                                 &        & $\PH \to \PGmp\PGmm$      & $2.8$ \\ \hline\hline
\end{tabular} 
\label{tubd}
\end{center}
\end{table}
Extrapolating to $13\UTeV$ a major improvement is expected, in particular when
differential distributions will be included.

Global constraints of the SMEFT have been developed in \Bref{Berthier:2015oma}, with results 
that show how the SMEFT theory uncertainties should not be neglected in future fits, see also 
\Bref{Berthier:2016tkq}.
Preliminary results of a Bayesian fit to the Wilson coefficients using data on EW precision 
observables and Higgs boson signal strengths have been presented in \Bref{deBlas:2014ula}.

\Bref{Englert:2015dlp} deals with Higgs production through weak boson fusion with 
subsequent decay to bottom quarks. By combining jet substructure techniques and matrix element 
methods in different limits the authors motivate this channel as a probe of the bottom-Yukawa 
interactions in the boosted regime. 

The possibility to separate, in gluon fusion, loop-induced Higgs boson production from 
point-like production has been examined in \Bref{Langenegger:2015lra}. The Higgs boson is 
reconstructed in the $\PH\PGg\PGg$ final state at very large transverse momentum. Using the 
Higgs boson yields (normalised to the overall rate) and the shape of the Higgs boson $\pT$ 
distribution, the two hypotheses can be separated with $2$ standard deviations with an 
integrated luminosity of about $500$\ifb. The largest experimental uncertainty affecting 
this estimate is the background event yield. The theoretical uncertainties from missing top 
mass effects are large, but can be decreased with dedicated calculations.

For the measured Higgs boson mass of $125\UGeV$ the limit of heavy top quarks provides a 
reliable approximation as long as the relative QCD corrections are scaled with the full 
mass-dependent LO cross section. In this limit the Higgs coupling to gluons can be described 
by an effective Lagrangian. The same approach has been applied to the coupling of more 
than one Higgs boson to gluons \cite{Spira:2016zna}, deriving the effective Lagrangian 
for multi-Higgs couplings to gluons up to N${}^{4}$LO thus extending previous results for 
more than one Higgs boson. 

The authors of \Bref{Bizon:2016wgr} have examined the constraints on the trilinear Higgs 
coupling that originate from associated and vector boson fusion Higgs production in the 
context of the SMEFT, showing that future LHC runs may be able to probe modifications of 
the coupling with a sensitivity similar to the one that is expected to arise from 
determinations of double-Higgs production. 

The authors of \Bref{Degrassi:2016wml} have proposed a method to determine the trilinear 
Higgs self coupling that is alternative to the direct measurement of Higgs pair production 
total cross sections and differential distributions. The method relies on the effects that 
electroweak loops featuring an anomalous trilinear coupling would imprint on single Higgs 
production at the LHC. It is found that the bounds on the self coupling are already competitive 
with those from Higgs pair production and will be further improved in the current and next 
LHC runs.

The authors of \Bref{Bishara:2016jga} have proposed a novel strategy to constrain the bottom 
and charm Yukawa couplings by exploiting LHC measurements of transverse momentum distributions 
in Higgs production. The method does not rely on the reconstruction of exclusive final states 
or heavy-flavour tagging. Compared to other proposals it leads to an enhanced sensitivity to 
the Yukawa couplings due to distortions of the differential Higgs spectra from emissions 
which either probe quark loops or are associated to quark-initiated production. 

The authors of \Bref{Gritsan:2016hjl} investigated anomalous interactions of the Higgs boson 
with heavy fermions, employing shapes of kinematic distributions, presenting applications of 
event generation, reweighting techniques for fast simulation of anomalous couplings, as well 
as matrix element techniques for optimal sensitivity. 

The authors of \Bref{Demartin:2016axk} have studied Higgs boson production in association with 
a top quark and a $\PW$ boson at the LHC. At NLO in QCD, $\PQt\PW\PH$ interferes with 
$\PAQt\PQt\PH$ and a procedure to meaningfully separate the two processes needs to be developed. 

The authors of \Bref{Hespel:2016qaf} analysed the production of a top quark pair through 
a heavy scalar at the LHC. While the background and the signal can be obtained at NNLO and NLO 
in QCD respectively, that is not the case for their interference, which is currently only 
approximately known at NLO. 
In order to improve the accuracy of the prediction for the interference term, the effects of 
extra QCD radiation are considered: as a result, it is found that the contribution of the 
interference is important both at the total cross-section level and, most importantly, for 
the line-shape of the heavy scalar. 

The main lesson from Run~1 of LHC is that, to first approximation, we have a (minimal)
SM-like scalar sector. To be more precise, the best precisions achieved are approximately 
$30\%$ for the ratio of cross sections VBF/ggF (vector boson fusion and gluon-gluon fusion) 
and for the ratios of branching fractions, 
$\mathrm{BR}(\PW\PW)/\mathrm{BR}(\PZ\PZ)$ and $\mathrm{BR}(\PGg\PGg)/\mathrm{BR}(\PZ\PZ)$. The 
ratios of coupling modifiers (kappa parameters) are measured with precisions of approximately 
$10 \to 20\%$.
The main message from Run~1: it is important to check the apparent minimality of the Higgs
sector as it is important to anticipate deviations. The improvements expected from Run~2 will 
come from greater statistics, greater kinematic range and improvement in 
theoretical uncertainties.
To be considered together with the LHC data, are the EW precision data (EWPD). For instance, 
measurements of the $\PW$ mass provide an important consistency check of the SM and constrain 
the possibility of physics beyond the SM. 

The work of \Bref{Bjorn:2016zlr} has shown that the extra error introduced in these 
measurements due to SMEFT higher dimensional operators is subdominant to the current 
experimental systematic errors. This means that the leading challenge to interpreting 
these measurements in the SMEFT is the pure theoretical uncertainty in how these measurements 
are mapped to Lagrangian parameters. 

Inclusion of EWPD in a global fit deserves additional comments. Usually bounds on the
coefficients are obtained in two ways: individual coefficients are switched one at a time, 
or marginalised in a simultaneous fit.
In \Bref{Berthier:2016tkq} the global constraint picture on SMEFT parameters has been updated 
with the conclusion that stronger constraints can be obtained by using some combinations of 
Wilson coefficients, when making assumptions on the UV completion of the SM. 
To summarise: global fits show that the degree of constraint on the SMEFT parameters is
strongly dependent on the assumptions made about possible UV physics matched onto the SMEFT. The
theoretical uncertainty, due to neglected terms in the SMEFT, is also UV dependent.

It is worth noting that fitting $\mrdim = 6$ Wilson coefficients to LHC Higgs data can be done 
and has been done for Run~1 data, but not by members of the ATLAS and CMS collaborations.
What has been learnt is that kinematic distributions can significantly improve the multi-dimensional 
parameter by resolving strong correlations present in total rate measurements.

As discussed in \Brefs{Brehmer:2015rna,Biekotter:2016ecg}, a few selected kinematic distributions 
can be used to collect information on modified Higgs couplings, for example in the gluon fusion
production process. In the top-gluon-Higgs sector one can compare three different analysis 
strategies: a modified $p_{\mrT}$ spectrum of boosted Higgs production in gluon 
fusion~\cite{Banfi:2013yoa}, off-shell Higgs production, and a measurement of the gluon fusion 
vs $\PAQt\PQt\PH$ production rates. Unfortunately, explicit threshold effects in boosted Higgs 
production are too small to be observable in the near future~\cite{Buschmann:2014twa}. 
Global analyses including kinematic information in all Higgs channels cannot 
rely on the kappa framework, but they could be based on a SMEFT. Such analyses provide 
potentialities and challenges at the same time~\cite{Brehmer:2015rna,Biekotter:2016ecg}.

%% file: NHCS53.tex
In \Bref{Kauer:2012hd} the off-shell production cross section has been shown to be sizeable 
at high $\PZ\PZ\,$-invariant mass in the gluon fusion production mode, with a ratio relative 
to the on-peak cross section of the order of $8\%$ at a center-of-mass energy of $8\UTeV$.
This ratio can be enhanced up to about $20\%$ when a kinematical selection used to extract 
the signal in the resonant region is taken into account~\cite{Kauer:2013cga}. This arises from 
the vicinity of the on-shell $\PZ$ pair production threshold, and is further enhanced at the 
on-shell top pair production threshold.

In \Bref{Caola:2013yja} the authors demonstrated that, with few assumptions and using
events with pairs of $\PZ$ particles, the high invariant mass tail can be used
to constrain the Higgs width. For a detailed description, see \Bref{Passarino:2013bha}.

Off-shell measurements are (much) more than consistency checks on $\Gamma_{\PH}$:
observing an excess in the off-shell measurement will be a manifestation of BSM physics,
which might or might not need to be in relation with the $\PH$ width.
We need to extend the SM with dynamics, representing an intermediate step toward the next SM, 
distancing the experimental analysis from repeated refinements due to ever-improving calculations.

\paragraph{How was off-shell production  used?}
First one introduces the notion of $\infty\,$-degenerate solutions for the Higgs couplings to 
SM particles, as done in \Bref{Dixon:2013haa,Caola:2013yja} and uses the fact that the enhanced 
tail is obviously $\Gamma_{\PH}\,$-independent and that this could be exploited to constrain 
the Higgs width model-independently. Finally, use a matrix element method to construct a 
kinematic discriminant to sharpen the constraint, see \Bref{Campbell:2014gha}.

More precisely, \Brefs{Caola:2013yja,Campbell:2013una} define the following scenario for 
on-shell $\infty\,$-degeneracy: there is invariance under a scaling of the Higgs couplings and 
of the total Higgs width defined by
\bq
\sigma_{i \to \PH \to f} = \lpar \sigma\cdot\mathrm{BR}\rpar = 
\frac{\sigma^{\myprod}_i\,\Gamma_f}{\Gamma_{\PH}}
\quad
\sigma_{i \to \PH \to f} \;\varpropto\;\frac{g^2_i g^2_f}{\Gamma_{\PH}}
\quad 
g_{i,f} = \xi\,g^{\mySM}_{i,f}, \;\; \Gamma_{\PH} = \xi^4\,\Gamma_{\PH}^{\mySM}
\label{ascal}
\eq

The gluon fusion production cross section as a function of $\PZ\PZ$ invariant mass can be written 
as:
\bq
\frac{d\sigma_{\Pg\Pg \to \PH \to \PZ\PZ}}{dm^2_{\PZ\PZ}}  
\sim 
\frac{g_{\Pg\Pg\PH}^2 g_{\PH\PZ\PZ}^2}{(m^2_{\PZ\PZ} - m^2_{\PH}) + m^2_{\PH} \Gamma^2_{\PH}}. 
\eq 
where $g_{\Pg\Pg\PZ}$ and $g_{\PH\PZ\PZ}$ are the couplings of the Higgs boson to gluons and 
$\PZ$ bosons, respectively. Integrating either in a small region around $m_{\PH}$, or above the 
mass threshold $m_{\PZ\PZ} > 2 m_{\PZ}$, where $(m_{\PZ\PZ} - m_{\PH}) >> \Gamma_{\PH}$, the 
cross sections are, respectively:
\bq 
\label{eq:resonnantregion}
\sigma^{\mathrm{on-shell}}_{\Pg\Pg \to \PH \to \PZ\PZ^*} \sim 
\frac{g_{\Pg\Pg\PH}^2 g_{\PH\PZ\PZ}^2}{m_{\PH} \Gamma_{\PH}}
\eq 
\bq 
\label{eq:offshell}
\sigma^{\mathrm{on-shell}}_{\Pg\Pg \to \PH \to \PZ\PZ^*} \sim 
\frac{g_{\Pg\Pg\PH}^2 g_{\PH\PZ\PZ}^2}{2 m_{\PH}^2}
\eq 
The cross section for the on-shell production will not change if the squared product of the 
coupling constants $g_{\Pg\Pg\PH}^2 g_{\PH\PZ\PZ}^2$ and the total width $\Gamma_{\PH}$ are scaled 
by a common factor $r$. On the contrary, away from the resonance the cross section is independent 
of the total width and therefore increases linearly with $r$.
Thus a measurement of the relative off-shell to on-shell production in the $\PH \to \PZ\PZ$ 
channel provides direct information on $\Gamma_{\PH}$, as long as the coupling ratios remain 
unchanged, \ie the gluon fusion production is dominated by the top-quark loop and there are no 
new particles contributing.

The final states $\PH \to \PZ\PZ \to 4\Pl$, where one $\PZ$ boson decays to an $\Pe$ or $\PGm$ 
pair and the other to either an $\Pe$ or $\PGm$ pair,  $\PH \to \PZ\PZ \to 2\Pl 2\PGn$ and  and
$\PH \to \PW\PW \to 2\Pl 2\PGn$ have been analysed in ATLAS \cite{Aad:2015xua}, and 
CMS \cite{Khachatryan:2014iha,Khachatryan:2016ctc}.

\begin{figure}[hbt]
\centering
\includegraphics[width=.48\textwidth]{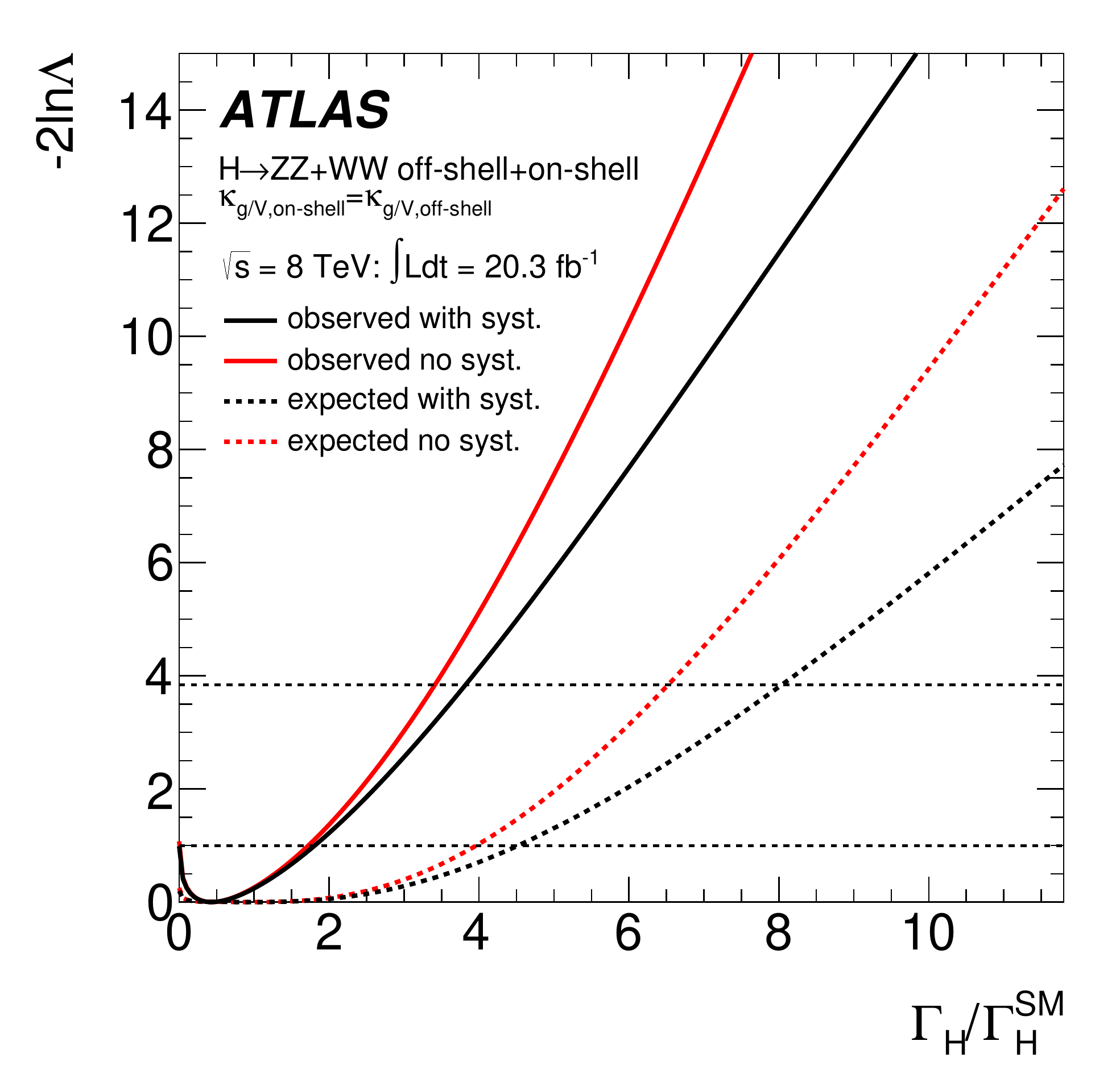}
\includegraphics[width=.48\textwidth]{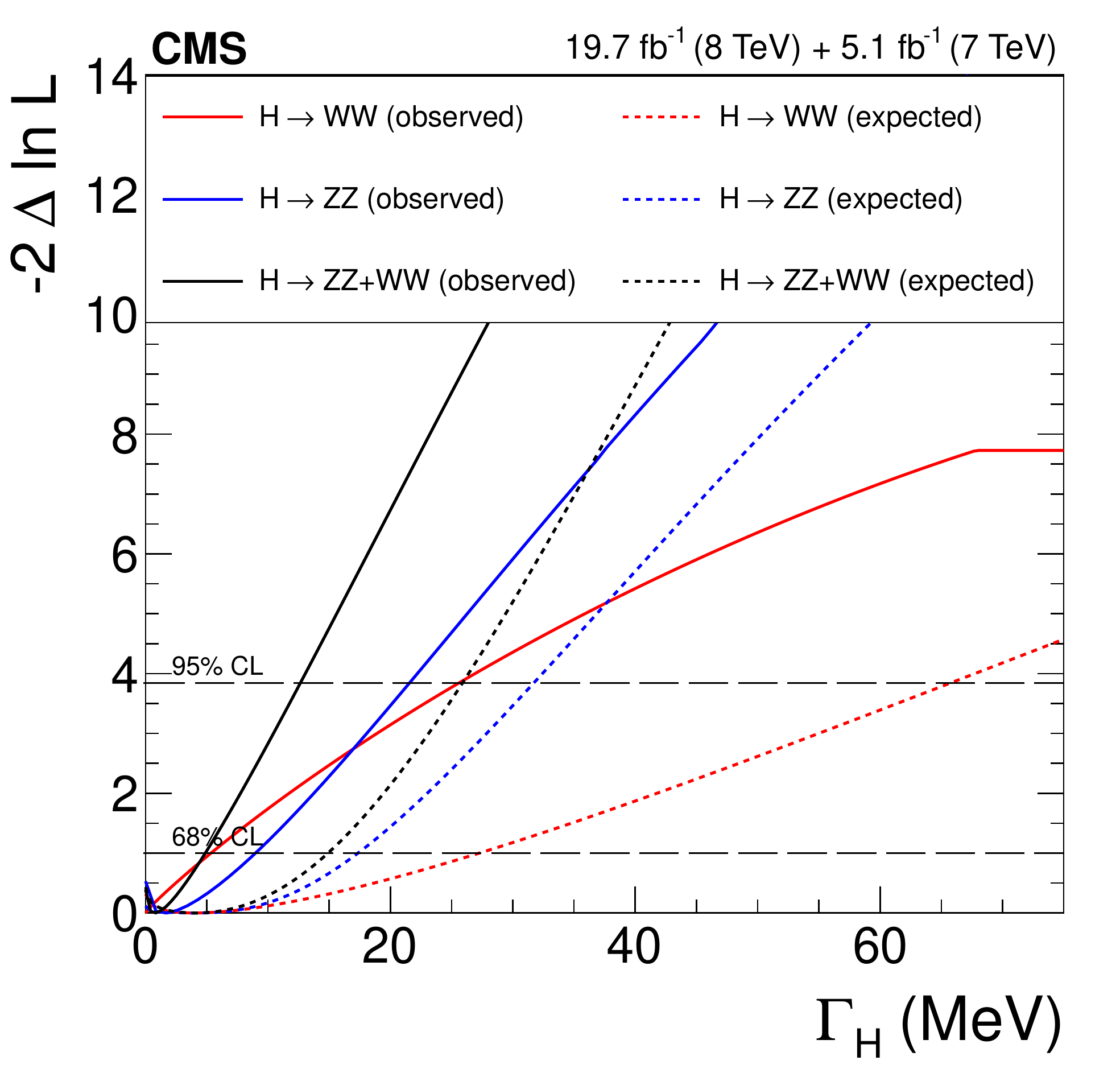}
\caption[]{Scan of the negative log-likelihood as a function of $\Gamma_{\PH} / \Gamma_{SM}$ 
(ATLAS)  and $\Gamma_{\PH}$ (CMS)  for the combined fit of the $\PH \to \PZ\PZ$ and  $\PH \to \PW\PW$  
channels at $7$ and $8\UTeV$ center of mass energy. The analyses assume the same ggF
and VBF ratio of signal strengths.}
\label{fig:width} 
\end{figure}

The results on the limit on the $\PH$ width from the analysis of the off-shell $\PH$ production 
for the ATLAS and CMS experiments are shown in Fig.~\ref{fig:width}. The observed $95\%$ CL 
upper limits on the width are $22.7$  and $13\UMeV$  for ATLAS and CMS respectively, while the 
expected $95\%$ CL upper limits are $33\UMeV$ and $26\UMeV$. 
These results have to be compared with the upper limit on the direct measurement of the width of 
the $\PH$ resonance in $\PGg\PGg$ and $4\Pl$ final states of about $2$ and $3\UGeV$ at $95\%$ CL. 
The analysis of the off-shell production improves the limits on the Higgs width by a factor of 
about $100$. 

After the experiments published their analysis, and with respect to \Bref{Passarino:2013bha} 
the following theoretical improvements  have been made:

\bei

\item full next-to-next-to-leading (NNLO) for 
$\PAQq\PQq \to \PV\PV$, \Bref{Cascioli:2014yka}, 

\item $2\,$-loop amplitudes for massless $\Pg\Pg \to \PV\PV$\, 
\Brefs{Caola:2015psa,vonManteuffel:2015msa}, 

\item $\PZ\PZ$ production in NNLO QCD, \Bref{Grazzini:2015hta}.

\eei
The $2\,$-loop amplitudes for massive $\Pg\Pg \to \PV\PV$ seem out of reach; NLO is known 
in $1/m_{\PQt}\,$-expansion, see \Bref{Melnikov:2015laa}. For off-shell studies in vector boson
fusion (VBF) see \Bref{Campbell:2015vwa}.
\paragraph{How should off-shell production be used?}
However, scaling couplings at the $\PH$ peak is not the same thing as scaling them off-peak; 
the consequence of this fact is that the kappa framework is not adequate and needs a 
generalisation, \eg using SMEFT.
Therefore, one should use SMEFT at a given order (possibly NLO) or any other consistent way
of describing SM deviations. Using SMEFT as an example, the strategy is: 
\bei

\item write any amplitude as a sum of deformed SM sub-amplitudes and 

\item add another sum of deformed, non-SM, sub-amplitudes.

\eei
The rationale for this course of action (with respect to the original kappa framework) is that
physics is symmetry plus dynamics; symmetry is quintessential (gauge invariance, \etc)
but symmetry without dynamics does not bring us this far.
At LEP the dynamics was SM, the unknowns were $\mh\,(\alphas(\mz), \dots )$, while at LHC  (post the Higgs discovery) 
the unknowns are the SM-deviations but, how to describe dynamics?
Selecting a BSM model is a choice but is there something more model independent?
The answer is positive: we need a decomposition where dynamics is controlled by amplitudes
with known analytical properties and the deviations (with a direct link to UV completions) are  ultimately connected to Wilson coefficients.

On-shell studies will tell us a lot, off-shell ones will tell us (hopefully) everything else.
We want to study the physics away from the $\PH$ peak with a SM-deformed theory (up to some
reasonable value $E \muchless \Lambda$) but we also need to reproduce (deformed) SM 
low-energy effects, \eg $\PV\PV$ and $\PQt\PQt$ thresholds; BSM loops will remain 
unresolved, just like SM loops are unresolved in the Fermi theory.
That is why one needs to expand SM deformations into a SM basis with the correct (low energy) 
behaviour: in the neighbourhood of the peak any parameterisation of the deviations will work
(certainly at LO), while running off-shell requires a deeper knowledge of the analytical properties 
of the deformed amplitude.
      
To summarise, the correct way of describing off-shell SM deviations is
\bq
\sigma_{\EFT}(s) = \mid \upkappa_{\myprod}(s)\,\upkappa_{\dec}(s) \mid^2\,\sigma_{\mySM}(s) \spc 
\eq
where the ``running'' kappa parameter ($s$ is the Higgs virtuality) is defined by
\bq
\upkappa_{\myprod}(s)\,\upkappa_{\dec}(s) = \mu^{\off}_{\PZ\PZ}(s) \spp
\eq
A discussion on $\mu^{\off}_{\PZ\PZ}(s)$ will be given in Sect.~\ref{Sect622}.
\paragraph{Off-shell gauge invariance}
A final comment on on-shell versus off-shell for LHC physics is needed.
There is an important role played by gauge invariance, thus due care is needed in the ``signal'' definition, i.e, the resonant part. In other words, we meet the usual problems 
that are present  when  unstable particles enter the game.
Certainly, LHC is not LEP, where only one resonant contribution was present;
for a light SM Higgs boson the $4\Pf$ decays are $40\%$ of the $2\Pf$ decays,
as a consequence we always face the problem of having to deal with many off-shell, unstable, 
particles, even at the $\PH$ peak. 

The question that we will readdress in Sect.~\ref{Sect6} is: how to interpret 
\bq
\Gamma\lpar \PH \to \PW\PW \to \PGn\Pl\PGn'\Pl'\rpar 
\quad \mbox{vs.} \,\,
\Gamma\lpar \PH \to \PW\PW\rpar \spc
\eq
or, stated differently, how to define $\Gamma\lpar \PH \to \PW\PW\rpar$.
The short answer is: never introduce quantities that are not well-defined; the Higgs couplings 
can be extracted from Green's functions in well-defined kinematical limits, \eg residues of 
the poles after extracting the parts which are one-particle irreducible. These are well-defined 
Quanptum-Field-Theory (QFT) objects, that we can probe both in production and in decays. From 
this perspective, $\PV\PH$ or $\mathrm{VBF}$ are on equal footing with $\Pg\Pg\PF$ and Higgs 
decays.

%% file: NNHCS6.tex
Run~1 at LHC has not shown any direct evidence for new physics and the available studies on 
the couplings of the $125\UGeV\,$ resonance show compatibility with the Higgs boson of the SM. 
What we need, in preparation for the final results of Run~2 and 3, requires a consistent theory of SM 
deviations and a consistent link between realistic (fiducial) observables and theory parameters, 
see \Brefs{Passarino:2016pzb,Passarino:2016saj} and \Bref{deFlorian:2016spz}.

As discussed in this review, the first attempt to characterise SM deviations is represented by 
the so-called kappa-framework, introduced in 
\Brefs{LHCHiggsCrossSectionWorkingGroup:2012nn,Heinemeyer:2013tqa}. 

A very general question that we want to address is: does it make sense to ``fit'' the EW core? 
Note that this problem (the LHC problem) is not confined to introducing a specific parametrisation 
for SM deviations.

At LEP the number of quantities was reduced, implying that some assumption was made on the 
behaviour of the primordial quantities. The validity of these assumptions was judged on 
statistical grounds. Within these assumptions, as QED deconvolution and resonance approach
(\ie restricting the process $\Pep\Pem \to \PAf\Pf$ to its $\PZ\,$-resonant component),  
the secondary quantities, the pseudo-observables (POs), were as ``observable'' as the first ones. 

However, Higgs physics at LHC is radically different from $\PZ$ physics at LEP, since we have 
to work with very different scales, \eg on-shell $\PH$ decays are very different from, 
large $Q^2$, $\PV\PH$ production. None of this ``complication'' existed at LEP.
Furthermore, at LHC the EW core (including the Higgs properties) is always embedded into a 
QCD environment, subject to large perturbative corrections and we certainly expect considerable 
progress in the ``evolution'' of these corrections. The same considerations apply to PDFs when 
studying high-mass (large $x$) final states. 
The consequence is that, also for LHC, one has to list the assumptions that are made in
studying Higgs couplings and judge them on statistical grounds. To repeat the main argument, 
the huge QCD background and the associated uncertainty are such that one can fit the data 
starting from a given parameterisation of the SM deviations but for each new QCD calculation 
the result will change substantially and not multiplicatively. 

The procedure that we are suggesting is: write the answer in terms of SM deviations, after 
which, certain combinations of the deviation parameters will define the LHC POs and will be 
fitted. Optimally, part of the factorising QCD corrections could enter the PO definition. 
As for the framework for deviations, the parametrisation should be as general as possible, with 
a priori no dropping of terms: this will allow to ``reweight'' when new (differential) 
$\mathrm{K}\,$-factors become available. Parton density functions (PDFs) changing is the most serious problem. At LEP 
the $\Pep\Pem$ structure functions were known to very high accuracy (the effect of using 
different QED radiators, differing by higher orders treatment, was tested). A change of PDFs 
at LHC will change the convolution.

Keeping in mind all the caveats in the interpretation of the Higgs couplings, we proceed with 
a description of POs at LHC since, in our opinion, it will be crucial to follow the path 
\begin{center}
LHC data $\,\to\,$ POs $\,\to\,$ SMEFT.
\end{center}
\subsection{LHC pseudo-observables \label{Sect61}}
Why going beyond the $\upkappa\,$-framework? We can repeat that the $\upkappa\,$-parameters  
are easy to understand in terms of how they change cross sections and partial decay widths
(the so-called ``peak'' observables), but there are theoretical limitations as discussed in 
Section \ref{Sect4}. In this Section we will emphasise that extending the framework is currently 
seen~\cite{Passarino:2016owu} as expressing the $\upkappa$-parameters in terms of Wilson 
coefficients in some Effective Field Theory (\ie SMEFT).
One important point is that electroweak corrections are not defined in the kappa framework if 
the kappa's are meant as coupling modifiers. In the SMEFT we will have the SM EW corrections 
plus higher-dimensional operators in EW loops.

The rationale in building a Quantum Field Theory (QFT) of SM-deviations is not so much in the 
numerical impact of higher orders (however, see \Bref{Hartmann:2015oia} for a nice discussion) 
but in promoting a phenomenological tool to the full status of QFT.
Another reason for having a complete formalism is to avoid a situation where experimenters
will have to go back and remove a provisional formalism from the analyses. 
However, for some of the Wilson coefficients it is hard to understand both qualitatively and 
quantitatively what some parameter value does to observables; therefore, one question that 
we would like to answer is the following: could we use and translate part of the LEP language 
(\eg the one of POs) to recast parameters into POs?

Any QFT describes dynamics in terms of its Lagrangian parameters. At the LHC (after the discovery) 
one would like to try a model-independent approach.
From LEP to LHC, does history repeat itself? Why should it? 
The rationale of using POs at LHC is because the POs are a platform 
between realistic observables and theory parameters~\cite{Bardin:1999gt,Bardin:1998nm}, allowing 
experimentalists and theorists to meet half way.
In principle, ATLAS and CMS should also publish their fiducial and template cross sections 
since they are alternative and not antithetic to the POs.
As already stated, LHC is much more complex than LEP since, as an example, in the study of 
off-shell Higgs physics, resonant and non-resonant parts are perfectly tied together and there 
are severe issues of gauge invariance that must be taken into account. Thus, how could we 
interpret objects like $\Gamma(\PH \to \PW\PW^*)$? 

Despite inherent technical difficulties the next physics goal for LHC is high precision study of 
SM-deviations. Ideally, this would require the following steps:
\begin{enumerate}

\item For each process write down some (QFT-compatible) amplitude allowing for SM-deviations, 
      both for the resonant part (usually called ``signal'') and the non-resonant component 
      (usually called  ``background''); NLO SMEFT is a good example. 

\item Insert the resonant part expressed through POs without neglecting terms and 
      without subtracting the SM background; the background, \ie the non resonant part, can  
      change as well.
      The idea is to project some ``primordial quantities'', say number of observed events in 
      some pre-defined set-up, into some ``secondary quantities'' closer to the 
      theoretical description of the phenomena. 
      In this step, the number of quantities should not be reduced a priori, unless
      the validity of the assumptions that have been made on the behavior of the primordial
      quantities is judged and justified on statistical grounds. 

\item Use fiducial/template observables and fit POs,  conventionally defined.

\end{enumerate}
At a later stage, we can derive Wilson coefficients or BSM Lagrangian parameters, publish the full 
list of fiducial/template (with modern RIVET and HEPDATA  technology) and POs {\it  \`a la} LEP (that will become the 
LHC legacy).

In the rest of this Section we will summarise the state of art in studying SM deviations.
\paragraph{A list of POs at LHC}
The complete and most updated reference for POs at LHC can be found in \Bref{POnote}.
A summary table for POs at LHC is the following: 
\bei

\item[$\bf{\mrL_1}$] An external layer of ``physical'' POs~\cite{David:2015waa} (similar to 
$\sigma^{\peak}_{\Pf}$ at LEP):
\bq
\sum_{\Pf}\,\Gamma_{\PV\Pf\Pf} \spc \quad 
\mathrm{N}^{4\,\Pl}_{\mathrm{off}} \spc \quad
\mathrm{A}^{\PZ\PZ}_{\mathrm{FB}} \spc \quad
\sigma(\PQq\PQq \to \PZ\PZ jj)
\label{extl}
\eq
etc, where $\Gamma_{\PV\Pf\Pf}$ is the partial decay width for $\PH \to [\mbox{on-shell}\,\PV] +
\PAf \Pf$, $\mathrm{N}^{4\,\Pl}_{\mathrm{off}}$ is the number of events for 
$[\mbox{off-shell}\,\PH] \to 4\,\Pl$ and $\mathrm{A}^{\PZ\PZ}_{\mathrm{FB}}$ is a 
forward-backward (FB) asymmetry: for instance, a FB-asymmetry in the angle between the $\Pe$ and 
the $\PW$ reconstructed from $\PQq\PQq$ pair in $\PH \to \Pe\PGn\PQq\PQq$ or an F-asymmetry 
($\pi/4$) w.r.t. $|\cos\phi|$, $\phi$ being the angle between the decay planes of the 
reconstructed $\PZ$ bosons, \eg in the decay $\PH \to \,\Pe\Pe\PQq\PQq$.
These variables are based on a multi pole expansion (MPE), \ie these are processes which 
include a resonance that is fully characterised by the position of the corresponding complex 
pole, its residue and the regular part in the Laurent expansion around the 
pole~\footnote{For technical details on MPE, see \Bref{David:2015waa}.}. Furthermore, 
these ``physical'' POs are computed without recurring to a zero width approximation or a 
truncated MPE. 
\item[$\bf{\mrL_2}$] An intermediate layer of POs (similar to LEP $g^{\Pe}_{\sPV\,\sPA}$), 
based on quantities that can be extracted from Green's functions in well-defined kinematical 
limits~\cite{Gonzalez-Alonso:2014eva,Bordone:2015nqa,Greljo:2015sla}, 
%
\bq
\varepsilon^{(1)}_{\PH\,\PV\PV} \spc \quad
\varepsilon^{(2)}_{\PH\,\PV\PV}\spc \quad
\varepsilon_{\PH\,\PGg\PGg}, \; \varepsilon_{\PH\,\PGg\PZ} \spc \quad
\varepsilon_{\PH\,\PAf\Pf} \spc
\label{intl}
\eq
once again based on MPE, analyticity, unitarity, and crossing symmetry. For instance, 
$\varepsilon^{(1,2)}_{\PH\,\PV\PV}$ will be introduced in \eqn{eqdefrho} and parametrises 
deviations from the SM in the decay $\PH \to \PV \PV$. Once it is ``measured'' we can
compare it with the corresponding SM value or any other specific BSM value.
For instance, $\varepsilon_{\PH\,\PGg\PGg}$ refers to the $\PH \to \PGg\PGg$ decay,  is zero
at LO, and is known up to two loops in the SM and up to one loop in many BSM models.
Finally, $\varepsilon_{\PH\,\PAf\Pf}$ refers to the decay $\PH \to \PAf\Pf$, \etc
Once the intermediate POs are ``measured'' their interpretation will be possible. 
In full generality, a $1$PI Green's function with $\mrN$ external (on-shell) lines, carrying 
Lorentz indices $\mu_i, i=1,\,\dots\,,\mrN$ will be decomposed as
\bq
\mrA_{\proc} = \sum_{l=1}^{\mrL}\,\varepsilon^{l}_{\proc}\,
\mrK^{l}_{\mu_1\,\dots\,\mu_{\mrN}} \spc
\eq
where the set $\mrK^l\; (l=1\,\dots\,\mrL$) contains independent tensor structures made up of 
external momenta, metric tensors, elements of the Clifford algebra, and Levi-Civita tensors.
Symmetries, including gauge invariance reduce the number of structures, \eg only one structure
in $\PH \to \PGg\PGg, \PZ\PGg$ (CP-even only) due to Ward identities.
\item[$\bf{\mrL_3}$] An internal layer of POs: the generalised kappas. A first example: consider 
the amplitude for a loop-initiated process, \eg $\PH \to \PGg\PGg$; in the SM it can be 
decomposed as
\bq
\mrA_{\PH \to \PGg\PGg} = \mrA^{\Pf}_{\PH \to \PGg\PGg} + \mrA^{\PW}_{\PH \to \PGg\PGg} \spc
\eq
where the superscript $\Pf$ denotes fermion loops and the superscript $\PW$ denotes the
bosonic part. The kappa-deformed amplitude is characterised by
\bq
\upkappa^{\PGg\PGg}_{\Pf} \spc \quad
\upkappa^{\PGg\PGg}_{\sPW} \spc\quad
\upkappa^{\PGg\PGg\,\mathrm{nf}}_i \spc \quad
\mbox{etc}
\label{intrl}
\eq
where the kappa parameters  
\bei

\item[-] with a $\Pf$ subscript multiply the SM fermion sub-amplitude 
         $\mrA^{\Pf}_{\PH \to \PGg\PGg}$,

\item[-] those with a $\PW$ subscript multiply the SM bosonic sub-amplitude
         $\mrA^{\PW}_{\PH \to \PGg\PGg}$,

\item[-] those with a $\mathrm{nf}$ superscript refer to a set of non-factorisable sub-amplitudes 
that are computable in SMEFT and are not present in the original kappa framework. 
These amplitudes are better defined in the context of the SMEFT, \eg see \eqn{SMEFTc}.

\eei
Note that these kappa parameters have an additional label specifying the process, therefore
$\upkappa^{\PGg\PGg}_{\Pf} \not= \upkappa^{\PZ\PGg}_{\Pf}$ \etc 
For tree initiated processes the generalised kappa parameters have another label, 
distinguishing LO from NLO, see Eq.~(141) and Eqs.~(173-175) of \Bref{Ghezzi:2015vva} for
details. The introduction of generalised kappas is not limited to decay processes; for instance,
consider $\PAQq \PQq \to \PZ \PH$, where we can write
\bqa
\mrA_{\PAQq \PQq \to \PH \PZ} &=& g^2\,\lpar 1 + \Delta\,\upkappa^{\PAQq \PQq \to \PH \PZ} \rpar\,
\mrA^{\myLO} _{\PAQq \PQq \to \PH \PZ} +
g^2\,\sum_{\stackrel{\lambda=\pm}{i=1,2}}\,{\overline{\mrv}}\,\gamma^{\mu}\,
\gamma_{\lambda}\,\mru\,
e^{\nu}\,\Delta\,\upkappa^{\PAQq \PQq \to \PH \PZ\,;\,\mathrm{nf}}_{\lambda\,i}\,
\mrA^i_{\mu\nu} \spc
\nl
\gamma_{\pm} &=& \frac{1}{2}\,(1  \pm \gamma^5) \spc
\qquad \mrA^1_{\mu\nu} = \delta_{\mu\nu} \spc
\quad \mrA^2_{\mu\nu} = p_{\sPZ\,\mu}\,p_{\sPH\,\nu} \spc
\eqa
and $e^{\mu}$ is the $\PZ$ polarisation vector and
\bq
\mrA^{\myLO} _{\PAQu \PQu \to \PH \PZ} = \frac{1}{8}\,\frac{M_{\sPW}}{\ctwc}\,
\sum_{\lambda= \pm}\,(1 + \lambda - \frac{8}{3}\,\stws)\,
{\overline{\mrv}}\,\sla{e}\,\gamma_{\lambda}\,\mru \spc \;\; \mbox{etc} \spp
\eq
Similar decompositions can be provided for $\PQq \PQq \to \PQq \PQq \PH$,
$\Pg \Pg \to \PAQt \PQt \PH$, \etc

To summarise, the generalised kappa parameters are a convenient way to combine Wilson
coefficients, \ie for a given group of observables the set of Wilson coefficients collapses
into a set of generalised kappas. Of course, the exact relation between generalised kappas and 
Wilson coefficients depends on the order of the SMEFT calculation, LO or NLO.
\item[$\bf{\mrL_4}$] Finally, the innermost layer: the Wilson coefficients in SMEFT (LO or NLO) 
or non-SM parameters in BSM (\eg mixing angles $\alpha, \beta$, soft-breaking scale, 
$M_{\mathrm{sb}}$, \etc, in THDMs).

\eei
As we will explain in more detail below, introducing the innermost layer is an operation that
can be eventually postponed. The original work on intermediate POs at LHC can be found in 
\Brefs{Gonzalez-Alonso:2014eva,Bordone:2015nqa,Greljo:2015sla} and in \Bref{David:2015waa}
for external POs. 
\paragraph{Intermediate layer ($\mrL_2$)}
In the following we give few examples, starting with the intermediate layer, \eqn{intl}.
The amplitude for the Higgs decay $\PH \to \PGg(p_1)\PGg(p_2)$ can be parametrised as
\bq
\mrA_{\PH \to \PGg\PGg} = \varepsilon_{\PH\,\PGg\PGg}\,
\frac{\spro{p_1}{p_2}\,g^{\mu\nu} - p^{\mu}_2 p^{\nu}_1}{\mh} \spc
\eq
Another example where we use MPE is the following:
\bq
\mrA_{\PH \to \PV \PV} = 
\varepsilon^{(1)}_{\PH\,\PV\PV}\,\mh^2\,g^{\mu\nu} + 
\varepsilon^{(2)}_{\PH\,\PV\PV}\,p^{\mu}_2\,p^{\nu}_1  \spp
\label{eqdefrho}
\eq
In all cases the Higgs production parametrization follows by using analyticity and crossing 
symmetry. In any process, the residues of the poles corresponding to internal, unstable particles 
are numbers, while the non-resonant part is a multivariate function that requires some basis, 
\ie a less model independent  
theory of SM deviations. 
That is to say, residues of the poles (MPE) can be intermediate POs by themselves.
Numerically speaking, the residue of the pole depends on the impact of the non-resonant part 
which is small in gluon-fusion but not in Vector Boson Scattering. 
Therefore, the focus for  reporting data should always be on real observables, fiducial/template 
cross sections, and pseudo-observables.
To give an idea of the general structure we consider the $\PH \to 4\,\Pf$ channels: including
CP-odd couplings, there are $3$ flavor-universal charged-current POs, 
$7$ flavor-universal neutral-current POs, and a set of flavor non-universal charged-current POs;
the number of non-universal POs depends on the fermion species we are interested in~\cite{POnote}.

Few additional definitions are needed to appreciate the consistency of the whole theoretical 
construction. Here, we only provide a brief summary, with the corresponding references.

\bei

\item[-] By on-shell $\mrS\,$-matrix for an arbitrary process (involving external unstable 
particles) we mean the corresponding (amputated) Green's function supplied with LSZ factors 
and sources, computed at the (complex) poles of the external 
lines~\cite{Grassi:2000dz,Goria:2011wa}. 
Processes that involve stable particles can be straightforwardly transformed into 
a physical PO.

\item[-] The connection of the $\PH\PV\PV, \PV = \PZ,\PW$ (on-shell) $\mrS\,$-matrix with the 
off-shell vertex $\PH \to \PVV$ and the full process $\Pp\Pp \to 4\,\Pf$ is more complicated and 
is discussed in some detail in Sect.~3 of \Bref{David:2015waa}.
The ``on-shell'' $\mrS\,$-matrix\ for $\PH\PV\PV$, being built with the the residue of the 
$\PH{-}\PV{-}\PV$ poles in $\Pp\Pp \to 4\, \Pf$ is gauge invariant by construction (it can be 
proved by using Nielsen identities) and represents one of the building blocks for the full 
process: in other words,  $\PH\PV\PV$ is a PO.

\item[-] Technically speaking the ``on-shell'' limit for external legs should be understood 
``to the complex poles'' (for a modification of the LSZ reduction formulas for unstable 
particles see \Bref{Weldon:1975gu}) but, as well known, at one loop we can use on-shell masses 
(for unstable particles) without breaking the gauge parameter independence of the result.

\item[-] It is worth noting that there are subtleties when the $\PH$ is off-shell. They are 
described in Appendix~C.1 of \Bref{Goria:2011wa}. Briefly, there is a difference between 
performing an analytical continuation ($\PH$ virtuality $\to$ $\PH$ on-shell mass) in the 
off-shell decay width and using leading-pole approximation (LPA) of \Bref{Denner:2005es}, 
\ie the doubly-resonant part, where the matrix element (squared) is projected but not the 
phase-space. Analytical continuation is a unique, gauge invariant procedure, the advantage of 
leading-pole approximation is that it allows for a straightforward implementation of experimental cuts.

\eei

\paragraph{External layer ($\mrL_1$)}
The external layer, \eqn{extl}, requires additional work. It would be ideal to stop reporting
non existing objects, \eg $\PH \to \PZ\PZ$ or non-gauge-invariant objects, \eg $\PH \to \PZ^*\PZ$.
The external layer, \ie the ``physical" POs, allows for experimental cuts and thus bypasses 
ad hoc  constructions like ``diagram removal'' (not gauge invariant) or ``diagram subtraction'' 
(with  ad hoc prefactors and Breit-Wigner profiles).

MPE is the tool for isolating gauge invariant parts of any amplitude, therefore it is enough to
construct quantities in the intermediate layer. Quantities in the external layer aim to
identify ``elementary'' subprocesses inside a ``realistic'' processes, \eg simple
production/decay processes inside realistic processes, \eg $\PAQt\PQt\,\PH$ production inside
$\PW\PQb\PW\PQb$ production. 
MPE, \ie gauge-invariant splitting of the amplitude, is not the same as ``factorisation'' of 
the process into sub-processes; let us explain what it is meant by requiring the pole to be 
inside the physical region. The propagator and phase-space are
\bqa
{}&{}&|\Delta|^2 = \frac{1}{\lpar s - M^2\rpar^2 + \Gamma^2\,M^2} =
\frac{\pi}{M\,\Gamma}\,\delta\lpar s - M^2\rpar +
\mathrm{PV}\,\left[ \frac{1}{\lpar s - M^2\rpar^2}\right] \spc
\nl
{}&{}& d\Phi_n\lpar P, p_1 \dots p_n\rpar =
\frac{1}{2\,\pi}\,dQ^2\,d\Phi_{n-(j+1)}\lpar P, Q, p_{j+1} \dots p_n\rpar\,
d\Phi_j\lpar Q, p_1 \dots p_j \rpar \spc
\label{fact}
\eqa
where PV denotes the principal value (understood as a distribution, see \Bref{Nekrasov:2007ta}).
%
%
It would be convenient to define POs through a factorisation of the process into sub-processes,
\eg by ``isolating'' the production and decay of a particle of momentum $Q$. A typical example 
is the ``isolation'' of $\PH \to \PZ\PGg$ in $\PH \to \PAf\Pf\PGg$, where $\PH \to \PZ\PGg$ will 
be the relevant PO~\cite{Passarino:2013nka}. However, to ``complete'' the $Q\,$-decay 
(requiring $d\Phi_j$), we need the $\delta\,$-function in \eqn{fact}. We can say that the 
$\delta\,$-part of the resonant (squared) propagator ``opens'' the corresponding line and allows 
us to define POs; however, this is not the case for $t\,$-channel propagators, which cannot be cut.
Note that the pole must be inside the allowed portion of phase space which is not always the case 
when experimental cuts are introduced or the kinematics of the process does not allow it. For instance, 
$\Gamma(\PH \to \PZ\PGg)$ can be promoted to be a ``physical'' PO whereas $\Gamma(\PH \to \PZ\PZ)$ 
cannot and we should introduce $\sum_{\Pf}\,\Gamma(\PH \to \PZ \PAf\Pf)$.
As an additional example, consider the process $\PQq\PQq \to \PAf_1\Pf_1\PAf_2\Pf_2 jj$:
given the structure of the resonant poles we can define different POs, \eg
\bqa
\sigma(\PQq\PQq \to \PAf_1\Pf_1\PAf_2\Pf_2 jj) &\stackrel{PO}{\longmapsto}&
\sigma(\PQq\PQq \to \PH jj)\,\mathrm{Br}(\PH \to \PZ\PAf_1\Pf_1)\,
\mathrm{Br}(\PZ \to \PAf_2\Pf_2) \spc
\nl
\sigma(\PQq\PQq \to \PAf_1\Pf_1\PAf_2\Pf_2 jj) &\stackrel{PO}{\longmapsto}&
\sigma(\PQq\PQq \to \PZ\PZ jj)\,\mathrm{Br}(\PZ \to \PAf_1\Pf_1)\,
\mathrm{Br}(\PZ \to \PAf_2\Pf_2) \spp
\eqa
There are additional fine points to be considered when factorising a process into ``physical'' 
sub-processes, and details have been discussed in \Bref{David:2015waa}. 

The internal and innermost layers require the introduction of SMEFT.
\subsection{Standard Model Effective Field Theory}
The basis for NLO calculus of the Standard Model effective field theory (SMEFT) has been 
developed in \Bref{Ghezzi:2015vva} and in \Brefs{Hartmann:2015oia,Hartmann:2015aia}.
The issue of theoretical uncertainty associated with missing higher orders has been raised
and discussed in \Bref{Berthier:2015oma}. The interplay between SMEFT and pseudo-observables
has been introduced in \Bref{David:2015waa}.

In \Bref{Ghezzi:2015vva} one can find all technical details as well as the full list
of counterterms and mixings that are relevant for Higgs physics. 
The SMEFT used is based on the assumption of one Higgs doublet with a linear representation;
for non-linear representations see \Brefs{Buchalla:2013rka,Brehmer:2015rna}. 
There are two different research approaches in any EFT, the top-down and bottom-up approach.
\subsubsection{The top-down approach \label{tda}}
The top-down approach starts with a known theory and systematically eliminates degrees of freedom 
associated with energies above some energy scale $\Lambda$. The construction of a low-energy 
effective theory is, in general, not simple for the high-energy degrees of freedom may be 
tangled up with the low-energy ones. For a distinction between Wilsonian and ``continuum''
EFT see \Bref{Bain:2013aca}.

The integration of heavy fields in a wide class of BSM models, containing more that one 
representation for scalars has been discussed in 
\Brefs{Henning:2014wua,delAguila:2016zcb,Boggia:2016asg,Henning:2016lyp,Fuentes-Martin:2016uol}.

The authors of \Bref{Contino:2016jqw} have discussed the conditions for an effective field 
theory to give an adequate low-energy description of the underlying physics beyond the SM. 
Starting from the EFT where the SM is extended by $\mrdim =6$ operators, experimental data can 
be used without further assumptions to measure (or set limits on) the EFT parameters. Their 
conclusion is that interpretation of these results requires instead a set of broad assumptions 
on the UV dynamics.

The work of \Brefs{Brehmer:2015rna,Biekotter:2016ecg} shows how the Higgs couplings analysis 
of Run~1 can be expanded into a Higgs operator analysis, based on a linearly-realised 
$\mrdim =6$ Lagrangian. The extraction of the corresponding Wilson coefficients from Run~1 data 
is discussed emphasising how kinematic distributions can be included and how they affect the
extraction of the Wilson coefficients. Even more important, they study how well weakly interacting
extensions of the Higgs and gauge sector are described by such a $\mrdim = 6$ Lagrangian. It turns
out that in spite of the lacking scale hierarchy the $\mrdim =6$ approximation quantitatively 
captures the relevant features of the full models, except for the appearance of new resonances.

SMEFT is a double expansion in $g^2\,\mrv^2/\Lambda^2$ and
$g^2\,E^2/\Lambda^2$, where $\mrv$ is the Higgs VEV and $E$ is the typical energy scale at which the
measurement is performed. Operators controlled by the VEV do not change kinematic
distributions. The general argument, discussed in \Brefs{Brehmer:2015rna}, is the following:
for a ``reasonable'' weakly interacting theory, when only studying total rates with an LHC
accuracy of $10\%$, the LO impact can be estimated according to:
\bq
\bmid\,\frac{\sigma\,\times\,\mathrm{BR}}{(\sigma\,\times\,\mathrm{BR})_{\mySM}} - 1\,\bmid =
\frac{g^2\,M^2_{\PH}}{\Lambda^2} > 10\%     
\eq
that implies
\bq
\Lambda < \frac{g\,M_{\PH}}{\sqrt{10\%}} \approx 400\UGeV \spp
\eq
This explains why differential distributions are essential, involving a new scale
$g\,p_{\mrT}/\Lambda$. There is always a huge improvement on the sensitivity when using 
differential distributions as can be seen by comparing the constraints obtained in the 
measurement of the signal strength with the ones including distributions.  

\Bref{Freitas:2016iwx} makes the point that, for the new physics scales accessible at the LHC, 
the effective operator expansion is only relatively slowly converging. For tree-level processes, 
the agreement between SMEFT and a range of UV-complete models depends sensitively on the 
appropriate definition of their matching.
To be more precise consider a theory with both light and heavy particles; the Lagrangian depends 
on $M$, the mass of the heavy degree of freedom. Next, introduce the corresponding effective 
theory, valid up to a scale $\Lambda = M$. Afterwards we renormalise the two theories, say in 
the $\overline{\rm{MS}}\,$-scheme (taking care that loop-integration and heavy limit are 
operations that do not commute), and impose ``matching'' conditions among renormalised ``light'' 
$1$ particle irreducible ($1$PI) Green's 
functions~\footnote{For a complete definition of the matching procedure see Sect.~3.1 of 
\Bref{Georgi:1994qn}.}, \ie 
\bq
\Gamma^{\ssR}(\mu)\mid_{\mathrm{full}}(\mu) = \Gamma^{\ssR}(\mu)\mid_{\mathrm{eff}} \spc
\qquad \mu \le M \spc
\eq
where $\mu$ is the renormalisation scale and $\Gamma_{\ssR}$ is any renormalised Green's function.
At the one-loop level, near the scale of EWSB the validity of SMEFT description can be 
systematically improved through an appropriate matching procedure. 
\subsubsection{The bottom-up approach \label{Sect622}}
Finding deviations is what we expect, at least experimentally; in that sense we should place 
more focus on measuring them than in just setting limits.
In any (strictly) renormalisable theory with $n$ parameters one needs to match $n$ data points
and the $(n+1)$th calculation becomes a prediction. 
In the SMEFT there are ($\mrN_6{+}\mrN_8{+}\,\dots = \infty)$ renormalised Wilson coefficients 
(where $\mrN_n$ is the number of Wilson coefficients of $\mrdim= n$) that have to be fitted, 
measuring SM deformations due to a single $\mathcal{O}^{(6)}$ insertion.
A basis for $\mrdim = 6$, the so-called Warsaw basis, has been provided in 
\Bref{Grzadkowski:2010es}.
Recent progress for $\mrdim = 8$ operators can be found in \Brefs{Lehman:2015coa,Henning:2015daa,Henning:2015alf}.
By $\mrdim = 6$ SMEFT basis we mean any set of gauge invariant, 
dimension $6$ operators, none of which is redundant. For a formal definition of redundancy 
see Sect.~3 of \Bref{Einhorn:2013kja}.

\paragraph{SMEFT Lagrangian}
The treatment of EFT was pioneered in~\cite{Weinberg:1980wa,Coleman:1969sm,Callan:1969sn};
the SMEFT approach is based on the following Lagrangian:
\bq
\mrA = \sum_{n=\ssN}^{\infty}\,g^n\,\mrA^{(4)}_n +
       \sum_{n=\ssN_6}^{\infty}\,\sum_{l=1}^n\,\sum_{k=1}^{\infty}\,
        g^n\,g^l_{4+2\,k}\,
        \mrA^{(4+2\,k)}_{n,l,k} \spc
\label{SMEFTLag}
\eq
where $g$ is the $SU(2)$ coupling constant and 
$g_{4+2\,k} = 1/(\sqrt{2}\,G_{\ssF}\,\Lambda^2)^k = g^k_6$;
$G_{\ssF}$ is the Fermi coupling constant and $\Lambda$ is the scale around which
new physics (NP) must be resolved. For each process, $N$ defines the $\mrdim = 4$ 
Lagrangian at LO (\eg $N = 1$ for $\PH \to \PV\PV$, \etc, but $N = 3$ for $\PH \to \PGg\PGg$). 
$N_6 = N$ for tree initiated processes and $N - 2$ for loop initiated ones. 

A simple SMEFT ordertable for tree-initiated $1 \to 2$ processes is available in
\Bref{Passarino:2016owu}, where the complete definition of LO SMEFT, NLO SMEFT,
and MHOU (missing higher order uncertainties) for SMEFT is presented. Here it will suffice to 
say that for a decay $1 \to 2$, 
$g\,g_6\,\mrA^{(6)}_{1,1,1}$ defines the LO SMEFT expression while $g^3\,g_6\,\mrA^{(6)}_{3,1,1}$ 
defines the NLO SMEFT amplitude in the perturbative expansion.

NLO corrections are a necessary consequence of the SMEFT being a well-defined field theory.
The {\it numerical size of the higher order terms} depends upon the high energy (UV) scenario 
dictating the coefficients and $\Lambda$, which is unknown.
Restricting to a particular UV case is not an integral part of a general SMEFT treatment and 
various cases can be chosen once the general calculation is performed. 
All explicit references to the underlying theory are introduced via the matching procedure in 
the standard approach to EFTs and power counting, see 
Refs.~\cite{Weinberg:1980wa,Coleman:1969sm,Callan:1969sn,Manohar:1983md,Georgi:1994qn,Kaplan:1995uv,Manohar:1996cq,Cohen:1997rt,Luty:1997fk,Polchinski:1992ed,Rothstein:2003mp,Skiba:2010xn,Burgess:2007pt,Jenkins:2013fya,Jenkins:2013sda,Buchalla:2014eca,Buchalla:2013eza,Gavela:2016bzc} for reviews.

SMEFT is a double expansion: in $g$ and $g_{_6} = \mrv^2_{\mrF}/\Lambda^2$ for
pole observables and in $g, g_{_6}\,E^2/\mrv^2_{\mrF}$ for off-shell ones. 

To summarize, LO SMEFT refers to $\mrdim = 6$ operators in tree diagrams, sometimes called
``contact terms'', while NLO SMEFT refers to one loop diagrams with a single insertion
of $\mrdim = 6$ operators. One can make additional assumptions by introducing classification 
schemes in SMEFT. One example of a classification scheme is the Artz-Einhorn-Wudka 
``potentially-tree-generated'' (PTG) scenario~\cite{Arzt:1994gp,Einhorn:2013kja}. In this scheme, 
it is argued that classes of Wilson coefficients for operators of $\mrdim = 6$ can 
be tree level, or loop level (suppressed by $g^2/16 \, \pi^2$)\footnote{This classification 
scheme corresponds only to a subset of weakly-coupled and renormalisable UV physics cases.}.
In these cases, the expansion is reorganised in terms of ``tree-generated'' (TG) (we assume a
BSM model where PTG is actually TG) and LG insertions (\ie ``loop-generated''), \ie LG 
contact terms and one loop TG insertions, one loop LG insertions and two loop SM \etc
It is clear that LG contact terms alone do not suffice.

Strictly speaking we are considering here the virtual part of SMEFT, under the
assumptions that LHC POs are defined {\it \`a la} LEP, \ie when QED and QCD corrections are deconvoluted.
Otherwise, the real (emission) part of SMEFT should be included and it can be shown that
the infrared/collinear part of the one-loop virtual corrections and of the real ones respect 
factorisation: the total = virtual ${+}$ real is IR/collinear finite at $\mcO(g^4\,g_{_6})$,
see \Bref{Passarino:2016owu}.

It is worth noting that SMEFT has limitations, obviously the scale should be such that
$E \muchless \Lambda$.
Understanding SM deviations in tails of distributions requires using SMEFT only up to the 
point where it stops to be valid, or using the kappa--BSM-parameters connection, \ie replace 
SMEFT with BSM models, optimally matched to SMEFT at lower scales.
It is worth noting the complementarity between ``pole'' vs. ``tail'' measurements:
operators that influence tail observables  are not the same affecting  pole observables, \ie derivative vs. 
non-derivative operators. Furthermore, the SMEFT does break down in the tails if you have a cutoff 
scale that is not outside the larger probe scale. 
We can say that, eventually, the SM will break in ``tails''; projecting into the SMEFT will have 
a large intrinsic uncertainty, \ie we do not know what exactly is going on because the SMEFT 
interpretation becomes a series expanded in a parameter larger than $1$.

To explain SMEFT in a nutshell (for a complete description see \Bref{NLOnote}) consider a 
process described by some SM amplitude
\bq
\mrA_{\mySM} = \sum_{i=1,n}\,\mrA^{(i)}_{\mySM} \spc
\eq
where $i$ labels gauge-invariant sub-amplitudes. 
In order to construct the SMEFT one has to select a set of higher-dimensional operators and
to start the complete procedure of renormalization. Of course, different sets of operators
can be interchangeable as long as they are closed under renormalization. It is a matter of
fact that renormalization is best performed when using the so-called Warsaw basis, see 
\Bref{Grzadkowski:2010es}. Moving from SM to SMEFT we obtain
\bq
\mrA^{\myLO}_{\SMEFT} = \sum_{i=1,n}\,\mrA^{(i)}_{\mySM} + i\,\gds\,\upkappa_{\mrc} \spc
\qquad
\mrA^{\myNLO}_{\SMEFT} = \sum_{i=1,n}\,\upkappa_i\,\mrA^{(i)}_{\mySM} +
                i\,\gds\,\upkappa_{\mrc} +
                \gds\,\sum_{i=1,\ssN}\,a_i\,\mrA^{(i)}_{\nfact} \spc
\label{nloeft}
\eq
where $g^{-1}_{_6} = \sqrt{2}\,\myGF\,\Lambda^2$. The last term in \eqn{nloeft} collects all 
loop contributions that do not factorize and the coefficients $a_i$ are Wilson coefficients. 
The $\upkappa_i$ start from one, \ie $\upkappa_i = 1 + \Delta\upkappa_i$, and the 
$\Delta\upkappa_i$ are linear combinations of the $a_i$.
We conclude that \eqn{nloeft} gives the correct generalization of the original
$\upkappa\,$-framework at the price of introducing additional, non-factorizable, terms in
the amplitude.

Finally, we would like to clarify certain points regarding the following question: do we 
necessarily have to make UV assumptions?
The work in \Bref{Contino:2016jqw} discusses the conditions for an EFT to give an adequate 
low-energy description of an underlying physics beyond the SM. The claim is that one has to 
make a set of broad assumptions on the UV dynamics as one cannot just use SMEFT without UV 
assumptions (if not for setting limits on the Wilson coefficients).
We agree with the view presented in \Bref{Gavela:2016bzc}: the pattern of suppressions for 
Wilson coefficients is not a SMEFT prediction but must be determined experimentally. Of course, 
it depends on the underlying UV completion but can be determined experimentally solely by 
using ``low-energy'' measurements that can be computed by using SMEFT, as was always done in 
the past. 
\paragraph{SMEFT vs. original kappa framework, the internal layer ($\mrL_3$)}
For a discussion of the internal layer, \eqn{intrl}, we consider the present implementation of 
coupling modifiers, as illustrated in \eqn{asdone}. In general, for $\Pg\Pg \to \PH$ we can 
use the following decomposition,
\bq
\mrA^{\Pg\Pg} = \sum_{\PQq=\PQt,\PQb}\,\upkappa^{\Pg\Pg}_{\PQq}\,
\mrA^{\Pg\Pg}_{\PQq} + \upkappa^{\Pg\Pg}_c \spc
\label{morek}
\eq
$\mrA^{\Pg\Pg}_{\PQt}$ being the SM $\PQt\,$-loop \etc The contact term (which is the 
LO SMEFT) is given by $\upkappa^{\Pg\Pg}_c$. Furthermore
\bq
\upkappa_{\PQq} = 1 + \Delta\,\upkappa_{\PQq} = 1 + g_{_6}\,\delta_{\PQq} \spp
\label{duk}
\eq

\paragraph{The innermost layer: Wilson coefficients ($\mrL_4$)}
In the context of NLO SMEFT \eqn{morek} is rewritten as
\bqa
\mrA^{\Pg\Pg}_{\EFT} &=& 
\sum_{\PQq=\PQt,\PQb}\,\upkappa^{\Pg\Pg}_{\PQq}\,\mrA^{\Pg\Pg}_{\PQq} +
\mrK\,\apG +
\sum_{\PQq=\PQt,\PQb}\,\mrA^{\mathrm{nf}\,;\,\Pg\Pg}_{\PQq}\,a_{\PQq \PG} 
\nl
{}&=&
 \frac{g\,g_{\mrS}}{\pi^2}\,\sum_{\PQq=\PQt,\PQb}\,\upkappa^{\Pg\Pg}_{\PQq}\,\mrA^{\Pg\Pg}_{\PQq} +
 2\,g_{\mrS}\,g_{_6}\,\frac{s}{\mws}\,\apG  + 
 \frac{g\,g^2_{\mrS}\,g_{_6}}{\pi^2}\,
  \sum_{\PQq=\PQt,\PQb}\,\mrA^{\mathrm{nf}\,;\,\Pg\Pg}_{\PQq}\,a_{\PQq \PG} \spc
\label{SMEFTc}
\eqa
where the $\upkappa$ are linear combination of Wilson coefficients, denoted by $a$ following
the Warsaw basis convention. At NLO, $\Delta \upkappa = g_{_6}\,\delta$ and
\bqa
{}&{}& g^{-1}_{_6} = \sqrt{2}\,\myGF\,\Lambda^2 \spc
\quad
4\,\pi\,\alphas = g^2_{\mrS}
\nl
{}&{}&
\delta^{\Pg\Pg}_{\PQt} = \apW + \atp + 2\,\apBox - \frac{1}{2}\,\apD \spc
\qquad
\delta^{\Pg\Pg}_{\PQb} = \apW - \abp + 2\,\apBox - \frac{1}{2}\,\apD \spp
\label{rhodef}
\eqa
In \eqn{rhodef} $\apW$ is LG, as well as $\apG$ and $a_{\PQq \PG}$. Therefore,
\bq
\delta^{\Pg\Pg}_{\PQt}\mid_{\mathrm{PTG}} = \atp + 2\,\apBox - \frac{1}{2}\,\apD \quad ,
\mbox{\etc} 
\eq
\paragraph{Remark} Instead of the two parameters appearing in \eqn{asdone} we now have 
$\PQt,\PQb$ modifiers (containing both PTG and LG), a (LG) contact term and (LG) non factorisable 
contributions. In order to compare with \eqn{asdone} we introduce
\bqa
{}&{}& \PX_{\PQt\,,\,\PQb} = \mid \mrA_{\PQt\,,\,\PQb} \mid^2 \spc \quad
       \PX_{\PQt\,\PQb} = 2\,\Re\,\Bigl[ \mrA^{\dagger}_{\PQt}\,\mrA_{\PQb} \Bigr] \spc \quad
       \PX_{\PQq} = \mid \sum_{\PQq=\PQt,\PQb}\,\mrA_{\PQq} \mid^2
\nl
{}&{}& \PY_{\PQt} = 2\,\Re\,\mrA_{\PQt} \spc \quad
       \PY_{\PQb} = 2\,\Re\,\mrA_{\PQb} \spc \quad
       \PY_{\PQq} = 2\,\sum_{\PQq=\PQt,\PQb}\,\Re\,\mrA_{\PQq} \spc
\eqa
and derive the results of Tab.~\ref{RSF}.
\begin{table}[htb]
\tbl{Resolved scaling factor in $\Pg\Pg \to \PH$. $\mrK$ is the coefficient of the contact term
as introduced in \eqn{SMEFTc}. }
{\begin{tabular}{@{}ll@{}} \hline\hline
Framework & Scaling factor \\ \colrule
$\upkappa$    &  $\upkappa^2_{\PQt}\,\PX_{\PQt} +
                  \upkappa^2_{\PQb}\,\PX_{\PQb} +
                  \upkappa_{\PQt}\,\upkappa_{\PQb}\,\PX_{\PQt,\PQb}$   \\ 
LO SMEFT      &  $\PX_{\PQq} + \mrK^2\,\aspG + \mrK\,\apG\,\PY_{\PQq}$ \\
NLO SMEFT PTG &  $\upkappa^2_{\PQt}\,\PX_{\PQt} +
                   \upkappa^2_{\PQb}\,\PX_{\PQb} +
                   \upkappa_{\PQt}\,\upkappa_{\PQb}\,\PX_{\PQt,\PQb} +
                   \mrK^2\,\aspG +
                   \mrK\,\apG\,\upkappa_{\PQt}\,\PY_{\PQt} +
                   \mrK\,\apG\,\upkappa_{\PQb}\,\PY_{\PQb}$ \\
\hline\hline
\end{tabular} \label{RSF}}
\end{table}

When going interpretational we make additional assumptions: for instance, use NLO SMEFT, adopt 
the Warsaw basis and eventually work in the Einhorn-Wudka PTG scenario. Options are:

\bei

\item[-] LO SMEFT: $\upkappa_{\PQq} = 1$ and $\apG$  being LG is scaled by 
$1/16\,\pi^2$;

\item[-] NLO PTG-SMEFT: $\upkappa_{\PQq} \not= 1$ but only PTG operators 
inserted in loops (non-factorisable terms absent), $\apG$ scaled as above;

\item[-] NLO full-SMEFT: $\upkappa_{\PQq} \not= 1$, LG/PTG operators 
inserted in loops (non-factorisable terms present), LG coefficients scaled as above.

\eei
Summarising: relaxing the PTG assumption introduces non-factorisable sub-amplitudes 
proportional to $\atG, \abG$ with a mixing among $\{\apG, \atG, \abG\}$. Meanwhile, 
renormalisation has made one-loop SMEFT finite, \eg in the $\myGF\,$-scheme, with a residual 
$\muR\,$-dependence.

A connection between intermediate POs, generalised kappas, and Wilson coefficients has been 
derived in Sect.~8 of \Bref{POnote}.

\begin{figure}[b]
\centerline{\includegraphics[width=15.cm,angle=90]{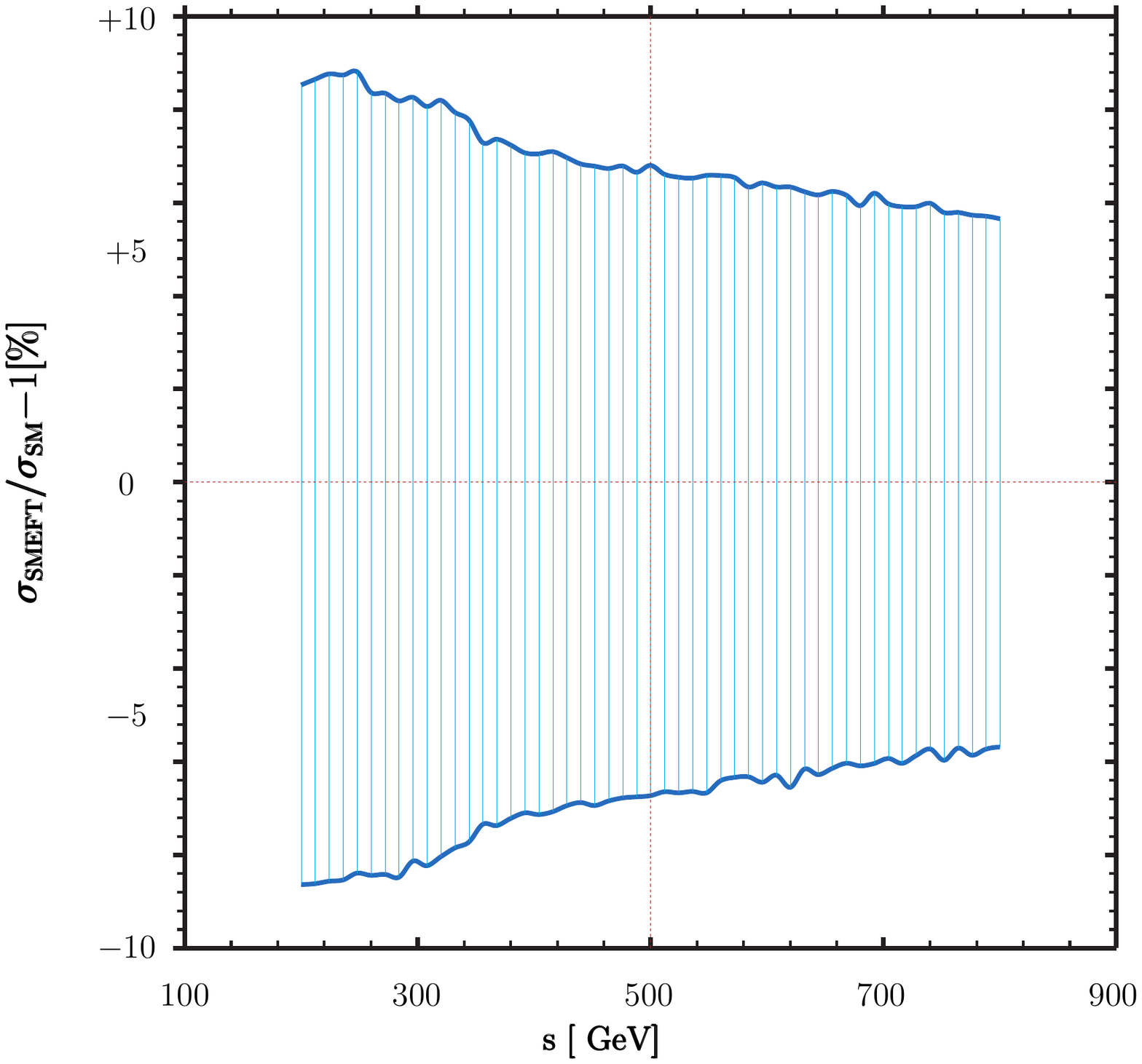}}
\vspace{-4.cm}
\caption{Off-shell $\Pg\Pg \to \PH$. The Higgs virtuality is $s$ and Wilson coefficient are 
generated with support $\mid a_i \mid\;\in\;[-1\,,\,+1]$ ($\mathrm{unif}(-1\,,\,1)$). 
$\Lambda = 3\UTeV$. 
\label{mfig}}
\end{figure}

\paragraph{Off-shell SMEFT}
Consider $\PH \to \PZ\PZ$ for a Higgs of virtuality $s$. The amplitude is
decomposed into Lorentz structures,
\bq
\mrA^{\mu\nu}_{\sPZZ} = \mcD_{\sPZZ}\,\delta^{\mu\nu} + \mcP_{\sPZZ}\,p^{\mu}_2\,p^{\nu}_1 \spc
\eq
with the following result:
\bq
\mcD_{\sPZZ} = g\,\upkappa^{\sPZZ}_{\myLO}\,\mcD^{\myLO}_{\sPZZ}
+ \frac{g^3}{16\,\pi^2}\,
  \sum_{i=\PQt,\PQb,\sPW}\,\upkappa^{\sPZZ\,;\,\sPD}_{\myNLO\,;\,i}\,\mcD^{\myLO\,;\,i}_{\sPZZ}
+ \frac{g^3 g_{_6}}{16\,\pi^2}\,
  \sum_{a\in \mathrm{A}_{\sPZZ}}\,\mcD^{\myLO\,;\,\nfact\,;\,a}_{\sPZZ}\,a \spc
\eq
\bq
\mcP_{\sPZZ} = 2\,\frac{g g_{_6}}{\mw}\,\aZZ
+ \frac{g^3}{16\,\pi^2}\,
  \sum_{i=\PQt,\PQb,\sPW}\,\upkappa^{\sPZZ\,;\,\sPP}_{\myNLO\,;\,i}\,\mcP^{\myLO\,;\,i}_{\sPZZ}
+ \frac{g^3 g_{_6}}{16\,\pi^2}\,
  \sum_{a\in \mathrm{A}_{\sPZZ}}\,\mcP^{\myLO\,;\,\nfact\,;\,a}_{\sPZZ}\,a \spp
\eq
Once again, $\upkappa^{\sPZZ}_{\myLO}$ \etc start from one, $\Delta \upkappa^{\sPZZ}_{\myLO}$ 
\etc parametrise SM deviations and $\mcD^{\myLO}_{\sPZZ}$ \etc are computed within the SM. 
Following \eqn{duk} the kappas are determined by
\bqa
\delta^{\sPZZ}_{\myLO} &=&
2\,\apBox + \stws\,\aAA + \stw\,\ctw\,\aAZ +
\Bigl[ 4 + \ctws\,( 1 - \frac{s}{\mws} ) \Bigr]\,\aZZ \spc
\nl
\delta^{\sPZZ\,;\,\sPD}_{\myNLO\,;\,\sPW} &=&
2\,\apBox + \frac{1}{12}\,\frac{1 + 4\,\ctws}{\ctws}\,\apD +
\stws\,\aAA + \frac{1}{3}\,\stw\,(\frac{5}{\ctw} + 9\,\ctw)\,\aAZ +
(4 + \ctws)\,\aZZ \spc
\nl
\delta^{\sPZZ\,;\,\sPD}_{\myNLO\,;\,\PQt} &=&
\atp + 2\,\apBox - \frac{1}{2}\,\apD + 2\,\aZZ + \stws\,\aAA \spc
\nl
\delta^{\sPZZ\,;\,\sPD}_{\myNLO\,;\,\PQb} &=&
- \abp + 2\,\apBox - \frac{1}{2}\,\apD + 2\,\aZZ + \stws\,\aAA \spc
\eqa
\bqa
\delta^{\sPZZ\,;\,\sPP}_{\myNLO\,;\,\PQt} &=&
\delta^{\sPZZ\,;\,\sPD}_{\myNLO\,;\,\PQt} \spc
\nl
\delta^{\sPZZ\,;\,\sPP}_{\myNLO\,;\,\PQb} &=&
\delta^{\sPZZ\,;\,\sPD}_{\myNLO\,;\,\PQb} \spc
\nl
\delta^{\sPZZ\,;\,\sPP}_{\myNLO\,;\,\sPW} &=&
4\,\apBox + \frac{5}{2}\,\apD + 3\,\stws\,\aAA + 12\,\aZZ \spp
\eqa
Results for the probability distribution of SM deviations are shown in Fig.~\ref{mfig}. Here
the Wilson coefficients are distributed according to $\mathrm{unif}(-1\,,\,1)$ for a scale
$\Lambda = 3\UTeV$.
Consider a measure of deviation from the SM prediction written as 
$\mathrm{R} = \Gamma_\mathrm{NLO SMEFT}/\Gamma_\mathrm{SM} - 1$, where 
$\Gamma = \Gamma(\PH \to \PGg\PGg)$.
Allowing each Wilson coefficient to have a uniform probability distribution in a interval 
$\mathrm{I}_n = [-n\,,\,+n]$, the result from sampling random values for the Wilson coefficients
and calculating $\mathrm{R}$ at LO and NLO produces results that are substantially different.
This demonstrates how the LO and NLO procedures inherently allow for and lead to different 
inferences on the Wilson coefficients.
\paragraph{Moving from kappa fits to PO, SMEFT fits}.

In LEP times, the experimental collaborations followed a complete path, \ie 
primordial quantities (PQs) $\,\to\,$ pseudo-observables (POs) $\,\to\,$ couplings. 
Primordial quantities are obtained through many assumptions, event classification, detector 
response, \etc 
On the observability of POs we can say that all quantities are equally ``observable'' provided 
one endorses the conceptual description of the phenomena that they are supposed to quantify.

It is foreseeable that ATLAS and CMS, given the more complex environment at LHC, will only do 
PQs $\,\to\,$ ``something'' and the big decision will be about ``something'' and ``how''. Let's 
assume that the analysis will concentrate on extracting Wilson coefficients ($\mrW$), as done 
in Run~1 for PQs $\,\to\,$ kappa ratios. 
Let's examine some of the consequences: assuming a (possible) experimental choice given by 
PQs $\,\to\,\mrW_{\myLO}$, we will miss the intermediate step PQs $\,\to\,$ POs $\,\to\,\mrW\,$;
therefore we will not be able to undo the analysis, \eg to derive NLO Wilson coefficients from 
LHC data.
%
%
Furthermore, the extraction at LO will be based on some ``LO'' procedure and little 
additional knowledge will be available because everything will be hard-coded somewhere inside 
some code. Once again, for this reason, nobody will be able to unfold the $\mrW_{\myLO}$,
\ie $\mrW_{\myLO}\,\to\,$ POs $\,\to\,$ something else. For this reason we want to emphasize that
\bei

\item[a)] PQs $\,\to\,$ POs is crucial while PQs $\,\to\,\mrW\,$ should be viewed as subsidiary, 

\item[b)] PQs $\to\,\mrW_{\myLO}\,$, if done directly, should be done with a 
general/consistent procedure, \ie  any LO procedure used should be extendible to NLO. 
Therefore the use of ``LO'' should be restricted to the lowest-order prediction of a, fully 
consistent, theory and not used as a synonym, essentially meaning ``phenomenological approach''.

\eei
The advantages in upgrading the (original) kappa fits are obvious: on the theory side the 
extension allows us to systematically include loop corrections, not only in QCD; on the 
experimental side it becomes possible to describe modified kinematic distributions, like for 
example the transverse momentum of the Higgs. Finally, on the theory and experimental sides, 
it allows us to combine measurements in the Higgs sector with any search for physics beyond the SM.
\subsection{Motivations for an EFT approach}
The lesson of the experiments from $1973$ to today is that it is extremely difficult to find 
a flaw in the SM, thus maybe the SM includes elements of a truly fundamental theory. But then 
how can one hope to make theoretical progress without experimental guidance? One should pay 
close attention to what we do not understand precisely about the SM even if the standard 
prejudice is that it is a hard technical problem, and solving it won't change anything.
The conventional vision is that some very different physics occurs at the Planck scale and
the SM is just an effective field theory. 
It is possible that at some very large energy scale, all non-renormalisable interactions 
disappear. This seems unlikely, given the difficulty with gravity. It is possible 
that the rules change drastically. It may even be possible that there is no end, simply 
more and more scales. 
This prompts the important question whether there is a last fundamental theory in this tower 
of EFTs which supersede each other as energy rises. Some people conjecture that this 
deeper theory could be a string theory, \ie a theory which is not a field theory any more. 
Or should one ultimately expect from physics theories that they are only valid as approximations 
and in a limited domain (see \Bref{Hartmann2001})?
Alternatively, the approach could be that one should not resort to arguments involving gravity. 
When looking for UV completions of the SM the following remarks are relevant:
there are several spin $1/2$ and spin $1$ degrees of freedom, but only one spin $0$. If there are more, the
present experimental evidence requires a hierarchy of VEVs which, once again, is a serious 
fine-tuning problem. Why are all mixings small? Is it accidental or systematic (\ie a new 
symmetry)? The real problem when dealing with UV completions is that one model is falsifiable, 
but an endless stream of them is not.

Here, we return to the ``more and more scales'' scenario. Even in this case it would be an error 
to believe that rigour is the enemy of simplicity. On the contrary, we find it confirmed by 
numerous examples that the rigorous method is at the same time simpler and more easily 
comprehended. 
We need a consistent theoretical framework in which deviations from the SM (or NextSM) predictions 
can be calculated (on average, for every $20$ bogus hypotheses you test, one of them will give you a $p$ 
of $< 0.05$). Such a framework should be applicable to comprehensively describe measurements 
in all sectors of particle physics: LHC Higgs measurements, past EWPD, \etc

In this section we revised a theory defined to be the SM augmented with the inclusion of higher 
dimensional operators ($\mrdim > 4$); this theory (let's call it Th$1$) is not strictly 
renormalisable since, order by order, the number of ultraviolet (UV) divergences increases even if, 
order by order, they can be removed.
Th$1$ is standalone and makes predictions, not only at lowest order (LO).
Although workable to all orders, Th$1$ fails above a certain scale, $\Lambda_1$, because a) the 
perturbative expansion, in $E/\Lambda$, breaks down and b) unitarity is violated. Above
$\Lambda_1$ we are forced to abandon Th$1$.

Consider any BSM model that is strictly renormalisable and respects unitarity and
call it Th$2$; its parameters can be fixed by comparison with data, while masses of heavy states 
are presently unknown. Th$1$ and Th$2$ differ in the UV but must have the same infrared (IR)
behaviour.

Consider now the whole set of data below $\Lambda_1$. Th$1$ should be able to explain them by 
fitting Wilson coefficients, Th$2$ should be able to explain the data adjusting the masses of heavy states 
(as SM did with the Higgs mass at LEP). Goodness of both explanations is
crucial in understanding how well Th$1$ and Th$2$ match and how reasonable it is to use Th$1$ instead 
of the full Th$2$, see \Bref{Brehmer:2015rna}. It is worth noting that, to a large extent Th$2$ 
is ``suggested'' by the comparison of Th$1$ and data. 

The next question is: does Th$2$ explain all observations? Possibly not, but it should be able to 
explain something more than Th$1$. In any case, both Th$1$ and Th$2$ should be formulated at
least at next-to-leading order (NLO) with UV divergences removed and, whenever possible, with
Lagrangian parameters traded for experimental data.

We could now define Th$3$ as Th$2$ augmented with its own higher dimensional operators, and 
valid up to a scale $\Lambda_2$.


The construction of the SMEFT, to all orders, is not based on assumptions on the size of the 
Wilson coefficients of the higher dimensional operators; restricting to a particular UV case 
is not an integral part of a general SMEFT treatment and various cases can be considered after 
the general calculation is performed. 
Of course, if the value of Wilson coefficients in broad UV scenarios could be inferred in 
general this would be of significant scientific value.

To summarise: constructing SMEFT is based on the fact that experiments occur at finite energy 
and ``measure'' an effective action $\mrS^{\mathrm{eff}}(\Lambda)$;
whatever QFT should give low energy
$\mrS^{\mathrm{eff}}(\Lambda)\,,\;\forall\,\Lambda < \infty$.
One also assumes that there is no fundamental scale above which $\mrS^{\mathrm{eff}(\Lambda)}$ 
is not defined~\cite{Costello2011} and $\mrS^{\mathrm{eff}}(\Lambda)$ loses its predictive power 
if a process at $E = \Lambda$ requires $\infty$ renormalised parameters~\cite{Preskill:1990fr}.

Extending the kappa framework should be seen as expressing the $\upkappa$-parameters in 
terms of coefficients in the SM effective field theory (SMEFT), remembering that SMEFT 
converges to SM in the limit of zero Wilson coefficients, but SMEFT and SM are different 
theories in the UV. 

\subsection{Theoretical uncertainties \label{THU}}
There is now an overall consensus on having a ``truncation'' error in SMEFT, and the recommendation
is to quote it separately, as for example is common practice among experimentalists for different 
kind of systematic or statistical uncertainties. The reason for doing that is that this error 
is strongly model dependent, differently from other ones. 

Let us refer to the case when we observe some deviation from the SM. Then, if 
the experimental precision allows, we could be able to test loop corrections and/or 
$\mrdim = 8$ effects.
In this case we can say that there is already strong evidence that NLO SMEFT provides 
non-negligible corrections, which are relevant for per-mille/few percent constraints. 
As it is always the case, if one works at LO and a tiny deviation is found, this deviation could 
be due to new physics or to missing higher orders.
For a given observable one can compute the deviation and the corresponding probability 
distribution function (pdf) with the result that the LO pdf differs from the NLO pdf at the 
level required by the projected precision.

Of course, one should also consider the case where no deviation from the SM is observed and 
limits on the $\mrdim = 6$ coefficients are set through a LO procedure. Also in this case loops 
should not be neglected if one wants to constrain a specific UV model, or if experimental data 
at widely different scales are to be combined into a single fit.

The SMEFT framework is useful because one can set limits on the effective coefficients in 
a model-independent way~\cite{Berthier:2015oma}. This is why SMEFT in the bottom-up approach 
is so useful: we do not know what the tower of UV completions is (or if it exists at all) but 
we can formulate the SMEFT and perform calculations with it without needing to know what happens 
at arbitrarily high scales.
%
%
On the other hand, in the top-down approach (Sect.~\ref{tda}), interpreting such limits as 
bounds on UV models does require some assumption of the UV dynamics~\cite{Contino:2016jqw}. 

Unless we start observing deviations, any phenomenological ``$\mrdim = 6$'' approach 
is a reasonable starting point to describe limits on SM deviations. While this outcome is much 
less desirable than dealing with a consistent SMEFT it is important to recognize that the 
difference relates to the possibility of including theory uncertainties. 
SMEFT can be used in LHC Higgs physics as long as we do not see new resonances. 
It allows us to describe kinematic distributions and to include EW corrections. 
An important question to answer is on theory uncertainties when linking SMEFT to full, (weakly 
interacting) BSM models.

%% file: HC_S8.tex
From the extraordinarily successful and well-established kappa-framework we know that measurements
of Higgs couplings at the level of several percent can be expected from the upcoming LHC run(s).
Towards higher luminosity the Higgs couplings to weak bosons will likely be the best-measured
also because the theoretical uncertainties linked to the corresponding LHC production
cross sections are well under control.

%% file: HC_run2_exp.tex

The increase in LHC beam energy from $8\UTeV$ to $13\UTeV$ (or more) will have a significant 
impact on the physics reach of ATLAS and CMS. In addition, the experiments will be accumulating 
$10$ or $100$ times more data. 
A multi-TeV particle produced via gluon fusion will see an increase in
the parton luminosity by one or two orders of magnitude relative to $7\UTeV$ collisions. The
jump in energy will  enable precision measurements of Higgs boson properties and SM
processes that will either help to elucidate the nature of the newly discovered particle or exclude
a large set of possible alternatives to the standard model.

Precise performance assessments at $13\UTeV$ center of mass energy and with the new upgraded 
detector (or with the envisaged upgrade that will happen in the future shutdown) do not exist.  
Only extrapolations from $7$ and $8\UTeV$ results with the detector and analyses
for the Run~1 data are available. 

In this Section only the results from CMS will be shown~\cite{CMS:2013xfa}. The CMS results 
are extrapolated to larger datasets of $300$ and $3000$ $fb^{-1}$ and a center-of-mass energy 
of $14\UTeV$ by scaling signal and background event yields accordingly. In order to study the 
precision of future measurements, a number of assumptions are made. As stated before, the 
extrapolations are done not considering the future detector upgrade, but  
assuming the same level of detector and trigger performances achieved with the detector in the 
$2012$ data-taking period. 
Extrapolations are presented for two uncertainty scenarios.
In Scenario 1, all systematic uncertainties are left unchanged. In Scenario 2, the theoretical
uncertainties are scaled by a factor of $1/2$, while other systematic uncertainties are scaled by
the inverse square root of the integrated luminosity. The comparison of the two uncertainty 
scenarios indicates a range of possible future measurements. The extrapolation without 
theoretical uncertainties is also presented as Scenario 3, to illustrate the importance of 
reducing those uncertainties in the future. Systematic uncertainties are inputs to the fits. 
They can be further constraint by the data when extracting the signal strength, coupling 
modifier or ratios of such. Similar extrapolations have been discussed in \Bref{CMS:2013xfa}.

\begin{figure}[hbt]
\centering
\includegraphics[width=0.7\textwidth]{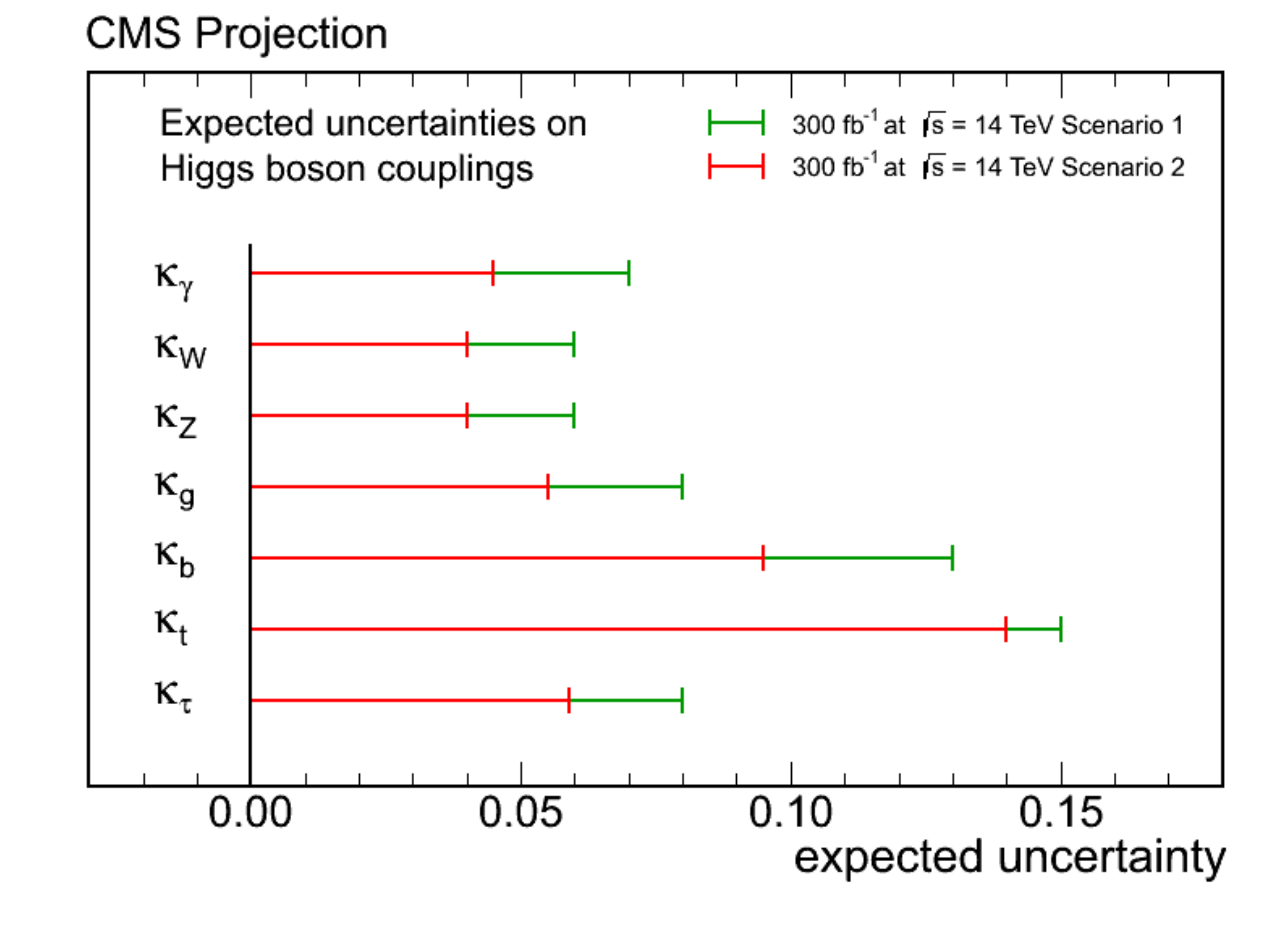}
\includegraphics[width=0.7\textwidth]{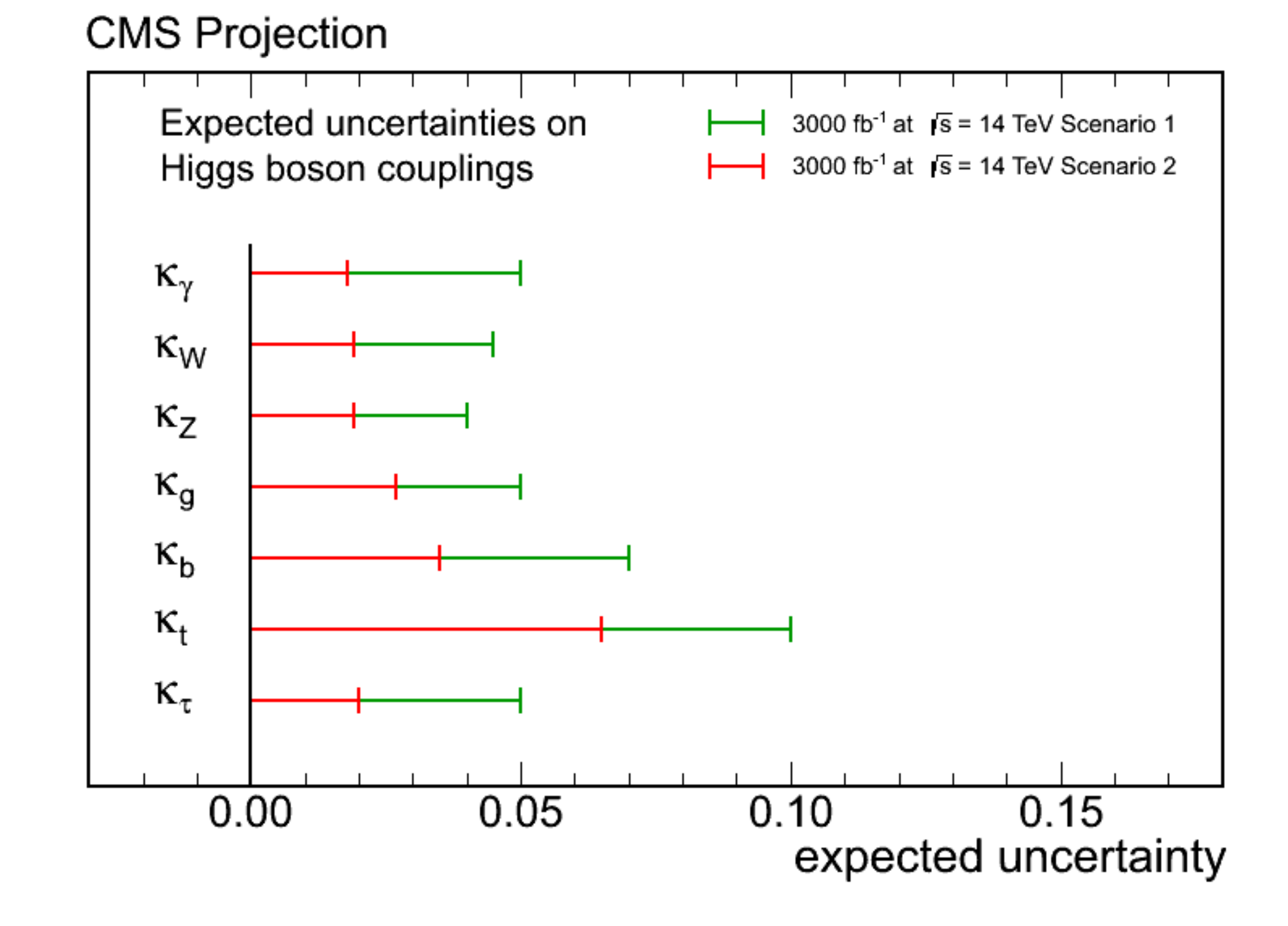}
\includegraphics[width=0.7\textwidth]{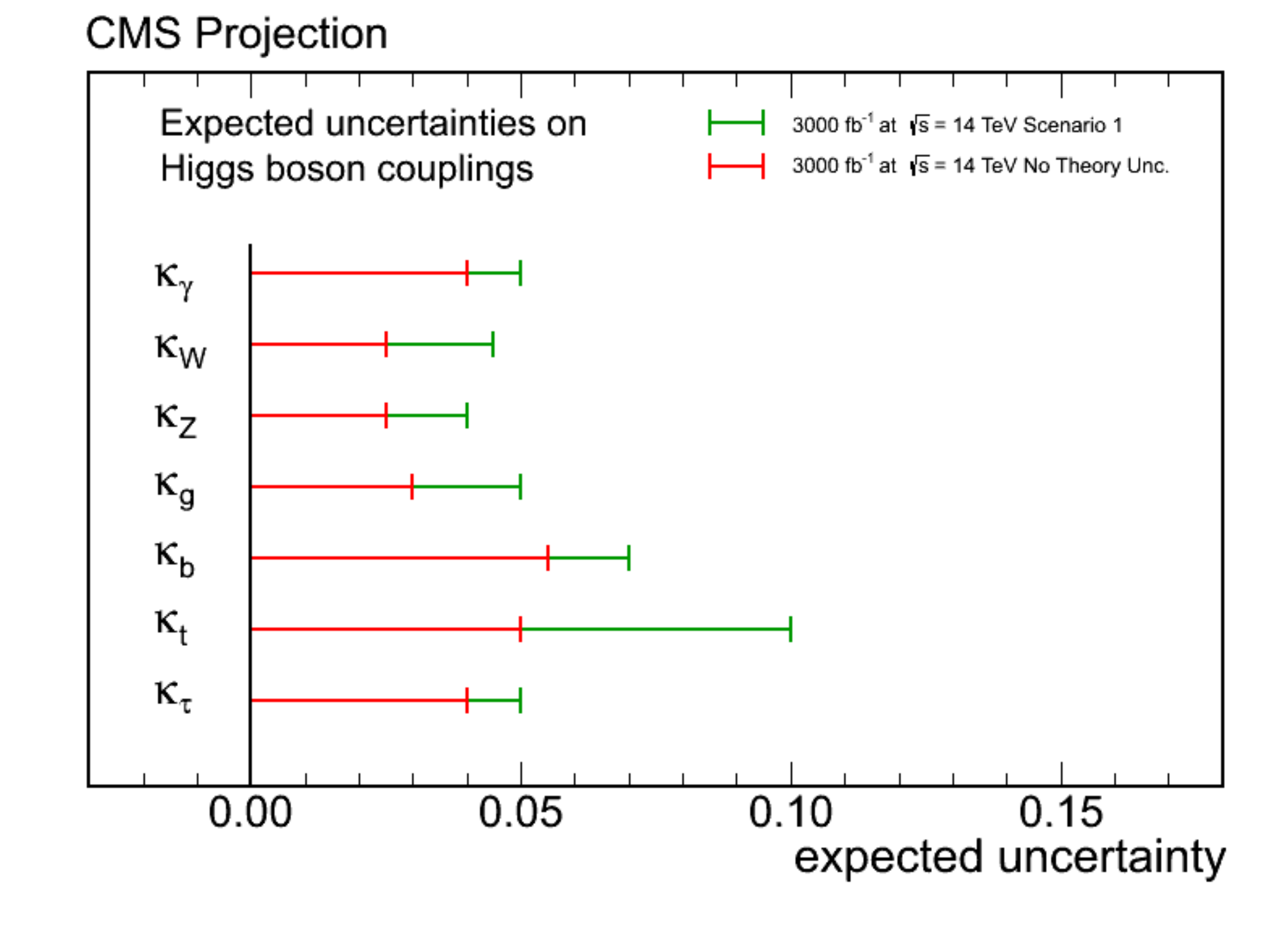}
\caption[]{Estimated precision on the measurements of the signal strength for a SM-like Higgs
boson. The projections assume $\sqrt{s} = 14 \UTeV$ and an integrated luminosity of 
$300$ fb$^{-1}$ (top), $3000$ fb$^{-1}$ (middle) and assuming no theoretical uncertainty 
(bottom).The projections are obtained with the three uncertainty scenarios described in the text.}
\label{fig:CMSextrapolations} 
\end{figure}

The estimated precision on the measurements of the signal strength for a SM-like Higgs boson 
is presented in Fig.~\ref{fig:CMSextrapolations}. 
The projections assume $\sqrt{s} = 14 \UTeV$ and an integrated luminosity of $300$ $fb^{-1}$ 
(top) and $3000$ $fb^{-1}$ (middle and bottom), with and without the theoretical uncertainty. 
The projections are obtained with the three uncertainty scenarios described above.

The huge improvements that the detector upgrades will provide in terms of detection efficiency 
and resolution (efficiencies could easily double for some of the processes and resolution will 
be improved considerably~\cite{CMS:2013xfa}), the new methods to control the background and 
analyse the data that will be introduced, and the addition of more Higgs channels in the 
extraction of the couplings. All these will contribute to reach a much smaller uncertainty on 
the Higgs couplings. 
We think that at least a factor of $2$ could be gained from the one presented in this section.
Moreover, an additional improvement will come from the combination of the data from ATLAS and 
CMS experiments, as it has been demonstrated in the first part of this report for the 
Run~1 analysis.

%% file: HC_conc.tex
Understanding the Higgs properties is a pillar of the present paradigm. In this review we
discussed the frameworks adopted during Run~1 at LHC as well as the experimental results.
Understanding that the main accent should be put on observables (\ie quantities related
to an $\mrS\,$-matrix) is important, and mapping those observables to a Lagrangian is a truly 
subtle affair,  that we must understand and that cannot really be demoted.

The Higgs boson production and decay rates measured by ATLAS and CMS have been combined within 
the context of different generic parameterisations, \eg one based on ratios of coupling modifiers. 
The data are consistent with the Standard Model predictions for all parameterisations considered.

Are we expecting new revolutions and do we want them out as soon as possible?
This attitude to physics is associated with the modern era and it is not new~\footnote{See
http://motls.blogspot.it/2016/08/modern-obsession-with-permanent.html}.
Most approaches are subjective, model-dependent, and uncertain. One simply cannot promote 
them to something that will reliably lead us to the next discovery in particle physics.

The desire to ``abandon all the existing knowledge'' is in no way new. It was surely common among 
physicists before the war, see \Bref{Gross:1994zu}. ``The task is, not so much to see what no one has 
yet seen; but to think what nobody has yet thought, about that which everybody sees'' (E. Schroedinger).
In our view the problem is not how to imagine wild scenarios, but rather how to arrive
at the correct scenario by making only small steps, without having to make unreasonable
assumptions (see Gerard 't Hooft's contribution in \Bref{GBU}). We have the Standard Model of 
particle physics with coupling strengths that we do not know how to derive, but which can be 
measured accurately. 

In this report  we have discussed not only the current framework for Run~1 studies,  but we have also reviewed a study
of potential generalisations, such as the SMEFT, with the belief that when it comes to the choice of the 
dynamics, quantum field theory will be the essential framework for effective theories.